\documentclass[12pt]{article}  
\usepackage{a41}
\usepackage{cite}
\usepackage{amsmath}               
\usepackage{amssymb}
\usepackage{placeins}
\usepackage{color}  

\usepackage[rflt]{floatflt}              
\usepackage{float}
\usepackage{slashed}

\def\b0{\beta_0}

\usepackage{array}

\newcommand{\HA}{{\rm H}}
\newcommand{\KA}{{\rm K}}

\usepackage[english]{babel}
\usepackage[latin1]{inputenc}
\usepackage[T1]{fontenc}
\usepackage{ae}

\usepackage{url}

\usepackage{amsmath, amsthm, amssymb}
\newtheorem{thm}{Theorem}[section]

\newtheorem{definition}[thm]{Definition}

 \newcommand{\GeV}{\mathrm{GeV}}

\newcommand{\Li}{{\rm Li}}

\newcommand{\Mvec}{{\rm\bf M}}

\newcommand{\ep}{\varepsilon}

\usepackage{rotating}

\usepackage{graphicx}

\newcounter{mmacnt}
\def\restartmma{\setcounter{mmacnt}{0}}
\restartmma \catcode`|=\active
\def|#1|{\mathrm{#1}}
\catcode`|=12
\newenvironment{mma}{
 \par\smallskip
 \catcode`|=\active
 \parskip=0pt\parindent=0pt 
 \small
 \def\In##1\\{%
\def\linebreak{\hfill\break\null\qquad}%
\refstepcounter{mmacnt}
\hangindent=2.5em\hangafter=0
\leavevmode
\llap{\tiny\sffamily n[\arabic{mmacnt}]:=\kern.5em}%
\mathversion{bold}\footnotesize$\displaystyle##1$\normalsize
\mathversion{normal}\par
 }%
 \def\Print##1\\{%
\def\linebreak{\hfill\break}%
\hangindent=2.5em\hangafter=0
\leavevmode ##1\par}%
 \def\Out##1\\{%
\def\linebreak{$\hfill\break\null\hfill$}%
\kern\abovedisplayskip\par
\hangindent=2.5em\hangafter=0
\leavevmode
\llap{\tiny\sffamily Out[\arabic{mmacnt}]=\kern.5em}
\footnotesize$\displaystyle##1$\normalsize\hfill\null\par
\kern\belowdisplayskip
 }%
 \def\Warning##1##2\\{%
\def\linebreak{\hfill\break}%
\hangindent=2.5em\hangafter=0
\leavevmode
{\scriptsize##1 : ##2}\par}%
}{%
 \par\smallskip
}

\usepackage{color}

\newenvironment{fshaded}{%
\MakeFramed {\FrameRestore}
}%
{\endMakeFramed}

\def\b0{\beta_0}

\def\Gp0{{\Gamma^{'}_0}}
\def\Gp1{{\Gamma^{'}_1}}
\def\Gp2{{\Gamma^{'}_2}}

\allowdisplaybreaks[4]

\begin{document}
\setlength{\baselineskip}{0.515cm}

\sloppy
\thispagestyle{empty}
\begin{flushleft}
DESY 19--038
\\
DO--TH 18/25\\
\end{flushleft}

\mbox{}
\vspace*{\fill}
\begin{center}

{\LARGE\bf The unpolarized two-loop massive}

\vspace*{3mm}
{\LARGE\bf pure singlet Wilson coefficients}

\vspace*{3mm}
{\LARGE\bf for deep-inelastic scattering}

\vspace{3cm}
\large
{\large 
J.~Bl\"umlein$^a$, 
A.~De Freitas$^a$,
C.G.~Raab$^b$,
and
K.~Sch\"onwald$^{a}$
}

\vspace{1.cm}
\normalsize
{\it   $^a$~Deutsches Elektronen--Synchrotron, DESY,}\\
{\it   Platanenallee 6, D--15738 Zeuthen, Germany}

\vspace*{2mm}
{\it  $^b$~Johannes Kepler Universit\"at Linz,}\\
{\it Altenberger Stra\ss{}e 69, A--4040 Linz, Austria}


\end{center}
\normalsize
\vspace{\fill}
\begin{abstract}
\noindent
We calculate the massive two--loop pure singlet Wilson coefficients for heavy quark production in the unpolarized 
case analytically in the whole kinematic region and derive the threshold and asymptotic expansions. We also recalculate
the corresponding massless two--loop Wilson coefficients. The complete expressions contain iterated integrals with elliptic
letters. The contributing alphabets enlarge the Kummer-Poincar\'e letters by a series of square-root valued letters. A
new class of iterated integrals, the Kummer-elliptic integrals, are introduced. For the structure functions $F_2$ 
and $F_L$ we also derive improved asymptotic representations adding power corrections. Numerical results are presented.
\end{abstract}

\vspace*{\fill}
\noindent
\newpage 

\section{Introduction}
\label{sec:1}

\vspace*{1mm}
\noindent
The complete massive two--loop Wilson coefficients for deep--inelastic scattering corresponding to the structure 
functions $F_2(x,Q^2)$ and $F_L(x,Q^2)$ were only available in numerical form 
\cite{Laenen:1992zk,Laenen:1992xs,Riemersma:1994hv}\footnote{Numerical results were also presented in 
\cite{Hekhorn:2018ywm}.} for a long time. Later the flavor non-singlet Wilson 
coefficients have been calculated analytically in \cite{Buza:1995ie} in the tagged-flavor case and recalculated for 
the inclusive case \cite{Blumlein:2016xcy} to obtain a representation consistent with the associated sum rules.

In the present paper we calculate the massive pure singlet two--loop Wilson coefficients analytically. Due to the
corresponding graphs, the formulae are structurally the same for the charm and the bottom contributions. In the
numerical illustrations we will concentrate on the charm contributions, considering the first three quarks as massless.
The knowledge of the complete analytic expressions allows to derive important limiting cases such as the limit
of large virtualities $Q^2 \gg m^2$, $m$ being the heavy quark mass, or the threshold expansion in a direct way. In the 
former case it is
possible to derive systematic expansions in $m^2/Q^2$ with coefficients represented in terms of harmonic 
polylogarithms, while the complete result depends on much more general functions. Harmonic polylogarithms can 
be easily calculated numerically \cite{Gehrmann:2001pz,Maitre:2005uu,Ablinger:2018sat}. Furthermore, they can be 
directly transformed to Mellin space \cite{Vermaseren:1998uu,Blumlein:1998if}. It has been observed numerically
in Ref.~\cite{Buza:1995ie} that the limit of large virtualities is approached beyond some process-dependent scale
$Q_0^2$. The Wilson coefficient in this limit can be calculated with the help of massive operator matrix 
elements (OMEs) and massless Wilson coefficients, cf. \cite{Buza:1995ie}. It is important to prove this analytically.
At three-loop order the massive Wilson coefficients are only known in the asymptotic region \cite{Blumlein:2006mh,
Bierenbaum:2009mv,Ablinger:2010ty,Ablinger:2014vwa,Ablinger:2014nga,Ablinger:2014lka,AGG,Behring:2014eya,
Ablinger:2017err,Ablinger:2017xml,Ablinger:2017ptf,Ablinger:2018brx}. We also recalculate the 
corresponding massless two--loop Wilson coefficients given in \cite{Kazakov:1987jk,Kazakov:1990fu,
SanchezGuillen:1990iq,vanNeerven:1991nn,Zijlstra:1992kj,Larin:1991fv,Moch:1999eb,Vermaseren:2005qc} before and compare 
to these results. 

The analytic calculation of the massive pure singlet Wilson coefficient has been envisaged by W.L. van Neerven 
and one of the authors (J.B.) 20 years ago, after the non-singlet contribution had been obtained in 
\cite{Buza:1995ie}. In retrospect, however, adequate mathematical techniques to perform this task have only
become available very recently. This includes the elimination of all functional relations in the final result 
and techniques to obtain a compact representation. The massive Wilson coefficient is given by a four-fold non-trivial 
phase space integral. Three of the integrals can be carried out using standard techniques. The integrand of the last 
integral is obtained as a polynomial of rational terms, logarithms  and polylogarithms \cite{DUDE,LEWIN} with an involved 
argument 
structure. Therefore, the last integral is performed after determining the contributing irreducible structure of letters 
of the contributing iterated integrals, using the techniques described in \cite{RaabRegensburger,Ablinger:2014bra}.
The Wilson coefficient can 
finally be obtained as a d'Alembertian integral over a finite alphabet. The analytic results allow to perform 
expansions in $m^2/Q^2$ including power corrections, which is of particular importance for the structure function
$F_L(x,Q^2)$. Here the corresponding expansion coefficients are then harmonic polylogarithms. Such a representation 
is easily envisaged for the two--loop non-singlet Wilson coefficients given in \cite{Buza:1995ie,Blumlein:2016xcy}, 
since 
there the whole Wilson coefficient depends at most on classical polylogarithms.

We also consider the limit $Q^2 \gg m^2$ of the Wilson coefficient and compare with the results given in 
Refs.~\cite{Buza:1995ie,Behring:2014eya,Bierenbaum:2007qe}. Furthermore, the threshold expansion of the 
Wilson coefficients are derived and numerical results are presented. In the present calculations, the packages
{\tt FORM} \cite{FORM}, {\tt Sigma} \cite{Schneider:2007a,Schneider:2013a}, {\tt 
EvaluateMultiSums}~\cite{Ablinger:2010pb,Schneider:2013zna} and {\tt HarmonicSums} 
\cite{Vermaseren:1998uu,Blumlein:1998if,Ablinger:2014rba,Ablinger:2010kw,Ablinger:2013hcp,Ablinger:2011te, 
Ablinger:2013cf,Ablinger:2014bra,Ablinger:2017Mellin} have been used.

The paper is organized as follows. In Section~\ref{sec:2} we first illustrate the asymptotic factorization
using the example of the $O(\alpha_s)$ calculation. The corresponding scattering cross sections will be used
in the two--loop massless and  massive calculation later. In Section~\ref{sec:3} the massless two--loop pure 
singlet Wilson coefficients are calculated. The mathematical method used to prepare for the last analytic integral 
in the massive case is described in Section~\ref{sec:4} and in Section~\ref{sec:5} we present the analytic results
for the massive Wilson coefficients. The asymptotic and threshold expansions are derived in Section~\ref{sec:6} and 
numerical results are presented in Section~\ref{sec:7}. Section~\ref{sec:8} contains the conclusions. Some technical 
aspects of the calculation  are given in the Appendix. 
\section{Asymptotic cross section factorization}
\label{sec:2}

\begin{figure}[H]
\centering
\includegraphics[width=0.6\textwidth]{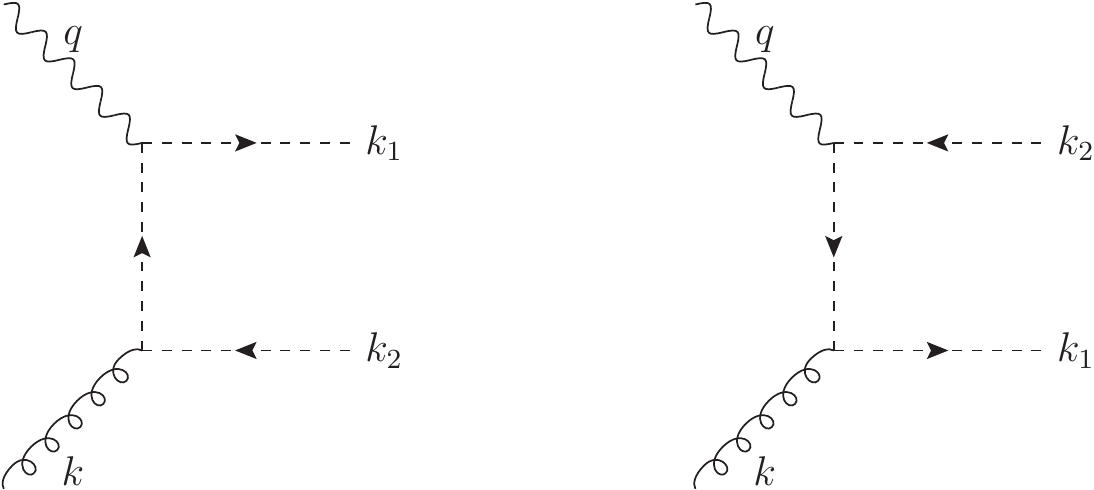}
\caption{\small \sf Diagrams of the $O(a_s)$ contributions to scattering cross section 
$\gamma^* +g \to q + \overline{q}$.}
\label{DIA1}
\end{figure}

\vspace*{1mm}
\noindent
The massive Wilson coefficients are calculated by factorizing the {\it massless} initial states (quarks and
gluons). In the unpolarized case and for longitudinal polarization the factorization is longitudinal, i.e.  by setting
$p = zP, z \in [0,1]$. Here $P$ denotes the incoming hadron momentum and $p$ the quark momentum.
In the transversal polarized case one has to use the covariant parton model \cite{Landshoff:1971xb}, 
see \cite{Jackson:1989ph,Roberts:1996ub,Blumlein:1996tp,Blumlein:2003wk}.
As an illustrative example we consider the unpolarized one--loop heavy flavor contribution to deep--inelastic scattering
\cite{Witten:1975bh,Babcock:1977fi,Shifman:1977yb,Leveille:1978px,Gluck:1980cp}. As for all the massive Wilson 
coefficients, it can be written in three parts: the massive operator matrix
element, the massless Wilson coefficient and a remainder part. The last one vanishes in the limit $Q^2/m^2 
\rightarrow \infty$ in the case of {\it asymptotic factorization}. A simple prediction on the structure of this 
term is not easily possible, but usually requires the calculation of the whole process followed by the expansion in 
$m^2/Q^2$. This term depends on the structure of the phase space and it is a process-dependent quantity. 
In Figure~\ref{DIA1} the contributing Feynman diagrams are shown.
The massive Wilson coefficients have the following series representation
\begin{eqnarray}
\label{eq:Hprinc}
H_{2(L),i}\left(z,\frac{Q^2}{\mu^2},\frac{m^2}{\mu^2} \right) = \sum_{k=1}^\infty a_s^k 
H_{2(L),i}^{(k)}\left(z,\frac{Q^2}{\mu^2},\frac{m^2}{\mu^2} \right),
\end{eqnarray}
where $i$ denotes the incoming parton and $2(L)$ refer to the associated structure functions  and $a_s \equiv 
a_s(\mu_R) =  g_s^2/(4\pi)^2$ denotes the strong coupling constant at the renormalization scale $\mu_R$.
We work in $d = 4 + \ep$  space-time dimensions.
Since we also need the $O(\ep)$ term of the LO result later on, we further define
\begin{eqnarray}
H_{2(L),i}^{(1)}\left(z,\frac{Q^2}{\mu^2},\frac{m^2}{\mu^2} \right) &=& h_{2(L),i}^{(1)} + \ep \bar{b}_{2(L),i}^{(1)},
\end{eqnarray}
where we dropped the arguments of the coefficient functions for brevity.

Let us consider the leading order contribution for the process $\gamma^* + g \rightarrow Q \overline{Q}$ as an 
example, cf.~\cite{Witten:1975bh,Babcock:1977fi,Shifman:1977yb,Leveille:1978px,Gluck:1980cp}. In the following 
we use the variable 
\begin{eqnarray}
\beta = \sqrt{1 - \frac{4m^2}{Q^2} \frac{z}{1-z}}.
\end{eqnarray}
The Wilson coefficients $H_{L,g}^{(1)}$ and $H_{2,g}^{(1)}$ are given by
\begin{eqnarray}
h_{L,g}^{(1)}\left(z,\frac{Q^2}{m^2}\right) &=& 16 T_F\left[\beta z(1-z) + 2 \frac{m^2}{Q^2} z^2 \ln 
\left(\frac{1-\beta}{1+\beta}\right) \right] 
\theta\left(a -z \right),
\\
h_{2,g}^{(1)}\left(z,\frac{Q^2}{m^2}\right) &=& 8 T_F \Biggl\{\beta \left[-\frac{1}{2} + 4 z(1-z) - 2 
\frac{m^2}{Q^2} z(1-z) \right]
\nonumber\\ && + \left[-\frac{1}{2} + z - z^2 + 2 \frac{m^2}{Q^2} z (3z-1) + 4 \left(\frac{m^2}{Q^2}\right)^2 
z^2\right] \ln \left(\frac{1-\beta}{1+\beta}\right) \Biggr\}
\nonumber\\ &&
\times \theta\left(a -z \right),
\end{eqnarray}
with $\theta(x)$ the Heaviside function, $ a = 1/(1+4m^2/Q^2)$ and
$T_F = 1/2$ for $SU(N_C)$. The coefficients at $O(\ep)$ read
\begin{eqnarray}
\bar{b}_{L,g}^{(1)} &=& T_F z(1-z) \Biggl\{
2 (1- \beta^2) \left[\HA_0^2\left(
                \frac{1-\beta }{1+\beta }\right)
        -2
        \HA_0\left(
                \frac{1-\beta }{1+\beta }
        \right) \left[1 + \HA_0 + \HA_1 - 2 \HA_0(\beta)\right]\right]
\nonumber\\ &&
        -8 \Biggl[
                \beta  (3 + \HA_0 + \HA_1 - 2 \HA_0(\beta))
                +(1-\beta^2) \Biggl[ \HA_{0,1}\left(\frac{1-\beta}
{1+\beta}\right)
+ [\ln(2) + \HA_0(\beta)
\nonumber\\ &&
- \HA_{-1}(\beta)] \HA_0\left(\frac{1-\beta}{1+\beta}\right) -
\zeta_2\Biggr] \Biggr]
\Biggr\} \theta(a-z),
\\
\bar{b}_{2,g}^{(1)} &=& T_F \Biggl\{
        2 (1-z) (1-\beta^2 )\left[
                \beta ^2
                -z \big(
                        3+\beta ^2\big)
        \right] \HA_0\left(
                \frac{1-\beta }{1+\beta }\right)
        -\frac{1}{2} \HA_0^2\left(
                \frac{1-\beta }{1+\beta }
        \right)
\nonumber\\ &&
\times \left[3
                -\beta ^4
                -2 z \big(
                        5-2 \beta ^2-\beta ^4\big)
                +z^2 \big(
                        9-4 \beta ^2-\beta ^4\big)
        \right]
        +2
                \beta  [
                        5
                        -2 \beta ^2 
\nonumber\\ &&
                        +2 z^2 \big(
                                12-\beta ^2\big)
                        -2 z \big(
                                13-2 \beta ^2\big)
                ]
                -2\bigg[
                        3
                        -\beta ^4
                        -2 z \big(
                                5-2 \beta ^2-\beta ^4\big)
\nonumber\\ &&
                        +z^2 \big(
                                9-4 \beta ^2-\beta ^4\big)
                \bigg] \Biggl[
-\HA_{0,1}\left(\frac{1-\beta}{1+\beta}\right) - [\ln(2) + \HA_0(\beta) - \HA_0(1+ \beta)]
\HA_0\left(\frac{1-\beta}{1+\beta}\right) + \zeta_2
\Biggr]
\nonumber\\ &&
        +\Biggl[
                2 \beta \big(
                        2
                        -\beta ^2
                        +z^2 (9-\beta^2 )
                        -2 z \big(
                                5-\beta ^2\big)
                \big)
                +\Bigl[
                        3
                        -\beta ^4
                        -2 z \big(
                                5-2 \beta ^2-\beta ^4\big)
\nonumber \\ &&
 +z^2 \big(
                                9-4 \beta ^2-\beta ^4\big)
                \Bigr]
\HA_0 \left(
                        \frac{1-\beta }{1+\beta }\right)\Biggr]
       \left[\HA_1 + \HA_0 - 2 \HA_0(\beta)\right]
\Biggr\} \theta(a-z).
\end{eqnarray}
Here we refer to the harmonic polylogarithms \cite{Remiddi:1999ew} defined by
\begin{equation}
\HA_{b,\vec{a}}(z) = \int_0^z dy f_b(y) \HA_{\vec{a}}(y),~~\HA_\emptyset = 1,~~b, a_i \in \{-1,0,1\},
\end{equation}
and the letters $f_c$ are
\begin{equation}\label{eq:HPL1}
f_0(z) = \frac{1}{z},~~~~
f_1(z) = \frac{1}{1-z},~~~~
f_{-1}(z) = \frac{1}{1+z}.
\end{equation}
Here and in the following we use the abbreviation $\HA_{\vec{a}}(z) \equiv \HA_{\vec{a}}$.

The expansion for large virtualities $Q^2 \gg m^2$ is given by
\begin{eqnarray}
H_{L,g}^{(1)}\left(z,\frac{Q^2}{m^2}\right) &=& 16 T_F \Biggl\{z(1-z) - 2 \frac{m^2}{Q^2} z^2 
\left[\ln\left(\frac{Q^2}{m^2}\right) 
+ 1 - \HA_1 -\HA_0\right]
+O\left(\left(\frac{m^2}{Q^2}\right)^2\right),
\nonumber\\
\\
H_{2,g}^{(1)}\left(z,\frac{Q^2}{m^2}\right) &=& 
4 T_F \Biggl\{
-1 + 8 z(1-z)  + [z^2 + (1-z)^2] \left[\ln\left(\frac{Q^2}{m^2}\right) - \HA_1 - \HA_0\right] 
\nonumber\\ &&
+ 4 \frac{m^2}{Q^2} \Biggl[- z (1 + 2 z) + 
    (1 - 3 z) z \Biggl[\ln\left(\frac{Q^2}{m^2}\right) - \HA_1 - \HA_0\Biggr]\Biggr]
+O\left(\left(\frac{m^2}{Q^2}\right)^2\right)
\Biggr\}
\nonumber\\
\end{eqnarray}  
for $z \in [0,a]$. 

In the asymptotic case, one has \cite{Buza:1995ie}
\begin{eqnarray}
\label{eq:AS1}
H_{L,g}^{(1)}\left(z,\frac{Q^2}{m^2}\right) &=& \tilde{C}_{g,L}^{(1)}(N_F+1),
\\
\label{eq:AS2}
H_{2,g}^{(1)}\left(z,\frac{Q^2}{m^2}\right) &=&  A_{Qg}^{(1)}(N_F+1) + \tilde{C}_{g,2}^{(1)}(N_F+1),
\end{eqnarray}  
using the definition
\begin{eqnarray}
\tilde{f}(N_F) = \frac{f(N_F)}{N_F},~~~~~\hat{f}(N_F+1) = {f}(N_F+1) - {f}(N_F).
\end{eqnarray}  
Note that Eqs.~(\ref{eq:AS1}, \ref{eq:AS2}) hold for $z \in [0,1]$.
Here ${C}_{g,{2(L)}}^{(1)}$ denote the massless two--loop Wilson coefficients and $A_{Qg}^{(1)}$ the massive 
one--loop operator matrix element (OME) with external gluons \cite{Buza:1995ie,Behring:2014eya,Bierenbaum:2007qe}
\begin{eqnarray}
A_{Qg}^{(1)} = - 4 T_F [z^2 + (1-z)^2] \ln\left(\frac{m^2}{\mu^2}\right).
\end{eqnarray}  
The massless one--loop Wilson coefficients read \cite{Zee:1974du,Furmanski:1981cw,Zijlstra:1992qd}
\begin{eqnarray}
\label{eq:MASL1a}
\tilde{C}_{g,L}^{(1)} &=& 16 T_F z(1-z),
\\
\label{eq:MASL1b}
\tilde{C}_{g,2}^{(1)} &=& 4 T_F [z^2 +(1-z)^2] \ln\left(\frac{Q^2}{\mu^2}\right),
\nonumber\\ &&
+ 4 T_F \left\{-1 + 8 z(1-z)  - [z^2 +(1-z)^2] \left[\HA_1 + \HA_0\right]\right\}, 
\end{eqnarray}  
where
\begin{eqnarray}
\hat{P}_{qg}(z) = 8 T_F [z^2 +(1-z)^2]
\end{eqnarray}  
is a one--loop splitting function \cite{Gross:1974cs,Georgi:1951sr}.\footnote{For earlier
references in QED, see \cite{Blumlein:2012bf}.}

It can now be seen that the massive Wilson coefficients can be decomposed in terms of the part obtained at 
large
virtualities $Q^2 \gg m^2$, Eqs.~(\ref{eq:AS1},\ref{eq:AS2}), consisting of massive OMEs and 
massless Wilson coefficients, and a remainder part vanishing in the limit $Q^2/m^2 \rightarrow \infty$.  
Whenever this is the case one calls the respective process {\it asymptotically factorizing}. The factorization scale 
$\mu$ cancels in the cross sections (\ref{eq:AS1}, \ref{eq:AS2}) since they are free of collinear singularities.
As a peculiarity in this case, the massive OME only contributes to the pure logarithmic term. This, however, is due to
its vanishing constant part  and is generally not the case.

Numerically it is interesting to see from which value of $Q^2_0/m^2$ onward the asymptotic representation 
holds, say at the accuracy of $O(2\%)$ or better, cf.~\cite{Buza:1995ie,Blumlein:2016xcy} and Section~\ref{sec:7}.
\section{The massless Wilson coefficients}
\label{sec:3}

\vspace*{1mm}
\noindent
The massless pure singlet Wilson coefficients obey the expansion
\begin{eqnarray}
C_{2(L)}^{\rm PS}\left(z,\frac{Q^2}{\mu^2}\right) &=& \delta(1-z) \delta_{2} +
\sum_{k=1}^\infty a_s^k C_{2(L)}^{(k),\rm PS}\left(z,\frac{Q^2}{\mu^2}\right), 
\end{eqnarray}
with $\delta_2 = 1$ for $C_2$ and $\delta_2 = 0$ for $C_L$. Throughout this paper we will identify
the factorization scale $\mu_F$ and the renormalization scale $\mu_R$.

In the following we also recalculate the massless Wilson coefficients $C_L^{\rm PS,(2)}$ and $C_2^{\rm PS,(2)}$
as a limiting case of the present massive calculation. They have been computed in Refs.~\cite{Kazakov:1987jk,
Kazakov:1990fu,
SanchezGuillen:1990iq,vanNeerven:1991nn,Zijlstra:1992kj,Larin:1991fv,Moch:1999eb} before.

The unrenormalized Wilson coefficients ${\cal F}_{L(2),q}$ are related to the hadronic tensor of deeply inelastic 
scattering in the partonic sub-system, $\hat{W}_{\mu\nu}$, by
\begin{eqnarray}
{\cal F}_{L,q} &=& - \frac{2q^2}{(p.q)^2} p_\mu p_\nu \hat{W}_{\mu\nu},
\\
{\cal F}_{2,q} &=& - \frac{2}{d-2} \left[\hat{W}^{\mu}_{\mu} +(d-1) \frac{q^2}{(p.q)^2} p^\mu p^\nu \hat{W}_{\mu\nu} 
\right].
\end{eqnarray}
Here $p$ denotes the incoming parton momentum and $q$ the space-like
momentum of the virtual photon with $q^2 = - Q^2$.

In the massive case we will also consider the Wilson coefficient 
\begin{eqnarray}
\label{eq:F1}
        {\cal F}_{1,q} &=& - 2 \hat{W}^{\mu}_{\mu}
\end{eqnarray}
as a subsidiary function in order to avoid redundancies in the calculation. Note that this Wilson coefficient does
not correspond to the structure function $F_1$, cf.~\cite{Blumlein:2012bf}.

The following expressions will be given in Mellin-$N$ space.
They are obtained from the momentum fraction $z$-space by a Mellin transform
\begin{eqnarray}
\Mvec[f(z)](N) = \int_0^1 dz z^{N-1} f(z)~.
\end{eqnarray}
The unrenormalized Wilson coefficients ${\cal F}^{(2),\rm PS}_{L(2),q}$ are given by \cite{Zijlstra:1992qd}
\begin{eqnarray}
\label{eq:UNR1}
\hspace*{-5mm}
{\cal F}^{(2),\rm PS}_{L,q} &=& N_F \hat{a}_s^2 S_\varepsilon^2 \left(\frac{Q^2}{\mu^2}\right)^\ep \left[
\frac{1}{\ep} P_{gq}^{(0)} c_{L,g}^{(1)} + c_{L,q}^{(2),\rm PS} +  P_{gq}^{(0)} a_{L,g}^{(1)} \right],
\\
\label{eq:UNR2}
\hspace*{-5mm}
{\cal F}^{(2),\rm PS}_{2,q} &=& N_F \hat{a}_s^2 S_\varepsilon^2 \left(\frac{Q^2}{\mu^2}\right)^\ep
\Biggl[\frac{1}{\ep^2} \frac{1}{2} P_{qg}^{(0)} P_{gq}^{(0)} + \frac{1}{\ep}\left( 
\frac{1}{2} P_{qq}^{(1),\rm PS} + P_{gq}^{(0)} 
c_{2,g}^{(1)}\right) 
+ c_{2,q}^{(2),\rm PS} + P_{gq}^{(0)} a_{2,g}^{(1)} \Biggr],
\end{eqnarray}
with $\hat{a}_s$ the unrenormalized coupling constant, the spherical factor
\begin{eqnarray}
S_\ep = \exp\left[\tfrac{\ep}{2}\left(\gamma_E - \ln(4\pi)\right) \right], 
\end{eqnarray}
and $\gamma_E$ the Euler--Mascheroni constant. We work in the $\overline{\sf MS}$-scheme and set $S_\ep = 1$ at the
end of the calculation.
Here the factors of $1/2$ in Eq.~(\ref{eq:UNR2}) emerge since for the splitting into the upper 
quark-antiquark pair, the quarks are produced correlated. Since the pure singlet contributions start 
at $O(a_s^2)$ only, the renormalized Wilson coefficients $C_{L,(2)}^{(2),\rm PS}$ are obtained after mass 
factorization
\begin{eqnarray}
{\cal F}^{(2),\rm PS}_{L,q} &=& C_{L,q}^{(2),\rm PS} + \Gamma_{gq}^{(0)} C_{L,q}^{(2),\rm PS},
\\
{\cal F}^{(2),\rm PS}_{2,q} &=& C_{2,q}^{(2),\rm PS}
+ \frac{1}{2} \Gamma_{qq}^{(1),\rm PS} C_{2,q}^{(2),\rm PS}
+\Gamma_{gq}^{(0)} C_{2,g}^{(1)},
\end{eqnarray}
with
\begin{eqnarray}
\Gamma_{gq}^{(0)} &=& \hat{a}_s S_\ep \left(\frac{\mu_F^2}{\mu^2}\right)^{\ep/2} \frac{1}{\ep} P_{gq}^{(0)},
\\
\Gamma_{qq}^{(1),\rm PS} &=& \hat{a}_s^2 S_\ep^2 \left(\frac{\mu_F^2}{\mu^2}\right)^{\ep}\left[
\frac{1}{\ep^2} P_{qg}^{(0)} P_{gq}^{(0)} + \frac{1}{\ep} P_{qq}^{(1),\rm PS}\right].
\end{eqnarray}

In $z$-space the functions in Eqs.~(\ref{eq:UNR1}, \ref{eq:UNR2}) read
\begin{eqnarray}
a^{(1)}_{L,g} &=& - 8 T_F z(1-z) \left[3+ \HA_1 + \HA_0\right],
\\
a^{(1)}_{2,g} &=& T_F \Bigl\{[z^2+(1-z)^2](\HA_1+\HA_0)^2 +2(1-8 z(1-z))(\HA_1+\HA_0)-3[z^2+(1-z)^2]\zeta_2
\nonumber\\ && +6 - 44z(1-z)\Bigr\}, 
\end{eqnarray}
see as well Eqs.~(\ref{eq:MASL1a}, \ref{eq:MASL1b}) for $\mu^2 = Q^2$.
The splitting functions are
\begin{eqnarray}
P_{qg}^{(0)} &=& N_F \hat{P}_{qg}^{(0)},
\\
P_{gq}^{(0)} &=& 4 C_F \frac{1+(1-z)^2}{z},
\\
P_{qq}^{(1), \rm PS} &=& 16 C_F T_F N_F \Biggl[
\frac{20}{9} \frac{1}{z} - 2 + 6z - 4\HA_0 +z^2 \left(\frac{8}{3} \HA_0 - \frac{56}{9}\right)
+(1+z)\left(5 \HA_0 - \HA_0^2\right)\Biggr].
\end{eqnarray}
The massless Wilson coefficients $C_L^{\rm PS,(2)}$ and $C_2^{\rm PS,(2)}$ are thus given by
\begin{eqnarray}
C_L^{\rm PS,(2)}\left(z,\frac{Q^2}{\mu_F^2}\right) &=& -32 C_F T_F N_F \Biggl\{
\left[z \HA_0 + \frac{1}{3}\left(3 - 2z^2 - \frac{1}{z}\right)\right] \ln\left(\frac{Q^2}{\mu_F^2}\right) 
\nonumber\\ &&
        \frac{(1-z) \big(
                1-2 z+10 z^2\big)}{9 z}
        - (1+z) (1-2 z) \HA_0
        - z \HA_0^2
\nonumber\\ &&
        +\frac{(1-z) \big(
                1-2 z-2 z^2\big)}{3 z} \HA_1
        - z \HA_{0,1}
        + z \zeta_2
\Biggr\},
\\
C_2^{\rm PS,(2)}\left(x,\frac{Q^2}{\mu_F^2}\right) &=&
C_F T_F N_F \Biggl\{
\left[8(1+z)\HA_0+\frac{4}{3}\left(3-4z^2-3z+\frac{4}{z}\right)\right] \ln^2\left(\frac{Q^2}{\mu_F^2}\right)
\nonumber\\ &&
+\Biggl[
16(1+z)[-\HA_{0,1} + \zeta_2 - \HA_0^2] + 32 z^2 \HA_0
- \frac{8}{3}\left(3 - 4z^2-3z + \frac{4}{z}\right) \HA_1
\nonumber\\ &&
- \frac{16}{9} \left(39 +4 z^2 -30z - \frac{13}{z} \right)
\Biggr] \ln\left(\frac{Q^2}{\mu_F^2} \right)
\nonumber\\ &&
+ \frac{4 (1-z) \big(
                172+409 z-224 z^2\big)}{27 z}
        +\frac{16}{9} \big(
                63-33 z-16 z^2\big) \HA_0
\nonumber\\ &&
        -\frac{32 (1+z)^3 \HA_{-1} \HA_0}{3 z}
        -\frac{2}{3} \big(
                3-45 z+32 x^2\big) \HA_0^2
        +\frac{20}{3} (1+z) \HA_0^3
\nonumber\\ &&
        +\Biggl[
                -\frac{16 (1-z) \big(
                        13-26 z+4 z^2\big)}{9 z}
                +\frac{8 \big(
                        4+3 z-6 z^2-4 z^3\big)}{3 z} \HA_0
        \Biggr] \HA_1
\nonumber\\ &&
        +\frac{4 \big(
                4+3 z-4 z^3\big) \HA_1^2}{3 z}
        +\Biggl[
                -\frac{8 (1+2 z) \big(
                        4-5 z+4 z^2\big)}{3 z}
                +16 (1+z) \HA_0
        \Biggr] \HA_{0,1}
\nonumber\\ &&
        +\frac{32 (1+z)^3 \HA_{0,-1}}{3 z}
        +16 (1+z) \HA_{0,1,1}
        -\Biggl[
                \frac{32 \big(
                        1+3 z^2-3 z^3\big)}{3 z}
\nonumber\\ &&
                +32 (1+z) \HA_0
        \Biggr] \zeta_2
        -16 (1+z) \zeta_3
\Biggr\}.
\end{eqnarray}
We agree with the results given in \cite{Moch:1999eb,Vermaseren:2005qc} and note a typo in \cite{vanNeerven:1991nn}, Eq.~(13), 
where the next-to-last term should read $(448/27) x^2$.
In Appendix~\ref{sec:A21} we present details of the calculation in the massless case.

The massless two-loop pure singlet contribution to the structure functions $F_{2(L)}$ for pure virtual photon exchange 
is given by
\begin{eqnarray}
F_{2(L)}^{(2),\rm PS}(x,Q^2) = a_s^2(Q^2) Q_H^2 x C_{2(L)}^{\rm PS,(2)}\left(\frac{Q^2}{\mu^2},x\right) \otimes 
\Sigma(x,\mu^2),
\end{eqnarray}
where $\mu$ denotes the factorization scale, $Q_H = 2/3$ for charm and $Q_H = -1/3$ for bottom, and
\begin{eqnarray}
\Sigma(x,\mu^2) =  \sum_{k=1}^3 \left[ q_k(x,\mu^2) + \overline{q}_k(x,\mu^2) \right]
\end{eqnarray}
denotes the quark singlet distribution for three light quarks.
\section{Systematic integration in the massive case}
\label{sec:4}

\vspace*{1mm}
\noindent
We will express the scattering cross sections in terms of a minimal number of special functions. In the case of
single scale quantities, various methods have been worked out in the past to achieve this; for a recent survey see
\cite{Blumlein:2018cms}. In the present case, we deal with a two-scale process, since the cross sections
depend on $z$ and $m^2/Q^2$ in a non-factorizing way.
The complete massive Wilson coefficients are represented in terms of four non-trivial integrals. The first three 
integrations are evaluated in terms of logarithms and polylogarithms at various complex arguments involving 
square-roots and trigonometric functions. What remains is a one-fold integral with respect to an angular variable 
$\varphi$ of a function 
that also depends on the parameters $z$ and $\beta$. The overall aim is to write this integral in terms of nested 
integrals. To this end, we  first write its integrand in terms of nested integrals. First, we apply the change of 
integration variables 
\begin{equation}
t=\sin(\varphi).
\end{equation}
As a result, we get rid of the trigonometric functions in the integrand. In addition, we introduce the quantity
\begin{equation}
 k:=\frac{\sqrt{z}}{\sqrt{1-(1-z)\beta^2}},
\end{equation}
which satisfies $\sqrt{z} < k <1$. We use it to express $\beta$ as $\frac{\sqrt{k^2-z}}{k\sqrt{1-z}}$. Altogether, 
the integrand is then an expression in terms of $z$, $k$, and $t$ as well as logarithms and dilogarithms with 
arguments expressed in terms of square-roots involving these quantities.

Next, we eliminate redundancies among square-root expressions to express the integrand using only the roots 
$\sqrt{1-k^2}$, $\sqrt{1-t^2}$, and $\sqrt{1-k^2 t^2}$. In order to facilitate the conversion of the logarithms 
and dilogarithms appearing in the integrand to nested integrals, we exploit the argument relations
\begin{align}
 \ln(z) &= \ln(-z) + i\pi & \text{for }&z<0\\
 \mathrm{Li}_2(z) &= -\mathrm{Li}_2(\tfrac{1}{z})-\tfrac{1}{2}\ln(z)^2-i\pi\ln(z)+2\zeta(2) & \text{for }&z>1 
\end{align}
to avoid arguments on branch cuts.

After these pre-processing steps, all the following steps for computing the integral are done by our code \cite{RAAB1} 
in {\tt Mathematica}, which also uses the routine \texttt{DSolveRational} of the package 
{\tt HolonomicFunctions}~\cite{KOUTSCHAN}; see \cite{RaabRegensburger,Raab} for the general theory 
underlying~\cite{RAAB1}. We also refer to \cite{GuoRegensburgerRosenkranz} for the simpler case when no singularities 
are 
present at the endpoints of integration, which, however, does not apply here.

First, the logarithms and dilogarithms are converted to nested integrals, which is based on repeated differentiation 
followed by expressing the integrands of these nested integrals in the form developed in (3.16)--(3.19) 
of~\cite{Ablinger:2014bra}. In fact, a generalized version of those forms is used to avoid the necessity of introducing 
new square-roots in terms of $z$ and $k$ in addition to $\sqrt{1-k^2}$ above. Then, a normal form of the integrand is 
computed. This affects all parts of the representation, also those that do not depend on $t$. For the nested integrals 
we use the shuffle relations and also for their coefficients we compute normal forms in terms of the logarithms and 
square-roots.

As a result, we obtain a representation of the integrand as a linear combination of nested integrals evaluated at $t$ 
whose integrands also depend on $z$ and $k$. Their coefficients only contain $z$, $k$, $t$, $\sqrt{1-t^2}$, 
$\sqrt{1-k^2 t^2}$, $\ln(z)$, $\ln(1-z)$, $\ln(k+z)$, and $\ln(k-z)$. The root $\sqrt{1-k^2}$, as well as all other 
logarithms and dilogarithms depending on $z$ and $k$, do not appear in this representation anymore. Moreover, since 
both the integrand as a whole and all integrands of the nested integrals in its representation are real, all complex 
expressions drop out of the coefficients as well and we have a completely real representation. This is ensured since 
the integrands in (3.16)--(3.19) of~\cite{Ablinger:2014bra}, and also their generalization used here, were designed so 
that the corresponding nested integrals all are linearly independent.

Finally, the integral over $t$ from $0$ to $\beta$ is computed as a linear combination of nested integrals evaluated 
at $\beta$, again in normal form. Like before, their integrands also depend on $z$ and $k$ and their coefficients only 
contain $z$, $k$, $t$, $\sqrt{1-t^2}$, $\sqrt{1-k^2 t^2}$, $\ln(z)$, $\ln(1-z)$, $\ln(k+z)$, and $\ln(k-z)$.

The following letters contribute in the present case:
\begin{eqnarray}
        f_{w_1}(t) &=& \frac{1}{1 - k t},
\\
        f_{w_2}(t) &=& \frac{1}{1 + k t},
\\
        f_{w_3}(t) &=& \frac{1}{\beta + t},
\\
        f_{w_4}(t) &=& \frac{1}{\beta - t},
\\
        f_{w_5}(t) &=& \frac{1}{k - z - ( 1 - z ) k t },
\\
        f_{w_6}(t) &=& \frac{1}{k + z - ( 1 - z ) k t },
\\
        f_{w_7}(t) &=& \frac{1}{k - z + ( 1 - z ) k t },
\\
        f_{w_8}(t) &=& \frac{1}{k + z + ( 1 - z ) k t },
\\
\label{eq:A1}
        f_{w_9}(t) &=& \frac{t}{k^2 \left(1 - t^2 \left(1 - z^2\right)\right)-z^2},
\\
        f_{w_{10}}(t) &=& \frac{1}{t \sqrt{1-t^2} \sqrt{1-k^2 t^2}},
\\
        f_{w_{11}}(t) &=& \frac{t}{\sqrt{1-t^2} \sqrt{1-k^2 t^2}},
\\
\label{eq:AA1}
        f_{w_{12}}(t) &=& \frac{t}{\sqrt{1-t^2} \sqrt{1-k^2 t^2} \left(k^2 \left(1 - t^2 \left(1 - 
z^2\right)\right)-z^2\right)} .
\end{eqnarray}
The set of letters
\begin{eqnarray}
\mathfrak{A} = \left\{ \left. \frac{1}{t-a} \right| a \in \mathbb{C} \right\}
\end{eqnarray}
span the Kummer-Poincar\'e iterated integrals \cite{KUMPO} defined as
\begin{eqnarray}
\KA_{b,\vec{a}}(z) = \int_0^z dy f_b(y) \KA_{\vec{a}}(y),~~\KA_\emptyset = 1, ~~f_c \in \mathfrak{A}.
\end{eqnarray}
The letter $f_{w_9}$ can be rewritten into Kummer-Poincar\'e letters \cite{KUMPO},
which we, however, avoid here. Some of the above letters contain the elliptic letter
\begin{eqnarray}
\frac{1}{\sqrt{1-t^2}}
\frac{1}{\sqrt{1- k^2 t^2}} 
\end{eqnarray}
as a factor. Therefore, one expects that in iterated integrals the incomplete elliptic integrals of the 1st, 2nd, and 3rd 
kind 
\begin{eqnarray}
F(x;k) &=& \int_0^x dt \frac{1}{\sqrt{1-t^2} \sqrt{1-k^2 t^2}},
\\
E(x;k) &=& \int_0^x dt \frac{\sqrt{1-k^2 t^2}}{\sqrt{1-t^2}},
\\
\Pi(n;x|k) &=& \int_0^x dt \frac{1}{1-nt^2} \frac{\sqrt{1-k t^2}}{\sqrt{1-t^2}},
\end{eqnarray}
cf.~\cite{ELLI1}, are emerging, over which further Kummer-Poincar\'e letters are iterated. We call iterated integrals 
of this type {\it Kummer-elliptic} integrals. Their alphabet is
\begin{eqnarray}
\mathfrak{A}' &=& \mathfrak{A} 
\cup 
\left\{\frac{1}{\sqrt{1-t^2}\sqrt{1-k^2t^2}},\frac{t}{\sqrt{1-t^2}\sqrt{1-k^2t^2}},\frac{1}{1-nt^2} \frac{\sqrt{1-m t^2}}{\sqrt{1-t^2}}
\right\}
\nonumber\\ 
&&\cup\left\{\frac{1}{(t-a)\sqrt{1-t^2}\sqrt{1-k^2t^2}}
\middle|a\in\mathbb{C}\setminus\{\pm1,\pm\tfrac{1}{k}\}\right\}.
\end{eqnarray}
Note that integrals of depth 1 over the letters $f_{w_1}$ to $f_{w_{12}}$ are (poly)logarithmic, since one may change 
variables $t \to \sqrt{t}$, cf.~Eqs.~(\ref{eq:A1}--\ref{eq:AA1}).

Yet Kummer-elliptic integrals appear in the iterated case.
Therefore, iterated integrals of depth 2 formed out of some of these letters will form results containing
incomplete elliptic integrals in part.
These iterative integrals cannot be reduced to the Kummer-Poincar\'e iterated integrals for general values of 
$k$. As also the incomplete elliptic integrals, they belong to the d'Alembert class, unlike the complete elliptic 
integrals \cite{ELLI1}, which also appear in various higher order calculations, cf. e.g.~\cite{ELLI2}, as letters 
in other iterated integrals.
\section{The massive Wilson coefficients}
\label{sec:5}

\vspace*{1mm}
\noindent
The unrenormalized two--loop massive pure singlet Wilson coefficients ${\cal H}_{i,q}$ with $i=1,2,L$, see also 
Eq.~(\ref{eq:F1}), are given in Mellin space by
\begin{eqnarray}
\label{eq:UNR3}
{\cal H}^{(2),\rm PS}_{i,q} &=&  \hat{a}_s^2 S_\varepsilon^2 \left(\frac{Q^2}{\mu^2}\right)^\ep \left[
\frac{1}{\ep} P_{gq}^{(0)} h_{i,g}^{(1)} + C_{i,q}^{(2),\rm PS,Q} +  P_{gq}^{(0)} \bar{b}_{i,g}^{(1)} \right] ~.
\end{eqnarray}
The functions $h_{1,g}^{(1)}$ and $\bar{b}_{1,g}^{(1)}$ are given by
\begin{eqnarray}
h_{1,g}^{(1)} &=& 2 h_{2,g}^{(1)} - 3 h_{L,q}^{(1)}
\\
\bar{b}_{1,g}^{(1)} &=& 
h_{2,g}^{(1)} -  h_{L,q}^{(1)} + 2 \bar{b}_{2,g}^{(1)} - 3 \bar{b}_{L,q}^{(1)}.
\end{eqnarray}
Since the two heavy quarks do not induce collinear divergences the mass factorization in the massive case
reads
\begin{eqnarray}
        {\cal H}^{(2),\rm PS}_{i,q} &=& {H}^{(2),\rm PS}_{i,q} + \Gamma_{gq} \otimes H_{i,g}^{(1)} ~.
\end{eqnarray}
Therefore, we find
\begin{eqnarray}
        {H}^{(2),\rm PS}_{i,q} &=&  \hat{a}_s^2 S_\varepsilon^2 \biggl\{ \left( \frac{Q^2}{\mu^2} \right)^\ep 
\biggl[ \frac{1}{\ep} P_{gq}^{(0)} h_{i,g}^{(1)} + C_{i,q}^{(2),\rm PS,Q} +  P_{gq}^{(0)} \bar{b}_{i,g}^{(1)} \biggr]
\nonumber \\ &&
- \left( \frac{\mu_F^2}{\mu^2} \right)^{\ep/2} \left( \frac{Q^2}{\mu^2} \right)^{\ep/2}  
\biggl[ \frac{1}{\ep} P_{gq}^{(0)} h_{i,g}^{(1)} + P_{gq}^{(0)} \bar{b}_{i,g}^{(1)} \biggr] \biggr\}.
\end{eqnarray} 
Identifying the renormalization and factorization scale, $\mu = \mu_F$, we finally obtain
\begin{eqnarray}
        H_{i,q}^{2,\rm PS} &=&  {a}_s^2 \biggl[ \frac{1}{2} P_{gq}^{(0)} h_{i,g}^{(1)} \ln \left( 
\frac{Q^2}{\mu_F^2} \right) + C_{i,q}^{(2),\rm PS,Q} \biggr] + O(\ep)
\nonumber \\ 
        &=&  {a}_s^2 \biggl[ \frac{1}{2} P_{gq}^{(0)} h_{i,g}^{(1)} \ln \left( \frac{m^2}{\mu_F^2} \right) 
- \frac{1}{2} P_{gq}^{(0)} h_{i,g}^{(1)} \ln \left( \frac{m^2}{Q^2} \right) + C_{i,q}^{(2),\rm PS,Q} \biggr] + O(\ep) ~.
\end{eqnarray} 
Note that in the pure singlet case the coupling constant is not renormalized at two--loop order.
To express our final result in terms of iterated integrals we refer to the letters given in 
Section~\ref{sec:4}, supplemented by the letters spanning the harmonic polylogarithms~(\ref{eq:HPL1}); for Eqs.~(\ref{eq:HFL}) 
and (\ref{eq:HF1}) we use the shorthand notation $\HA_{\vec{a}}(\beta) \equiv \HA_{\vec{a}}$.
One obtains
\begin{eqnarray}
\label{eq:HFL}
        H_{L,q}^{(2),\text{PS}} &=& 
C_F T_F \Biggl\{-\frac{8 P_1}{3 z} \biggl\{
  k \biggl[ \HA_{w_{1}}^2 - \HA_{w_{2}}^2
+ (1-z) \bigl( \HA_{w_{5},w_{1}} 
      + \HA_{w_{6},w_{2}}
      - \HA_{w_{7},w_{2}}
\nonumber \\ &&
      - \HA_{w_{8},w_{1}}
      - \HA_{w_{5}} \HA_{w_{1}}
      + \HA_{w_{8}} \HA_{w_{1}}
      - \HA_{w_{6}} \HA_{w_{2}}
      + \HA_{w_{7}} \HA_{w_{2}}
\bigr) \biggr]
+ 2 \bigl(  \HA_{w_{1},w_{4}}
      + \HA_{w_{2},w_{4}}
      + \HA_{w_{3},w_{1}}
\nonumber \\ &&
      + \HA_{w_{3},w_{2}}
\bigr)
-\bigl(
          2 \HA_{w_{3}}        
        - 6 \ln(k)
        + \ln\big(1-k^2\big)
        - \ln(k^2 - z^2)
        + 2 \ln \big(k^2-z\big)
\bigr) \bigl[ \HA_{w_{1}} 
\nonumber \\ &&
+ \HA_{w_{2}} \bigr]
\biggr\}
-\frac{16 (1-z) \beta P_2}{3 z} \ln (k^2 - z^2 )
-\frac{16 (1-z) \beta  P_3}{9 k^2 z}
+\frac{8 (1-k^2) (1-z) P_4}{3 k^4 z} 
\biggl[
        \HA_{w_{5},0}
\nonumber \\ &&
      - \HA_{w_{6},0}
      + \HA_{w_{7},0}
      - \HA_{w_{8},0}
      - \bigl(
        \HA_{w_{5}}
      - \HA_{w_{6}}
      + \HA_{w_{7}}
      - \HA_{w_{8}}
      \bigr) \HA_{0}
\biggr]
+ \frac{16 (1-k^2) P_4}{3 k^4 z} \bigl(
        \HA_{w_{1}}
\nonumber \\ &&
      + \HA_{w_{2}}
\bigr) \HA_0
+\frac{32  P_5}{3 k^2} \bigl( \HA_{-1} \HA_1 - 2 \HA_{-1,1} \bigr)
+\frac{32 P_6}{3 k^4 z} \bigl( \HA_{w_{1},0} + \HA_{w_{2},0} \bigr)
+\frac{16 P_7}{3 k^4} \bigl( \HA_1 \HA_{w_{1}} 
\nonumber \\ &&
- \HA_{-1} \HA_{w_{2}} \bigr)
+\frac{16 P_8}{3 k^4} \bigl( \HA_1 \HA_{w_{2}} - \HA_{-1} \HA_{w_{1}} \bigr)
-\frac{64 P_9}{3 k^2 z \beta } \HA_{w_{3}}
-\frac{16 (1-k^2) (1-z^2) P_{10}}{3 k^2} 
\biggl[
      \HA_{w_{9},1} 
\nonumber \\ &&
        + \HA_{w_{9},-1}
      - (1-z) k \bigl(
              \HA_{w_{9},w_{5}}
            + \HA_{w_{9},w_{6}}
            + \HA_{w_{9},w_{7}}
            + \HA_{w_{9},w_{8}}
      \bigr)
\biggr]
-\frac{16 P_{11}}{3 k^2} \bigl( \HA_1^2 - \HA_{-1}^2 \bigr)
\nonumber \\ &&
- \frac{(1-z) P_{12}}{3 z^{3/2} k^3} \biggl[
        \HA_{w_{10},w_{5}}
      - \HA_{w_{10},w_{6}}
      + \HA_{w_{10},w_{7}}
      - \HA_{w_{10},w_{8}}
      - k \bigl(
              \HA_{w_{5},w_{11}}
            + \HA_{w_{6},w_{11}}
            + \HA_{w_{7},w_{11}}
\nonumber \\ &&
            + \HA_{w_{8},w_{11}}
      \bigr)
      + k \bigl(
              \HA_{w_{5}}
            + \HA_{w_{6}}
            + \HA_{w_{7}}
            + \HA_{w_{8}}
      \bigr) \HA_{w_{11}}
      -\frac{2}{1-z} \bigl( \HA_{w_{10},w_{1}} + \HA_{w_{10},w_{2}} \bigr)
\biggr]
\nonumber \\ &&
+\frac{4 (1+k) (1-z) P_{13}}{3 k^4} \bigl(
        \HA_{w_{6},-1}
      - \HA_{w_{8},1}
      + \HA_{w_{8}} \HA_1
      - \HA_{w_{6}} \HA_{-1}
\bigr)
\nonumber \\ &&
+\frac{4 (1-k) (1-z) P_{14}}{3 k^4} \bigl(
        \HA_{w_{5},-1}
      - \HA_{w_{7},1}
      + \HA_{w_{7}} \HA_1
      - \HA_{w_{5}} \HA_{-1}
\bigr)
+\frac{8 P_{15}}{3 k^4 z} \bigl( \HA_{w_{1},1} - \HA_{w_{2},-1} \bigr)
\nonumber \\ &&
-\frac{4 (1-z) P_{16}}{3 k^4} \bigl(
        \HA_{w_{6},1}
      - \HA_{w_{8},-1}
      - \HA_{w_{6}} \HA_1
      + \HA_{w_{8}} \HA_{-1}
\bigr)
-\frac{4 (1-z) P_{17}}{3 k^4} \bigl(
        \HA_{w_{5},1}
      - \HA_{w_{7},-1}
\nonumber \\ &&
      - \HA_{w_{5}} \HA_1
      + \HA_{w_{7}} \HA_{-1}
\bigr)
-\frac{2 (1-k^2) P_{18}}{3 \sqrt{z} k^3}
\biggl[
        \HA_{w_{12},1}
      + \HA_{w_{12},-1}
      + (1-z) k \bigl(
              \HA_{w_{5},w_{12}}
            + \HA_{w_{6},w_{12}}
\nonumber \\ &&
            + \HA_{w_{7},w_{12}}
            + \HA_{w_{8},w_{12}}
      \bigr)
      - (1-z) k \bigl(
              \HA_{w_{5}}
            + \HA_{w_{6}}
            + \HA_{w_{7}}
            + \HA_{w_{8}}
      \bigr) \HA_{w_{12}}
\biggr]
-\frac{8 P_{19}}{3 k^4 z} \bigl( \HA_{w_{1},-1} 
\nonumber \\ &&
- \HA_{w_{2},1} \bigr)
+\frac{2 P_{20}}{9 k^2 z (1-k \beta)} \HA_{w_{1}}
-\frac{2 P_{21}}{9 k^2 z (1+k \beta)} \HA_{w_{2}}
+\frac{(1-z) P_{22}}{3 k^3 z (k (z - 2) + z ) (1-k \beta)} \HA_{w_{5}}
\nonumber \\ &&
+\frac{2 P_{23}}{9 k^4 z \big(k^2 (z-2)^2-z^2\big)} \HA_1
-\frac{2 P_{24}}{9 k^4 z \big(k^2 (z-2)^2-z^2\big)} \HA_{-1}
\nonumber \\ &&
-\frac{(1-z) P_{25}}{3 k^3 z (k (z-2)-z) (1+k \beta)} \HA_{w_{6}}
+\frac{(1-z) P_{26}}{3 k^3 z (k (z-2)+z) (1+k \beta)} \HA_{w_{7}}
\nonumber \\ &&
+\frac{(1-z) P_{27}}{3 k^3 z (k (z-2)-z) (1-k \beta)} \HA_{w_{8}}
-32 (1-z)^2 z ( \ln (z) + \ln (1-z) ) \bigl( 2 \beta - \HA_1 - \HA_{-1} \bigr)
\nonumber \\ &&
- 64 z \big(3-z+\frac{z}{k^2}\big) \ln (k) \bigl( \HA_1 + \HA_{-1} \bigr)
+ \frac{16 (-1+z) \beta }{3 z} \big(3-k^2-4 z-4 z^2\big) \bigl( 6 \ln (k) 
\nonumber \\ &&
- \ln \big(1-k^2\big) - 2 \ln \big(k^2-z\big) - 2 \HA_0 \bigr)
-\frac{64 z \big(k^2 (z-3)-z\big)}{3 k^2} \biggl[
        \HA_1 \HA_0
      + \HA_{-1,0}
      - \HA_{0,1}
\nonumber \\ &&
      - \HA_{1,w_{4}}
      - \HA_{-1,w_{4}}
      - \HA_{w_{3},1}
      - \HA_{w_{3},-1}
      + \biggl(
            \frac{1}{2} \ln \big(1-k^2\big)
            + \ln \big(k^2-z\big)
            + \HA_{w_{3}}
      \biggr) 
\nonumber \\ && \times \bigl( \HA_1 + \HA_{-1} \bigr)            
\biggr]
-\frac{32 z}{3 k^2} \big(z+k^2 \big(6-7 z+3 z^2\big)\big) \ln (k^2-z^2) \bigl( \HA_1 + \HA_{-1} \bigr)\Biggr\}
\nonumber \\ &&
+\frac{1}{2} P_{gq}^{(0)} \otimes \bar{h}_{L,g}^{(1)} \ln \left( \frac{Q^2}{\mu_F^2} \right)
- P_{gq}^{(0)} \otimes \bar{b}_{L,g}^{(1)}  ~,
\\
\label{eq:HF1}
        H_{1,q}^{(2),\text{PS}} &=& 
C_F T_F \Biggl\{-\frac{4 (1-z) P_{28}}{k^2} 
\bigl(
        \HA_{w_{6},-1}
      - \HA_{w_{8},1}
      + \HA_1 \HA_{w_{8}}
      - \HA_{-1} \HA_{w_{6}}
\bigr)
\nonumber \\ &&
-\frac{8 P_{29}}{3 k^3} \bigl( \HA_1 \HA_{w_{1}} - \HA_{-1} \HA_{w_{2}} \bigr)
-\frac{8 P_{30}}{3 k^3} \HA_1 \HA_{w_{2}} 
+\frac{8 \big(k^2-z\big) P_{30}}{3 k^5 (1-z) \beta ^2} \HA_{w_{1}} \HA_{-1}
\nonumber \\ &&
+\frac{4 (1-z) P_{31}}{k^2} 
\bigl(  
        \HA_{w_{5},-1}
      - \HA_{w_{7},1}
      + \HA_1 \HA_{w_{7}}
      - \HA_{-1} \HA_{w_{5}}
\bigr)
+\frac{8 P_{32}}{3 z} 
\biggl[
      k \bigl( \HA_{w_{1}}^2 - \HA_{w_{2}}^2 \bigr)
\nonumber \\ &&
      + 2 \bigl( \HA_{w_{1},w_{4}} + \HA_{w_{2},w_{4}} + \HA_{w_{3},w_{1}} + \HA_{w_{3},w_{2}} \bigr)
      + \bigl( \HA_{w_{1}} + \HA_{w_{2}} \bigr) \bigl[ 6 \ln (k) + \ln (k^2-z^2) \bigr]
\nonumber \\ &&
      + k (1-z) \bigl(
              \HA_{w_{5},w_{1}}
            + \HA_{w_{6},w_{2}}
            - \HA_{w_{7},w_{2}}
            - \HA_{w_{8},w_{1}}
            - \HA_{w_{1}} \HA_{w_{5}}
            - \HA_{w_{2}} \HA_{w_{6}}
            + \HA_{w_{2}} \HA_{w_{7}}
\nonumber \\ &&
            + \HA_{w_{1}} \HA_{w_{8}}
      \bigr)
      -\bigl( \HA_{w_{1}} + \HA_{w_{2}} \bigl) \bigl[ \ln \big(1-k^2\big) + 2 \ln \big(k^2-z)
        + 2 \HA_{w_{3}} \bigr]
\biggr]
\nonumber \\ &&
+\frac{16 (1-z) \beta  P_{33}}{9 k^2 z}
+\frac{32 P_{34}}{3 k^4}
\biggl[
        \HA_{0,1}
      - \HA_{-1,0}
      - \HA_0 \HA_1
      + \HA_{1,w_{4}}
      + \HA_{w_{3},1}
      + \HA_{w_{3},-1}
      + \HA_{-1,w_{4}}
\nonumber \\ &&
      - \bigl( \HA_1 + \HA_{-1} \bigr) \bigl( \frac{1}{2} \ln \big(1-k^2\big) + \ln \big(k^2-z\big) + \HA_{w_{3}} \bigr)
\biggr]
-\frac{32 (1-z^2) P_{35}}{3 k^2}
\biggl[
        \HA_{w_{9},1}
\nonumber \\ &&
      + \HA_{w_{9},-1}
      - (1-z) k \bigl( \HA_{w_{9},w_{5}} + \HA_{w_{9},w_{6}} + \HA_{w_{9},w_{7}} + \HA_{w_{9},w_{8}} \bigr)
\biggr]
+\frac{4 (1-z) P_{36}}{3 k^3} \bigl(  \HA_{w_{5},1} 
\nonumber \\ &&
- \HA_{w_{7},-1} - \HA_1 \HA_{w_{5}} + \HA_{-1} \HA_{w_{7}} \bigr)
+\frac{4 (1-z) P_{37}}{3 k^3} \bigl( \HA_{w_{6},1} - \HA_{w_{8},-1} - \HA_1 \HA_{w_{6}} + \HA_{-1} \HA_{w_{8}} \bigr)
\nonumber \\ &&
+\frac{16 P_{38}}{3 k^4} \bigl( \HA_{-1} \HA_1 - 2 \HA_{-1,1} \bigr)
-\frac{16 (1-z) \beta P_{39}}{3 k^2 z} \ln (k^2-z^2)
-\frac{8 P_{40}}{3 k^3 z} \bigl( \HA_{w_{1},1} - \HA_{w_{2},-1} \bigr)
\nonumber \\ &&
-\frac{8 P_{41}}{3 k^3 z} \HA_{w_{2},1}
-\frac{16 (1-z) \beta P_{42}}{3 k^2 z}
\biggl[
        \ln \big(1-k^2\big)
      + 2 \ln \big(k^2-z\big)
      - 6 \ln (k)
      + 2 \HA_0
\nonumber \\ &&
      + 4 \HA_{w_{3}}
\biggr]
-\frac{16 P_{43}}{3 k^2 z} \bigl( \HA_{w_{1},0} + \HA_{w_{2},0} \bigr)
-\frac{8 P_{44}}{3 k^4} \bigl( \HA_1^2 - \HA_{-1}^2 \bigr)
+\frac{16  P_{45}}{3 k^2 z} \bigl( \HA_{w_{1}} + \HA_{w_{2}} \bigr) \HA_0
\nonumber \\ &&
+\frac{8 (1-z) P_{45}}{3 k^2 z}
\biggl[
        \HA_{w_{5},0}
      - \HA_{w_{6},0}
      + \HA_{w_{7},0}
      - \HA_{w_{8},0}
      - \bigl(
              \HA_{w_{5}}
            - \HA_{w_{6}}
            + \HA_{w_{7}}
            - \HA_{w_{8}}
      \bigr) \HA_0
\biggr]
\nonumber \\ &&
+\frac{4 P_{46}}{3 z^{3/2} k^3}
\biggl[
        2 \HA_{w_{10},w_{1}}
      + 2 \HA_{w_{10},w_{2}}
      - (1-z) \biggl(
              \HA_{w_{10},w_{5}}
            - \HA_{w_{10},w_{6}}
            + \HA_{w_{10},w_{7}}
            - \HA_{w_{10},w_{8}}
\nonumber \\ &&
            - k \bigl(
                    \HA_{w_{5},w_{11}}
                  + \HA_{w_{6},w_{11}}
                  + \HA_{w_{7},w_{11}}
                  + \HA_{w_{8},w_{11}}
            \bigr)
            + k \bigl(
                    \HA_{w_{5}}
                  + \HA_{w_{6}}
                  + \HA_{w_{7}}
                  + \HA_{w_{8}}
            \bigr) \HA_{w_{11}}
      \biggr)
\nonumber \\ &&
      + 2 k ( 1 - k^2 ) z (1 - z) \biggl(
              \HA_{w_{5},w_{12}}
            + \HA_{w_{6},w_{12}}
            + \HA_{w_{7},w_{12}}
            + \HA_{w_{8},w_{12}}
            - \bigl(
                    \HA_{w_{5}}
                  + \HA_{w_{6}}
                  + \HA_{w_{7}}
\nonumber \\ &&
                  + \HA_{w_{8}}
            \bigr) \HA_{w_{12}}
      \biggr)
      + 2 (1-k^2) z \bigl( \HA_{w_{12},1} + \HA_{w_{12},-1} \bigr)
\biggr]
+\frac{8 P_{47}}{9 k^2 z (1+k \beta)} \HA_{w_{2}}
\nonumber \\ &&
-\frac{8 P_{48}}{9 k^2 z (1-k \beta)} \HA_{w_{1}}
-\frac{4 (1-z)^2 P_{49}}{3 k^3 z (k (z-2)-z)} \HA_{w_{6}}
-\frac{4 (1-z)^2 P_{50}}{3 k^3 z (k (z-2)+z)} \HA_{w_{5}}
\nonumber \\ &&
-\frac{4 (1-z)^2 P_{51}}{3 k^3 z (k (z-2)+z)} \HA_{w_{7}}
-\frac{4 (1-z)^2 P_{52}}{3 k^3 z (k (z-2)-z)} \HA_{w_{8}}
-\frac{8 P_{55}}{3 k^5 (1-z) z \beta ^2} \HA_{w_{1},-1}
\nonumber \\ &&
-\frac{8 P_{53}}{9 k^4 z (1+\beta ) \big(k^2 (z-2)^2-z^2\big)} \HA_1
+\frac{8 P_{54}}{9 k^4 z (1-\beta ) \big(k^2 (z-2)^2-z^2\big)} \HA_{-1}
\nonumber \\ &&
-\biggl[
         \frac{16 \big(1+k^2\big)\big(1-3 k^2\big) z^2}{3 k^4} \ln (k^2-z^2)
        +16 (1-z) \bigl( \ln (1-z) + \ln (z) \bigr)
\nonumber \\ &&
        +32 \biggl(3(1-z)+\frac{ \big(1+k^2\big)\big(1-3 k^2\big) z^2}{k^4}\biggr) \ln (k)
\biggr] \bigl( \HA_1 + \HA_{-1} \bigr)
\nonumber \\ &&
- 8 \frac{2k^2 +\big( 3 k^2 - 1 \big) z}{k^2}
\biggl[
      4  \HA_{0,1,1}
      + 4 \HA_{0,-1,1}
      - 20 \HA_{1,1,1}
      - 4 \HA_{1,1,w_{4}}
      - 4 \HA_{1,-1,w_{4}}
\nonumber \\ &&
      + 4 \HA_{w_{3},1,1}
      - 4 \HA_{w_{3},1,-1}
      + 4 \HA_{w_{3},-1,1}
      - 4 \HA_{w_{3},-1,-1}
      - 4 \HA_{-1,1,0}
      - 16 \HA_{-1,1,1}
      + 4 \HA_{-1,1,w_{4}}
\nonumber \\ &&
      - 4 \HA_{-1,-1,0}
      - 16 \HA_{-1,-1,1}
      + 4 \HA_{-1,-1,w_{4}}
      - 20 \HA_{-1,-1,-1}
+2 \bigl(
          \HA_1^2
         - 2 \HA_{-1,1}
\bigr) \HA_0
\nonumber \\ &&
+2\bigl(       
        -4 \HA_{-1,1}
        + \HA_1^2
        - \HA_{-1}^2
        +2 \HA_1 \HA_{-1}
\bigr) \HA_{w_{3}}
+\bigl(
         4 \HA_{-1,1}
        -5 \HA_{-1}^2
        +5 \HA_{1}^2
        -4 \HA_{0,1}
\nonumber \\ &&
        -4 \HA_{0,-1}
        -4 \HA_{w_{3},1}
        -4 \HA_{w_{3},-1}
\bigr) \HA_1
+\bigl(
         4 \HA_0 \HA_1
        -  \HA_1^2
        +4 \HA_{w_{3},1}
        +4 \HA_{w_{3},-1}
        +12 \HA_{-1,1}
\nonumber \\ &&
        +5 \HA_{-1}^2
\bigr) \HA_{-1}
      - \bigl[ \ln \big(1-k^2\big) - \ln (k^2-z^2) + 2 \ln \big(k^2-z\big) - 6 \ln (k) \bigr]
\nonumber \\ && \times
      \bigl( 4 \HA_{-1,1} + \HA_{-1}^2 - \HA_1^2 - 2 \HA_{-1} \HA_1 \bigr)
\biggr] 
-\frac{16 (1-z) \big(z-k^2 (2+3 z)\big)}{k} 
\biggl[
        \HA_{1,w_{4},w_{5}}
\nonumber \\ &&
      + \HA_{1,w_{4},w_{6}}
      + \HA_{1,w_{4},w_{7}}
      + \HA_{1,w_{4},w_{8}}
      - \HA_{w_{5},1,1}
      + \HA_{w_{5},1,-1}
      - \HA_{w_{5},w_{3},1}
      + \HA_{w_{5},w_{3},-1}
\nonumber \\ &&
      - \HA_{w_{6},1,1}
      + \HA_{w_{6},1,-1}
      - \HA_{w_{6},w_{3},1}
      + \HA_{w_{6},w_{3},-1}
      - \HA_{w_{7},w_{3},1}
      + \HA_{w_{7},w_{3},-1}
      + \HA_{w_{7},-1,1}
\nonumber \\ &&
      - \HA_{w_{7},-1,-1}
      - \HA_{w_{8},w_{3},1}
      + \HA_{w_{8},w_{3},-1}
      + \HA_{w_{8},-1,1}
      - \HA_{w_{8},-1,-1}
      - \HA_{-1,w_{4},w_{5}}
      - \HA_{-1,w_{4},w_{6}}
\nonumber \\ &&
      - \HA_{-1,w_{4},w_{7}}
      - \HA_{-1,w_{4},w_{8}}
      + k \bigl(
              \HA_{w_{2},w_{4},w_{5}}
            + \HA_{w_{2},w_{4},w_{6}}
            + \HA_{w_{2},w_{4},w_{7}}
            + \HA_{w_{2},w_{4},w_{8}}
\nonumber \\ &&
            - \HA_{w_{1},w_{4},w_{5}}
            - \HA_{w_{1},w_{4},w_{6}}
            - \HA_{w_{1},w_{4},w_{7}}
            - \HA_{w_{1},w_{4},w_{8}}
            + \HA_{w_{5},1,w_{1}}
            - \HA_{w_{5},1,w_{2}}
            + \HA_{w_{5},w_{3},w_{1}}
\nonumber \\ &&
            - \HA_{w_{5},w_{3},w_{2}}
            + \HA_{w_{6},1,w_{1}}
            - \HA_{w_{6},1,w_{2}}
            + \HA_{w_{6},w_{3},w_{1}}
            - \HA_{w_{6},w_{3},w_{2}}
            + \HA_{w_{7},w_{3},w_{1}}
            - \HA_{w_{7},w_{3},w_{2}}
\nonumber \\ &&
            - \HA_{w_{7},-1,w_{1}}
            + \HA_{w_{7},-1,w_{2}}
            + \HA_{w_{8},w_{3},w_{1}}
            - \HA_{w_{8},w_{3},w_{2}}
            - \HA_{w_{8},-1,w_{1}}
            + \HA_{w_{8},-1,w_{2}}
      \bigr)
\nonumber \\ &&
      + \bigl\{
              \HA_{w_{3},1}
            - \HA_{w_{3},-1} 
            + \HA_{-1,1}
            + k \bigl[
                    \HA_{w_{1},1}
                  - \HA_{w_{2},1}
                  - \HA_{w_{3},w_{1}}
                  + \HA_{w_{3},w_{2}}
            \bigr]
      \bigr\} \bigl( \HA_{w_{5}} + \HA_{w_{6}} \bigr)
\nonumber \\ &&
      + \bigl\{
              \HA_{w_{3},1}
            - \HA_{w_{3},-1}
            - \HA_{-1,1}
            - \HA_{-1,-1}
            + k \bigl[
                    \HA_{w_{2},-1}
                  - \HA_{w_{1},-1}
                  - \HA_{w_{3},w_{1}}
                  + \HA_{w_{3},w_{2}}
            \bigr]
      \bigr\}
\nonumber \\ && \times
      \bigl( \HA_{w_{7}} + \HA_{w_{8}} \bigr)
      + \bigl(
              \HA_{w_{5},1}
            + \HA_{w_{5},w_{3}}
            + \HA_{w_{6},1}
            + \HA_{w_{6},w_{3}}
            + \HA_{w_{7},w_{3}}
            - \HA_{w_{7},-1}
            + \HA_{w_{8},w_{3}}
\nonumber \\ &&
            - \HA_{w_{8},-1}
            - \bigl[    
                    \HA_{w_{5}}
                  + \HA_{w_{6}}
                  + \HA_{w_{7}}
                  + \HA_{w_{8}}
            \bigr] \HA_{w_{3}}
      \bigr) \bigl( \HA_1 - \HA_{-1} \bigr)
      - k \bigl(
              \HA_{w_{5},1}
            + \HA_{w_{5},w_{3}}
\nonumber \\ &&
            + \HA_{w_{6},1}
            + \HA_{w_{6},w_{3}}
            + \HA_{w_{7},w_{3}}
            - \HA_{w_{7},-1}
            + \HA_{w_{8},w_{3}}
            - \HA_{w_{8},-1}
            - \bigl[    
                    \HA_{w_{5}}
                  + \HA_{w_{6}}
                  + \HA_{w_{7}}
\nonumber \\ &&
                  + \HA_{w_{8}}
            \bigr] \HA_{w_{3}}
      \bigr) \bigl( \HA_{w_{1}} - \HA_{w_{2}} \bigr)
      + \bigl( \HA_{w_{7}} + \HA_{w_{8}} \bigr) \HA_1 \HA_{-1}
      - \frac{1}{2} \bigl( \HA_{w_{5}} + \HA_{w_{6}} \bigr) \HA_1^2
\biggr]
\nonumber \\ &&
+ 16 \big(z-k^2 (2+3 z)\big) \bigl[ \HA_{w_{1},1} + \HA_{w_{1},-1} - \HA_{w_{2},1} - \HA_{w_{2},-1} \bigr] \bigl( \HA_{w_{1}} - \HA_{w_{2}} \bigr)
\nonumber \\ &&
+ \frac{ 32 ( k^2 (2+3 z) - z ) }{k} 
\biggl[
        \HA_{w_{1},1,0}
      + \HA_{w_{1},1,1}
      - \HA_{w_{1},1,w_{4}}
      - \HA_{w_{1},1,-1}
      + \HA_{w_{1},-1,0}
      + \HA_{w_{1},-1,1}
\nonumber \\ &&
      - \HA_{w_{1},-1,w_{4}}
      - \HA_{w_{1},-1,-1}
      - \HA_{w_{2},1,0}
      - \HA_{w_{2},1,1}
      + \HA_{w_{2},1,w_{4}}
      + \HA_{w_{2},1,-1}
      - \HA_{w_{2},-1,0}
\nonumber \\ &&
      - \HA_{w_{2},-1,1}
      + \HA_{w_{2},-1,w_{4}}
      + \HA_{w_{2},-1,-1}
      + \HA_{w_{3},1,w_{1}}
      - \HA_{w_{3},1,w_{2}}
      + \HA_{w_{3},-1,w_{1}}
      - \HA_{w_{3},-1,w_{2}}
\nonumber \\ &&
      + \frac{1}{2} \bigl[
              \HA_{w_{1},1}
            + \HA_{w_{1},-1}
            - \HA_{w_{2},1}
            - \HA_{w_{2},-1}
      \bigr] \bigl( 2 \HA_{w_{3}} + \HA_1 - \HA_{-1} \bigr)
      + \frac{1}{4} \bigl[ 
              \HA_1^2 
            - 4 \HA_{w_{3},-1} 
\nonumber \\ &&
            - 4 \HA_{w_{3},1} 
            - 4 \HA_{-1,1} 
            - \HA_{-1}^2  
            + 2 \HA_{-1} \HA_1 
      \bigr] \bigl( \HA_{w_{1}} - \HA_{w_{2}}  \bigr)
      + \frac{1}{2} \bigl[
              \HA_{w_{2},-1}
            - \HA_{w_{1},1}
            - \HA_{w_{1},-1}
\nonumber \\ &&
            + \HA_{w_{2},1}
      \bigr] \bigl( 
              6 \ln (k) 
            - \ln \big(1-k^2\big) 
            + \ln (k^2-z^2) 
            - 2 \ln \big(k^2-z\big) 
      \bigr)
\biggr]
\nonumber \\ &&
+32 (1-z) \beta  \bigl( \ln (1-z) + \ln (z) \bigr)\Biggr\}
+\frac{1}{2} P_{gq}^{(0)} \otimes \bar{h}_{1,g}^{(1)} \ln \left( \frac{Q^2}{\mu_F^2} \right) 
 - P_{gq}^{(0)} \otimes \bar{b}_{1,g}^{(1)}  ~,
\end{eqnarray}
with the polynomials
\begin{eqnarray}
P_1&=&k^4+k^2 (2-6 z)-12 z^2+6 z-3, 
\\ 
P_2&=&-k^2+12 z^3-16 z^2-4 z+3, 
\\
P_3&=&8 k^4+k^2 \left(-25 z^2-28 z+12\right)+9 z^2, 
\\
P_4&=&k^6+k^4 \left(3-6 z^2\right)-4 z^4, 
\\
P_5&=&k^2 \left(z^2-3 z-1\right)-z^2-3 z+1, 
\\
P_6&=&k^8+k^6 \left(-3 z^2-3 z+2\right)-3 k^4 \left(z^2-z+1\right)-2 k^2 z^4+2 z^4, 
\\
P_7&=&3 k^6 (z-1)-2 k^5 z \left(3 z^2-7 z+6\right)+k^4 (3-9 z)-2 k^3 z^2+2 k^2 z^3-2 z^3, 
\\
P_8&=&3 k^6 (z-1)+2 k^5 z \left(3 z^2-7 z+6\right)+k^4 (3-9 z)+2 k^3 z^2+2 k^2 z^3-2 z^3, 
\\
P_9&=&k^4+k^2 \left(4 z^2+3 z-3\right)+z \left(-4 z^2-4 z+3\right), 
\\
P_{10}&=&k^2 \left(5 z^2-2\right)+3 z^2, 
\\
P_{11}&=&k^2 \left(5 z^2-15 z+1\right)-5 z^2+3 z-1, 
\\
P_{12}&=&k^4 \left(-80 z^3+35 z^2+30 z-9\right)+2 k^2 z \left(19 z^2-10 z-9\right)+3 z^2 \left(5 z^2+2 z+1\right), 
\\
P_{13}&=&6 k^5 (z-1)+k^4 \left(-4 z^3+21 z^2-30 z+8\right)+k^3 \left(4 z^3-21 z^2+12 z-2\right)+3 k^2 z^2\nonumber \\ &&
+k z^2 (4 z-3)-4 z^3, 
\\
P_{14}&=&6 k^5 (z-1)+k^4 \left(4 z^3-21 z^2+30 z-8\right)+k^3 \left(4 z^3-21 z^2+12 z-2\right)-3 k^2 z^2\nonumber \\ &&
+k z^2 (4 z-3)+4 z^3, 
\\
P_{15}&=&3 k^8-6 k^6 \left(z^2+2 z-1\right)+k^5 z \left(12 z^3-25 z^2+6\right)-3 k^4 \left(6 z^2-4 z+3\right)-2 k^3 z \bigl(z^2
\nonumber \\ &&
-6 z+3\bigr)-4 k^2 z^4+3 k z^3+4 z^4, 
\\
P_{16}&=&6 k^6 (z-1)+k^5 \left(20 z^3-35 z^2+24 z+2\right)+k^4 (6-18 z)+2 k^3 \left(2 z^3-5 z^2+6 z-1\right)
\nonumber \\ &&
+4 k^2 z^3-3 k z^2-4 z^3, 
\\
P_{17}&=&-6 k^6 (z-1)+k^5 \left(20 z^3-35 z^2+24 z+2\right)+6 k^4 (3 z-1)
\nonumber \\ &&
+2 k^3 \left(2 z^3-5 z^2+6 z-1\right)-4 k^2 z^3-3 k z^2+4 z^3, 
\\
P_{18}&=&k^4 \left(80 z^3-35 z^2-30 z+9\right)+2 k^2 z \left(-19 z^2+10 z+9\right)-3 z^2 \left(5 z^2+2 z+1\right), 
\\
P_{19}&=&3 k^8-6 k^6 \left(z^2+2 z-1\right)+k^5 \left(-12 z^4+25 z^3-6 z\right)-3 k^4 \left(6 z^2-4 z+3\right)
\nonumber \\ &&
+2 k^3 z \left(z^2-6 z+3\right)-4 k^2 z^4-3 k z^3+4 z^4, 
\\
P_{20}&=&16 \beta  k^7-40 k^6+8 \beta  k^5 \left(18 z^2+3 z-5\right)+8 k^4 \left(36 z^3-66 z^2-15 z+17\right)
\nonumber \\ &&
+3 \beta  k^3 \left(192 z^4-344 z^3+69 z^2+82 z-31\right)-3 k^2 \left(192 z^4-248 z^3-59 z^2+50 z-7\right)
\nonumber \\ &&
+3 \beta  k z \left(25 z^2-6 z-3\right)+3 z \left(-25 z^2+6 z+3\right), 
\\
P_{21}&=&16 \beta  k^7+40 k^6+8 \beta  k^5 \left(18 z^2+3 z-5\right)-8 k^4 \left(36 z^3-66 z^2-15 z+17\right)
\nonumber \\ &&
+3 \beta  k^3 \left(192 z^4-344 z^3+69 z^2+82 z-31\right)+3 k^2 \left(192 z^4-248 z^3-59 z^2+50 z-7\right)
\nonumber \\ &&
+3 \beta  k z \left(25 z^2-6 z-3\right)+3 z \left(25 z^2-6 z-3\right), 
\\
P_{22}&=&8 k^8 (z-2) (\beta  (z-1)+1)-8 k^7 \left(-2 \beta +\beta  z^3+(1-8 \beta ) z^2+(9 \beta -4) z+2\right)
\nonumber \\ &&
+k^6 \left(-66 \beta +(68 \beta -96) z^4+(328-186 \beta ) z^3+(17 \beta -288) z^2+(167 \beta -24) z+48\right)
\nonumber \\ &&
+k^5 \bigl(-30 \beta -192 \beta  z^5+4 (207 \beta -41) z^4+(314-935 \beta ) z^3+3 (47 \beta +5) z^2+
\nonumber \\ &&
(188 \beta -199) z+66\bigr)
+k^4 \bigl(-192 (\beta -1) z^5+4 (94 \beta -183) z^4-15 (9 \beta -41) z^3
\nonumber \\ &&
+(83-52 \beta ) z^2+(3 \beta -100) z-18\bigr)
+k^3 z \bigl(-6 \beta +192 z^4+7 (\beta -40) z^3+(7-18 \beta ) z^2
\nonumber \\ &&
+(17 \beta +20) z+21\bigr)
+k^2 (z-1) z \left((4 \beta -7) z^2+(3 \beta +11) z-6\right)
\nonumber \\ &&
-k (z-1) z^2 ((3 \beta +4) z+3)+3 (z-1) z^3, 
\\
P_{23}&=&
72 k^8 (z-2)^2 (\beta  (z-1)+1)
+k^6 \bigl(108 (8 \beta -7)+8 (36 \beta +29) z^5-2 (576 \beta +539) z^4
\nonumber \\ &&
+(576 \beta +1807) z^3+3 (768 \beta -563) z^2-1440 (2 \beta -1) z\bigr)
+k^4 z \bigl(-16 (18 \beta +17) z^4
\nonumber \\ &&
+208 z^3+(504 \beta +95) z^2-3 (72 \beta +145) z+360\bigr)
+k^2 z^2 \bigl(43 z^3+99 z^2-150 z+36\bigr)
\nonumber \\ &&
-3 z^4 (z+3), 
\\
P_{24}&=&72 k^8 (z-2)^2 (\beta  (z-1)-1)
+k^6 \bigl(108 (8 \beta +7)+8 (36 \beta -29) z^5-2 (576 \beta -539) z^4
\nonumber \\ &&
+(576 \beta -1807) z^3+3 (768 \beta +563) z^2-1440 (2 \beta +1) z\bigr)
-k^4 z \bigl(16 (18 \beta -17) z^4
\nonumber \\ &&
+208 z^3+(95-504 \beta ) z^2+3 (72 \beta -145) z+360\bigr)
-k^2 z^2 \bigl(43 z^3+99 z^2-150 z+36\bigr)
\nonumber \\ &&
+3 z^4 (z+3), 
\\
P_{25}&=&8 k^8 (z-2) (\beta  (z-1)+1)
+8 k^7 \bigl(-2 \beta +\beta  z^3+(1-8 \beta ) z^2+(9 \beta -4) z+2\bigr)
\nonumber \\ &&
+k^6 \bigl(-66 \beta +(68 \beta -96) z^4+(328-186 \beta ) z^3+(17 \beta -288) z^2+(167 \beta -24) z+48\bigr)
\nonumber \\ &&
+k^5 \bigl(30 \beta +192 \beta  z^5+(164-828 \beta ) z^4+(935 \beta -314) z^3-3 (47 \beta +5) z^2
\nonumber \\ &&
+(199-188 \beta ) z-66\bigr)
+k^4 \bigl(-192 (\beta -1) z^5+4 (94 \beta -183) z^4-15 (9 \beta -41) z^3
\nonumber \\ &&
+(83-52 \beta ) z^2+(3 \beta -100) z-18\bigr)
-k^3 z \bigl(-6 \beta +192 z^4+7 (\beta -40) z^3
\nonumber \\ &&
+(7-18 \beta ) z^2+(17 \beta +20) z+21\bigr)
+k^2 (z-1) z \bigl((4 \beta -7) z^2+(3 \beta +11) z-6\bigr)
\nonumber \\ &&
+k (z-1) z^2 ((3 \beta +4) z+3)+3 (z-1) z^3, 
\\
P_{26}&=&-8 k^8 (z-2) (\beta  (z-1)-1)
+8 k^7 \bigl(-2 (\beta +1)+\beta  z^3-(8 \beta +1) z^2+(9 \beta +4) z\bigr)
\nonumber \\ &&
-k^6 \bigl(-6 (11 \beta +8)+(68 \beta +96) z^4-2 (93 \beta +164) z^3+(17 \beta +288) z^2+(167 \beta +24) z\bigr)
\nonumber \\ &&
+k^5 \bigl(30 \beta +192 \beta  z^5-4 (207 \beta +41) z^4+(935 \beta +314) z^3-3 (47 \beta -5) z^2
\nonumber \\ &&
-(188 \beta +199) z+66\bigr)
+k^4 \bigl(192 (\beta +1) z^5-4 (94 \beta +183) z^4+15 (9 \beta +41) z^3
\nonumber \\ &&
+(52 \beta +83) z^2-(3 \beta +100) z-18\bigr)
+k^3 z \bigl(6 \beta +192 z^4-7 (\beta +40) z^3
\nonumber \\ &&
+(18 \beta +7) z^2+(20-17 \beta ) z+21\bigr)
-k^2 (z-1) z \bigl((4 \beta +7) z^2+(3 \beta -11) z+6\bigr)
\nonumber \\ &&
+k (z-1) z^2 ((3 \beta -4) z-3)+3 (z-1) z^3, 
\\
P_{27}&=&8 k^8 (z-2) (\beta  (z-1)-1)
+8 k^7 \bigl(-2 (\beta +1)+\beta  z^3-(8 \beta +1) z^2+(9 \beta +4) z\bigr)
\nonumber \\ &&
+k^6 \bigl(-6 (11 \beta +8)+(68 \beta +96) z^4-2 (93 \beta +164) z^3+(17 \beta +288) z^2+(167 \beta +24) z\bigr)
\nonumber \\ &&
+k^5 \bigl(30 \beta +192 \beta  z^5-4 (207 \beta +41) z^4+(935 \beta +314) z^3-3 (47 \beta -5) z^2
\nonumber \\ &&
-(188 \beta +199) z+66\bigr)
+k^4 \bigl(-192 (\beta +1) z^5+4 (94 \beta +183) z^4-15 (9 \beta +41) z^3
\nonumber \\ &&
-(52 \beta +83) z^2+(3 \beta +100) z+18\bigr)
+k^3 z \bigl(6 \beta +192 z^4-7 (\beta +40) z^3
\nonumber \\ &&
+(18 \beta +7) z^2+(20-17 \beta ) z+21\bigr)
+k^2 (z-1) z \bigl((4 \beta +7) z^2+(3 \beta -11) z+6\bigr)
\nonumber \\ &&
+k (z-1) z^2 ((3 \beta -4) z-3)-3 (z-1) z^3
\\
P_{28}&=&3 k^4 (z-2)+k^3 (20-14 z)+6 k^2 (z+1)+2 k z-z,
\\
P_{29}&=&9 k^5 (z-2)-6 k^4 z^2+18 k^3 (z+1)-4 k^2 z^2-3 k z+2 z^2,
\\
P_{30}&=&9 k^5 (z-2)+6 k^4 z^2+18 k^3 (z+1)+4 k^2 z^2-3 k z-2 z^2,
\\
P_{31}&=&3 k^4 (z-2)+2 k^3 (7 z-10)+6 k^2 (z+1)-2 k z-z,
\\
P_{32}&=&3 k^4-2 k^2 (9 z+2)+18 z-7,
\\
P_{33}&=&30 k^4+k^2 \bigl(-60 z^2+63 z+28\bigr)+16 z^2,
\\
P_{34}&=&3 k^4 \bigl(z^2+z-1\bigr)+2 k^2 z^2-z^2,
\\
P_{35}&=&3 k^4 \bigl(z^2+3\bigr)+k^2 \bigl(2 z^2+3\bigr)-z^2,
\\
P_{36}&=&-9 k^5 (z-2)+6 k^4 \bigl(2 z^2-7 z+10\bigr)-18 k^3 (z+1)+2 k^2 z (4 z+3)+3 k z-4 z^2,
\\
P_{37}&=&9 k^5 (z-2)+6 k^4 \bigl(2 z^2-7 z+10\bigr)+18 k^3 (z+1)+2 k^2 z (4 z+3)-3 k z-4 z^2,
\\
P_{38}&=&3 k^4 (z-8) z+k^2 \bigl(2 z^2+9 z-3\bigr)-z^2,
\\
P_{39}&=&3 k^4-k^2 \bigl(6 z^2+7\bigr)+2 z^2,
\\
P_{40}&=&9 k^7-3 k^5 \bigl(3 z^2+12 z+4\bigr)-6 k^4 z^2 (2 z+11)-3 k^3 \bigl(6 z^2-12 z+7\bigr)
\nonumber \\ &&
-2 k^2 z \bigl(4 z^2-9 z+6\bigr)+3 k z^2+4 z^3,
\\
P_{41}&=&9 k^7-3 k^5 \bigl(3 z^2+12 z+4\bigr)+6 k^4 z^2 (2 z+11)-3 k^3 \bigl(6 z^2-12 z+7\bigr)
\nonumber \\ &&
+2 k^2 z \bigl(4 z^2-9 z+6\bigr)+3 k z^2-4 z^3,
\\
P_{42}&=&-3 k^4+k^2 \bigl(6 z^2+6 z+7\bigr)-2 z^2,
\\
P_{43}&=&6 k^6-k^4 \bigl(9 z^2+18 z+8\bigr)-2 k^2 \bigl(9 z^2-9 z+7\bigr)+3 z^2,
\\
P_{44}&=&3 k^4 \bigl(5 z^2+14 z-6\bigr)+k^2 \bigl(10 z^2-9 z+3\bigr)-5 z^2,
\\
P_{45}&=&3 k^6-k^4 \bigl(9 z^2+4\bigr)-k^2 \bigl(18 z^2+7\bigr)+3 z^2,
\\
P_{46}&=&3 k^4 \bigl(6 z^3+9 z^2-z+2\bigr)+k^2 z \bigl(3 z^2+8 z+9\bigr)-z^2 (3 z+1),
\\
P_{47}&=&6 \beta  k^7+24 k^6+2 \beta  k^5 \bigl(27 z^2+27 z+28\bigr)+2 k^4 \bigl(9 z^2+27 z-2\bigr)
\nonumber \\ &&
-\beta  k^3 \bigl(36 z^3+27 z^2-93 z+52\bigr)+k^2 \bigl(-36 z^3+21 z^2+93 z-10\bigr)
\nonumber \\ &&
+3 \beta  k z \bigl(4 z^2+z-1\bigr)+3 z \bigl(4 z^2-3 z-1\bigr),
\\
P_{48}&=&6 \beta  k^7-24 k^6+2 \beta  k^5 \bigl(27 z^2+27 z+28\bigr)-2 k^4 \bigl(9 z^2+27 z-2\bigr)
\nonumber \\ &&
-\beta  k^3 \bigl(36 z^3+27 z^2-93 z+52\bigr)+k^2 \bigl(36 z^3-21 z^2-93 z+10\bigr)
\nonumber \\ &&
+3 \beta  k z \bigl(4 z^2+z-1\bigr)+3 z \bigl(-4 z^2+3 z+1\bigr),
\\
P_{49}&=&-6 (\beta -1) k^7 (z-2)+6 k^6 z (\beta +z-6)+k^5 \bigl(-28 \beta +3 (4 \beta -3) z^3-3 (8 \beta -5) z^2
\nonumber \\ &&
+2 (7 \beta -22) z+40\bigr)+k^4 \bigl((9-12 \beta ) z^3-8 z^2+(30-14 \beta ) z+12\bigr)
\nonumber \\ &&
+2 k^3 z \bigl(-2 \beta  z^2+(4 \beta +2) z+7\bigr)+2 k^2 z \bigl(2 \beta  z^2+z-1\bigr)+k (z-3) z^2-z^3,
\\
P_{50}&=&-6 (\beta -1) k^7 (z-2)-6 k^6 z (\beta +z-6)+k^5 \bigl(-28 \beta +3 (4 \beta -3) z^3-3 (8 \beta -5) z^2
\nonumber \\ &&
+2 (7 \beta -22) z+40\bigr)+k^4 \bigl(3 (4 \beta -3) z^3+8 z^2+2 (7 \beta -15) z-12\bigr)
\nonumber \\ &&
+2 k^3 z \bigl(-2 \beta  z^2+(4 \beta +2) z+7\bigr)-2 k^2 z \bigl(2 \beta  z^2+z-1\bigr)+k (z-3) z^2+z^3,
\\
P_{51}&=&6 (\beta +1) k^7 (z-2)-6 k^6 z (-\beta +z-6)+k^5 \bigl(28 \beta -3 (4 \beta +3) z^3+3 (8 \beta +5) z^2
\nonumber \\ &&
-2 (7 \beta +22) z+40\bigr)-k^4 \bigl(3 (4 \beta +3) z^3-8 z^2+2 (7 \beta +15) z+12\bigr)
\nonumber \\ &&
+2 k^3 z \bigl(2 \beta  z^2+(2-4 \beta ) z+7\bigr)+2 k^2 z \bigl(2 \beta  z^2-z+1\bigr)+k (z-3) z^2+z^3,
\\
P_{52}&=&6 (\beta +1) k^7 (z-2)+6 k^6 z (-\beta +z-6)+k^5 \bigl(28 \beta -3 (4 \beta +3) z^3+3 (8 \beta +5) z^2
\nonumber \\ &&
-2 (7 \beta +22) z+40\bigr)+k^4 \bigl(3 (4 \beta +3) z^3-8 z^2+2 (7 \beta +15) z+12\bigr)
\nonumber \\ &&
+2 k^3 z \bigl(2 \beta  z^2+(2-4 \beta ) z+7\bigr)-2 k^2 z \bigl(2 \beta  z^2-z+1\bigr)+k (z-3) z^2-z^3,
\\
P_{53}&=&54 \beta  k^8 (z-2)^2 z-3 k^6 \bigl(-24 (\beta +1)+(\beta -35) z^5+(5 \beta +113) z^4-(47 \beta +125) z^3
\nonumber \\ &&
+6 (15 \beta +31) z^2-240 z\bigr)+k^4 z \bigl(72 (3 \beta -4)+(59 \beta -193) z^4+(187-173 \beta ) z^3
\nonumber \\ &&
+2 (82 \beta -143) z^2-6 (17 \beta +5) z\bigr)-k^2 z^2 \bigl(12 (\beta +1)+3 (23 \beta -37) z^3
\nonumber \\ &&
+(11-25 \beta ) z^2+(103 \beta -167) z\bigr)+z^4 (3 \beta +13 \beta  z-23 z+3),
\\
P_{54}&=&54 \beta  k^8 (z-2)^2 z-3 k^6 \bigl(-24 (\beta -1)+(\beta +35) z^5+(5 \beta -113) z^4+(125-47 \beta ) z^3
\nonumber \\ &&
+6 (15 \beta -31) z^2+240 z\bigr)+k^4 z \bigl(72 (3 \beta +4)+(59 \beta +193) z^4-(173 \beta +187) z^3
\nonumber \\ &&
+2 (82 \beta +143) z^2-6 (17 \beta -5) z\bigr)-k^2 z^2 \bigl(12 (\beta -1)+3 (23 \beta +37) z^3
\nonumber \\ &&
-(25 \beta +11) z^2+(103 \beta +167) z\bigr)+z^4 (3 \beta +13 \beta  z+23 z-3),
\\
P_{55}&=&9 \beta ^2 k^9 (z-1)+k^7 \bigl(12 \beta ^2+\bigl(9-54 \beta ^2\bigr) z^2+6 \bigl(7 \beta ^2-3\bigr) z\bigr)+6 k^6 z^2 \bigl(-11 \beta ^2
\nonumber \\ &&
+3 \beta ^2 z^2+8 \beta ^2 z+z\bigr)+k^5 \bigl(21 \beta ^2-9 z^3+18 \bigl(3 \beta ^2+2\bigr) z^2+\bigl(18-75 \beta ^2\bigr) z\bigr)
\nonumber \\ &&
+2 k^4 z \bigl(-6 \beta ^2+\bigl(6 \beta ^2-3\bigr) z^3+\bigl(2-15 \beta ^2\bigr) z^2+15 \beta ^2 z\bigr)
\nonumber \\ &&
-3 k^3 z^2 (6 z+7)-2 k^2 z^3 \bigl(-3 \beta ^2+\bigl(3 \beta ^2+2\bigr) z+1\bigr)+3 k z^3+2 z^4.
\end{eqnarray}

The remaining Mellin convolutions in Eqs.~(\ref{eq:HFL},\ref{eq:HF1}) are given in Appendix~\ref{sec:A3}, with
\begin{eqnarray}
A(x) \otimes B(x) = \int_0^1 dx_1 \int_0^1 dx_2 \delta(x - x_1 x_2) A(x_1) B(x_2).
\end{eqnarray}
The Wilson coefficient $H_{2,q}^{(2),\text{PS}}$ is given by
\begin{eqnarray}
\label{eq:HF2}
        H_{2,q}^{(2),\text{PS}} = \frac{1}{2} \left( H_{1,q}^{(2),\text{PS}} + 3 H_{L,q}^{(2),\text{PS}} \right)~.
\end{eqnarray}
In summary, the two--loop massive Wilson coefficients are represented in terms of iterated integrals over the alphabets
given in Section~\ref{sec:4}. 
The integrals can be arranged such that only the last integral contains elliptic letters and
all other integrals can be expressed in terms of classical polylogarithms with involved arguments.
Some details are discussed in Appendix~\ref{sec:A4}.
Similar structures are expected also for other physical processes depending on two scales, $z$ and
$m^2/Q^2$, in a non-factorizing manner. Even more involved structures will emerge in the case of more scales.
The two--loop heavy flavor contributions to the structure functions $F_{2(L)}$ are given by
\begin{eqnarray}
F_{2(L)}^{(2),\rm PS, heav.}(x,Q^2) = a_s^2(Q^2) Q_H^2 x H_{2(L)}^{\rm PS,(2)}\left(\frac{Q^2}{\mu^2},x\right) \otimes 
\Sigma(x,\mu^2).
\end{eqnarray}
\section{The asymptotic and threshold expansions}
\label{sec:6}

\vspace*{1mm}
\noindent
The complete expressions calculated in Section~\ref{sec:5} allow now to perform the asymptotic expansion
for $Q^2 \gg m^2$ and the threshold expansion for $\beta \ll 1$.
In the asymptotic limit $Q^2 \gg m^2$ the massive pure singlet Wilson coefficient have the following representations
\cite{Buza:1995ie,Bierenbaum:2007qe} 
\begin{eqnarray}
\label{eq:AL}
H_{L,q}^{(2),\rm PS}\left(z,\frac{Q^2}{m^2}\right) &=& \tilde{C}_{q,L}^{(2),\rm PS}(N_F+1),
\\
\label{eq:A2}
H_{2,q}^{(2),\rm PS}\left(z,\frac{Q^2}{m^2}\right) &=&  A_{Qq}^{(2),\rm PS}(N_F+1) + \tilde{C}_{q,2}^{(2),\rm 
PS}(N_F+1).
\end{eqnarray}  
Here the massless Wilson coefficients $\tilde{C}_{q,L}^{(2),\rm PS}(N_F+1)$ are the ones given in 
Section~\ref{sec:3} normalized by $N_F+1$. The massive two--loop operator matrix element  $A_{Qq}^{(2),\rm PS}$
in Mellin space in the $\overline{\sf MS}$ scheme \cite{Buza:1995ie,Bierenbaum:2007qe} reads
\begin{eqnarray}
A_{Qq}^{(2),\rm PS} &=& -\frac{1}{8} \hat{P}_{qg}^{(0)} P_{gq}^{(0)} \ln^2\left(\frac{m^2}{\mu^2}\right) 
- \frac{1}{2} \hat{P}_{qq}^{(1), \rm PS} \ln\left(\frac{m^2}{\mu^2}\right) + \frac{1}{8} \hat{P}_{qg}^{(0)} P_{gq}^{(0)}
\zeta_2 + a_{Qq}^{(2),\rm PS}.
\end{eqnarray}  
The constant part of the unrenormalized OME $a_{Qq}^{(2),\rm PS}$ is given by
\begin{eqnarray}
a_{Qq}^{(2),\rm PS}(z) &=& C_F T_F \Biggl\{
        - \frac{4 (1-z) \big(
                112+121 z+400 z^2\big)}{27 z}
        -\left(
                \frac{8}{9} \big(
                        21+33 z+56 z^2\big)
                +8 (1+z) \zeta_2
        \right) \HA_0
\nonumber\\ &&
        +\frac{2}{3} \big(
                3+15 z+8 z^2\big) \HA_0^2
        -\frac{4}{3} (1+z) \HA_0^3
        +\frac{8 (1-z) \big(
                4+7 z+4 z^2\big)}{3 z} \HA_0 \HA_1
\nonumber\\ &&
        -\Biggl[
                \frac{8 (1-z) \big(
                        4+7 z+4 z^2\big)}{3 z}
                -16 (1+z) \HA_0
        \Biggr] \HA_{0,1}
\nonumber\\ && 
       -32 (1+z) \HA_{0,0,1}
        -\frac{4 (1-z) \big(
                4+7 z+ 4 z^2\big)}{3 z} \zeta_2
        +32 (1+z) \zeta_3
\Biggr\}
\end{eqnarray}  
in $z$-space.

Expanding the fully massive result given in Section~\ref{sec:5} in the asymptotic limit $Q^2 \gg m^2$
and setting $\mu^2 = Q^2$
we find
\begin{eqnarray}
        H_{L,q}^{2,\rm PS} &=&
-32 C_F T_F  \Biggl\{
        \frac{(1-z) \big(1-2 z+10 z^2\big)}{9 z}
        - (1+z) (1-2 z) \HA_0
        - z \HA_0^2
\nonumber\\ &&
        +\frac{(1-z) \big(1-2 z-2 z^2\big)}{3 z} \HA_1
        - z \HA_{0,1}
        + z \zeta_2
+ \textcolor{blue}{\frac{m^2}{Q^2}} \biggl[
        -\frac{(1-z) \big(2-z+2 z^2\big) }{3 z} \ln^2 \left( \textcolor{blue}{\frac{m^2}{Q^2}} \right)
\nonumber \\ &&
        +\frac{(1-z) \big(-22+4 z+29 z^2\big)}{9 z}
        -\biggl(
         \frac{(1-z) \big( 20 - 7 z-25 z^2\big)}{9 z}
        +\frac{2}{3} \big(3-6 z
\nonumber \\ &&
        -2 z^2\big) \HA_0 
        \biggr) \ln \left( \textcolor{blue}{\frac{m^2}{Q^2}} \right)
        +\biggl(
        \frac{2}{9} \big(-6+3 z+13 z^2\big)
        +\frac{2 (1+z) \big(-2+z+2 z^2+2 z^3\big)}{3 z} 
\nonumber \\ &&
\times \HA_{-1}
        \biggr) \HA_0
        -\frac{2}{3} z^3 \HA_0^2
        +\biggl(
        -\frac{(1-z)^2 (14+13 z)}{9 z}
        +\frac{4 (1-z) \big(2-z+2 z^2\big)}{3 z} \HA_0
        \biggr) \HA_1
\nonumber \\ &&
        +\frac{(1-z) \big(2-z+2 z^2\big)}{3 z} \HA_1^2
        -\frac{2 \big(4-3 z-4 z^3\big)}{3 z} \HA_{0,1}
\nonumber \\ &&
        +\frac{2 (1+z) \big(2-z-2 z^2-2 z^3\big)}{3 z} \HA_{0,-1}
        -\frac{2 (1-z) \big(2-z+2 z^2+2 z^3\big)}{3 z} \zeta_2
\biggr]
\nonumber \\ &&
+ \left( \textcolor{blue}{\frac{m^2}{Q^2}} \right)^2 \biggl[
        \frac{1}{2z} \big(4-2 z-z^2-2 z^3
+4 z^4\big) \ln^2 \left( \textcolor{blue}{\frac{m^2}{Q^2}} \right)
        +\biggl(
         2 \big(2-3 z+4 z^3\big) \HA_0
\nonumber \\ &&
        + \frac{ (1-z) \big(28-20 z+13 z^2+21 z^3\big)}{6 z}
        + \big(2-3 z-2 z^2+4 z^3\big) \HA_1
        \biggr) \ln \left( \textcolor{blue}{\frac{m^2}{Q^2}} \right)
\nonumber \\ &&
        + \frac{1}{1152 z} \big(16027-13011 z-6267 z^2+7571 z^3+4320 z^4\big)        
        +\biggl(
        \frac{1}{3} \big(24-21 z+16 z^2
\nonumber \\ &&
        -21 z^3\big)
        +\frac{4\big(1-z^2+z^3+2 z^4\big)}{z} \HA_{-1}
        \biggr) \HA_0
        -\biggl(
        \frac{1}{6 z} \big(4-15 z^2-16 z^3+21 z^4\big)
\nonumber \\ &&
        +\frac{4 \big(2-2 z+z^2\big)}{z} \HA_0
        \biggr) \HA_1
        -\frac{1}{2z} \big(4-6 z+5 z^2+2 z^3-4 z^4\big) \HA_1^2
\nonumber \\ && 
        +\frac{2\big(4-2 z-z^2+4 z^4\big)}{z} \HA_{0,1}
        -\frac{4 \big(1-z^2+z^3+2 z^4\big)}{z} \HA_{0,-1}
\nonumber \\ &&
        +\frac{2 \big(2-2 z+z^2\big) }{z} \zeta_2
\biggr]
\Biggr\}
        + O \left( \left(\frac{m^2}{Q^2}\right)^3 \ln^2\left(\frac{m^2}{Q^2}\right)\right)~,
        \\
        H_{2,q}^{2,\rm PS} &=& C_F T_F  \Biggl\{ 
        -\biggl(
        \frac{4 (1-z) \big(4+7 z+4 z^2\big)}{3 z}
        +8 (1+z) \HA_0
        \biggr) \ln^2 \left( \textcolor{blue}{\frac{m^2}{Q^2}} \right)
\nonumber \\ &&
        -\biggl(
        \frac{16 (1-z) \big(10+z+28 z^2\big)}{9 z}
        +\frac{8}{3} \big(3+15 z+8 z^2\big) \HA_0
\nonumber \\ &&
        -8 (1+z) \HA_0^2
        \biggr) \ln \left( \textcolor{blue}{\frac{m^2}{Q^2}} \right)
        +\frac{16 (1-z) \big(5+24 z-52 z^2\big)}{9 z}
\nonumber \\ &&
        +\biggl(
         \frac{8}{9} \big(105-99 z-88 z^2\big)
        -\frac{32 (1+z)^3 }{3 z} \HA_{-1}
        \biggr) \HA_0
        +8 z (5-2 z) \HA_0^2
        +\frac{16}{3} (1+z) \HA_0^3
\nonumber \\ &&
        -\biggl(
        \frac{16 (1-z) \big(13-26 z+4 z^2\big)}{9 z}
        -\frac{16 (1-z) \big(4+7 z+4 z^2\big)}{3 z} \HA_0
        \biggr) \HA_1
\nonumber \\ &&
        +\frac{4 (1-z) \big(4+7 z+4 z^2\big)}{3 z} \HA_1^2
        +\biggl(
        -\frac{16 \big(4+3 z-3 z^2+2 z^3\big)}{3 z}
        +32 (1+z) \HA_0
        \biggr) \HA_{0,1}
\nonumber \\ &&
        +\frac{32 (1+z)^3 }{3 z} \HA_{0,-1}
        -32 (1+z) \HA_{0,0,1}
        +16 (1+z) \HA_{0,1,1}
        -\biggl(
        \frac{32 \big(1+3 z^2-3 z^3\big)}{3 z}
\nonumber \\ &&
        +32 (1+z) \HA_0
        \biggr) \zeta_2
        +16 (1+z) \zeta_3       
        + \textcolor{blue}{\frac{m^2}{Q^2}} \biggl[
        \biggl(
        \frac{16 (1-z) \big(1+2 z^2\big)}{z}
        +16 z \HA_0
        \biggr) \ln^2 \left( \textcolor{blue}{\frac{m^2}{Q^2}} \right)
\nonumber \\ &&
        +\biggl(
        \frac{64 (1-z) \big(2-z-4 z^2\big)}{3 z}
        +32 \big(1-3 z-2 z^2\big) \HA_0
        -16 z \HA_0^2
        \biggr) \ln \left( \textcolor{blue}{\frac{m^2}{Q^2}} \right)
\nonumber \\ &&
        +\frac{8 \big(76-24 z-102 z^2+59 z^3\big)}{9 z}
        +\biggl(
         \frac{32 (1+z) \big(1-z-2 z^2-2 z^3\big)}{z} \HA_{-1}
\nonumber \\ &&
        + \frac{16}{3} \big(6+27 z-20 z^2\big)
        \biggr) \HA_0
        +32 z \big(1+z^2\big) \HA_0^2
        -\frac{32}{3} z \HA_0^3
        -\frac{16 (1-z) \big(1+2 z^2\big) }{z} \HA_1^2
\nonumber \\ &&
        +\biggl(
        \frac{16 \big(4-6 z-9 z^2+8 z^3\big)}{3 z}
        -\frac{64 (1-z) \big(1+2 z^2\big) }{z} \HA_0
        \biggr) \HA_1
\nonumber \\ &&
        +\biggl(
         \frac{32 \big(2-z+z^2-4 z^3\big)}{z}
        -64 z \HA_0
        \biggr) \HA_{0,1}
        -\frac{32 (1+z) \big(1-z-2 z^2-2 z^3\big)}{z} \HA_{0,-1}
\nonumber \\ &&
        +64 z \HA_{0,0,1}
        -32 z \HA_{0,1,1}
        +\biggl(
         \frac{32 (1+z) \big(1-2 z+2 z^2-2 z^3\big)}{z}
        +64 z \HA_0
        \biggr) \zeta_2
        -32 z \zeta_3        
        \biggr]
\nonumber \\ &&
        + \left( \textcolor{blue}{\frac{m^2}{Q^2}} \right)^2 \biggl[
        -\frac{4 P_{61}}{3 z} \ln^2 \left( \textcolor{blue}{\frac{m^2}{Q^2}} \right)
        -\biggl(
        \frac{4 P_{65}}{9 (1-z) z}
        +\frac{16}{3} \big(9-33 z-16 z^2+72 z^3\big) \HA_0
\nonumber \\ &&
        +8 \big(3-11 z-12 z^2+24 z^3\big) \HA_1
        \biggr) \ln \left( \textcolor{blue}{\frac{m^2}{Q^2}} \right)
        +\frac{64 P_{59}}{3 z} \HA_{0,-1}
        -\frac{4 P_{60}}{3 z} \HA_1^2      
        -\frac{16 P_{62}}{3 z} \HA_{0,1}
\nonumber \\ && 
        -\frac{P_{66}}{72 (1-z)^2 z}
        -\biggl(
         \frac{64 P_{59}}{3 z} \HA_{-1}
        +\frac{16 P_{63}}{9 (1-z)}
        \biggr) \HA_0
        +64 z^2 \HA_0^2
        -\biggl(
         \frac{4 P_{64}}{9 (1-z) z}
\nonumber \\ && 
        -\frac{32 \big(16-9 z-3 z^2+8 z^3\big)}{3 z} \HA_0
        \biggr) \HA_1
        -\frac{16 \big(16-9 z-3 z^2+24 z^3\big)}{3 z} \zeta_2
        \biggr]
\Biggr\}
\nonumber \\ &&
        + O \left( \left(\frac{m^2}{Q^2}\right)^3 \ln^2\left(\frac{m^2}{Q^2}\right)\right)~,
\end{eqnarray}
with the polynomials
\begin{eqnarray}
        P_{59} &=& 18 z^4+7 z^3-9 z^2+4 ~,
\\
        P_{60} &=& 72 z^4-52 z^3-27 z^2+27 z-32 ~,
\\
        P_{61} &=& 72 z^4-20 z^3-39 z^2-9 z+32 ~,
\\
        P_{62} &=& 72 z^4-8 z^3-39 z^2-9 z+32 ~,
\\
        P_{63} &=& 180 z^4-391 z^3+265 z^2-111 z+66 ~,
\\
        P_{64} &=& 360 z^5-898 z^4+667 z^3-132 z^2+118 z-88 ~,
\\
        P_{65} &=& 360 z^5-826 z^4+529 z^3+180 z^2-362 z+128 ~,
\\
        P_{66} &=& 12816 z^6-6615 z^5-51371 z^4+62178 z^3+7650 z^2-43867 z+17673 ~.
\end{eqnarray}
We note that the asymptotic terms are exactly reproduced, cf.~\cite{Buza:1995ie,Blumlein:2006mh,Bierenbaum:2007qe}, 
proving the asymptotic factorization in this process.
The additional power suppressed terms can be used to obtain fast numerical implementations for the
heavy quark Wilson coefficients which are valid for lower values of $Q^2$.
The reach of this approximations is discussed in Section~\ref{sec:7}. 

The threshold expansion of the Wilson coefficients for $\beta \ll 1$ is given 
by
\begin{eqnarray}
H_{L,g}^{(1)}\left(z,\frac{Q^2}{m^2}\right) &=& 32 T_F z (1-z) \beta^3 \Biggl\{
\frac{1}{3} +\frac{\beta ^2}{15} +\frac{\beta ^4}{35} +\frac{\beta ^6}{63} 
\Biggl\} + O(\beta^{11}),
\\
H_{2,g}^{(1)}\left(z,\frac{Q^2}{m^2}\right) &=& 4 T_F \beta \Biggl\{
1 +\frac{2}{3} (3 - 2 z) \beta ^2 -\frac{2}{15} \left(3 - 10 z + 4 z^2 \right) \beta ^4
+ \frac{2}{105} \bigl(5 + 2 z 
\nonumber \\ && 
+ 8 z^2 \bigr)\beta ^6 +\frac{2}{315} \bigl( 21 - 22 z + 36 z^2\bigr) \beta ^8
\Biggr\} + O(\beta^{11}),
\\
H_{L,q}^{(2), \rm PS}\left(z,\frac{Q^2}{m^2}\right) &=& C_F T_F z (1-z)^2 \beta^5 \biggl[
-\frac{9856}{225}
+\frac{128}{15} \bigl[ \ln (1-z) -  \ln (z) + 4 \ln (2 \beta ) \bigr]
\nonumber \\ &&
-\beta ^2 \biggl(
         \frac{256}{11025} (2785-2186 z)
        -\frac{256}{105} (5-4 z) \bigl[ \ln (1-z) 
\nonumber \\ && 
        -  \ln (z) + 4 \ln (2 \beta ) \bigr]
\biggr)
-\beta ^4 \Biggl(
         \frac{256}{297675} \big( 93721-162830 z+73888 z^2 \big)
\nonumber \\ &&
        -\frac{128}{945} \big(121-200 z+88 z^2\big) \bigl[ \ln (1-z) -  \ln (z) + 4 \ln (2 \beta ) \bigr]     
\Biggr) 
\Biggr]
\nonumber \\ && 
+ O(\beta^{11})~,
\\
H_{2,q}^{(2), \rm PS}\left(z,\frac{Q^2}{m^2}\right) &=& C_F T_F (1-z) \beta^3 \Biggl[
-\frac{208}{9}
+\frac{16}{3} \bigl[ \ln (1-z) -  \ln (z) + 4 \ln (2 \beta ) \bigr]
\nonumber \\ &&
-\beta ^2 \biggl(
         \frac{16}{225} ( 817 - 496 z)
        -\frac{16}{15} ( 11 - 8 z) \bigl[ \ln (1-z) -  \ln (z) + 4 \ln (2 \beta ) \bigr]
\biggr)
\nonumber \\ &&
-\beta ^4 \biggl(
         \frac{64}{11025} \big(10649-11942 z+2358 z^2+1260 z^3\big)
        -\frac{16}{105} \big(79-112 z
\nonumber \\ &&
        +48 z^2\big) \bigl[ \ln (1-z) -  \ln (z) + 4 \ln (2 \beta ) \bigr]
\biggr)
-\beta ^6 \Biggl(
        \frac{32}{297675} \big(673297
\nonumber \\ && 
        -1361520 z
        +934476 z^2-13048 z^3-120960 z^4\big)
        -\frac{16}{945} \big(817-1800 z
\nonumber \\ && 
+1536 z^2-448 z^3\big) 
         \bigl[ \ln (1-z) -  \ln (z) + 4 \ln (2 \beta ) \bigr]
\Biggr)
\biggr] + O(\beta^{11})~.
\end{eqnarray}

\section{Numerical results}
\label{sec:7}

\begin{figure}[h!]
\centering
\includegraphics[width=0.7 \linewidth]{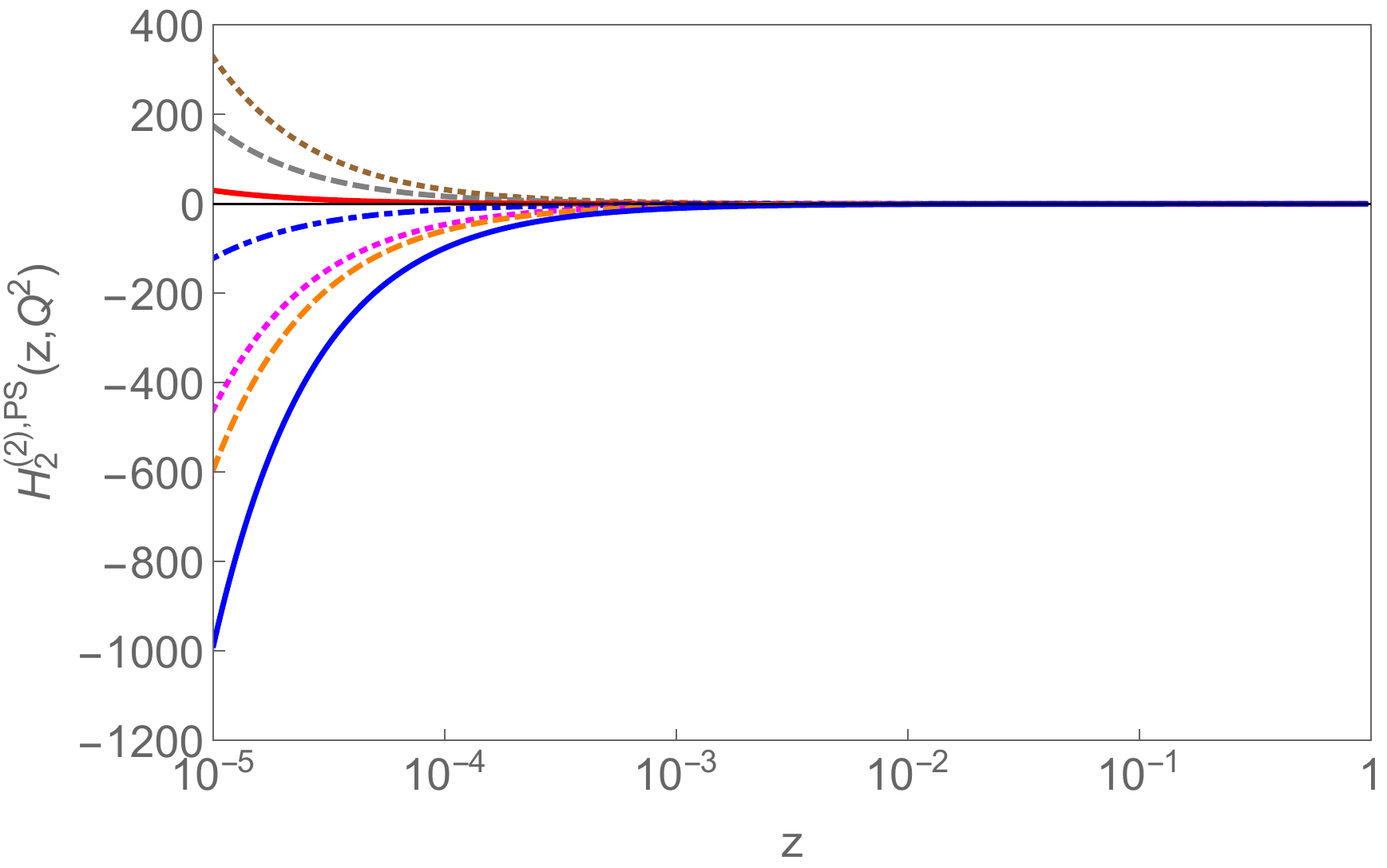}
\includegraphics[width=0.7 \linewidth]{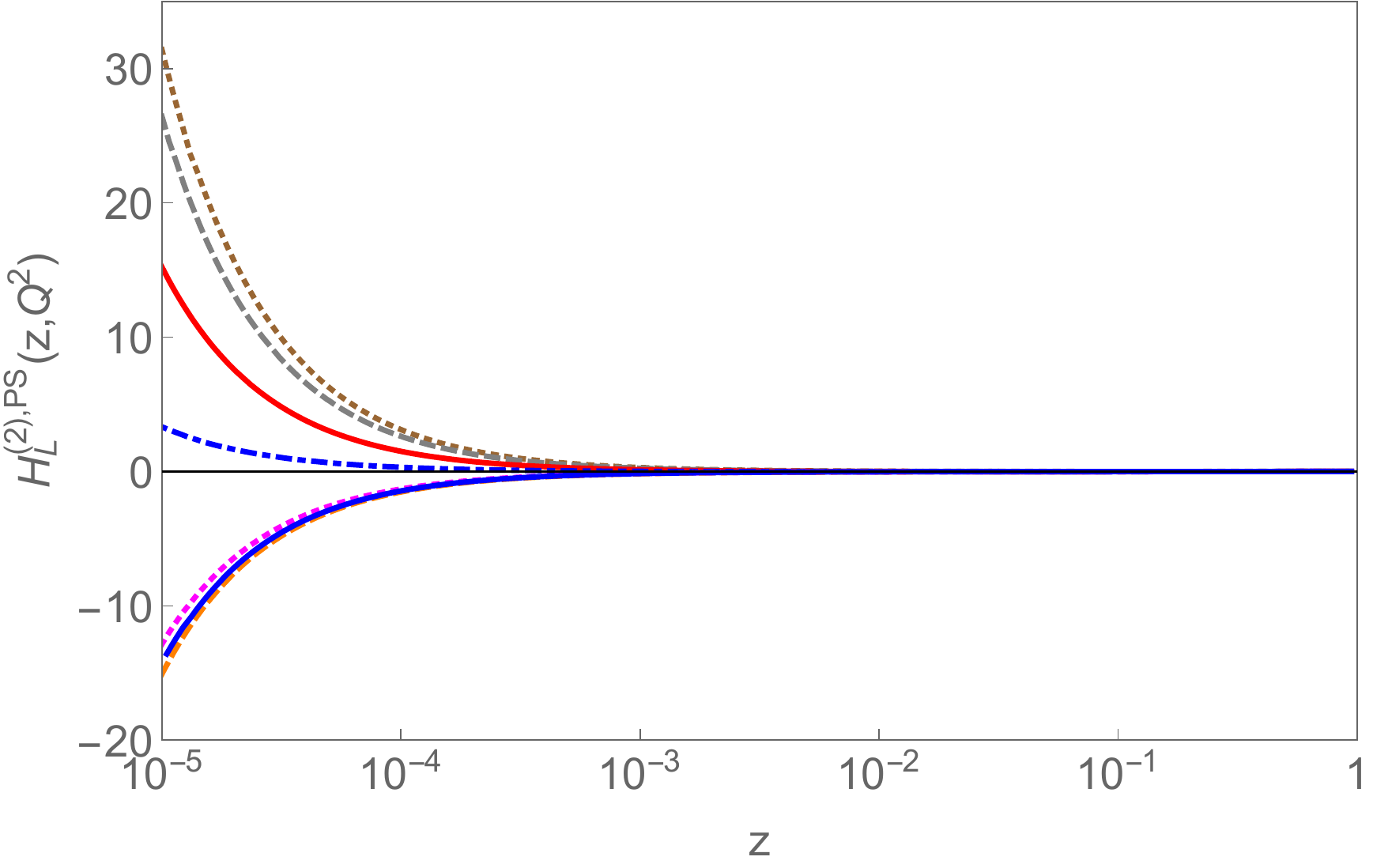}
\caption{\small \sf The Wilson coefficients $H_{2,q}^{2,\rm PS}$ (upper panel) and $H_{L,q}^{2,\rm PS}$ (lower panel) 
as a function of $z$ for 
different values of $Q^2$ and the scale choice $\mu^2 = \mu_F^2 = Q^2$. 
Lower full line (Blue):      $Q^2=10^4~\GeV^2$; 
lower dashed line (Orange):  $Q^2=10^3~\GeV^2$; 
lower dotted line (Magenta): $Q^2=500~\GeV^2$; 
dash-dotted line (Blue):     $Q^2=100~\GeV^2$; 
upper full line (Red):       $Q^2=50~\GeV^2$; 
upper dashed line (Gray):    $Q^2=25~\GeV^2$; 
upper dotted line (Brown):   $Q^2=10~\GeV^2$.}
\label{fig:HLandH2}
\end{figure}

\vspace*{1mm}
\noindent
Let us now illustrate the analytic results numerically. In Figure~\ref{fig:HLandH2} the two--loop heavy flavor
Wilson coefficients  are illustrated  as a function of $z$ for different values of $Q^2 \in [10, 10^4]~\GeV^2$,
setting the charm quark mass to $m_c = 1.59~\GeV$, 
cf.~\cite{Ablinger:2014vwa},].
For large values of $Q^2$ these results compare to Ref.~\cite{Ablinger:2014nga} for $H_{2,q}^{2,\rm PS}$.

\begin{figure}[h!]
\includegraphics[width=0.49 \linewidth]{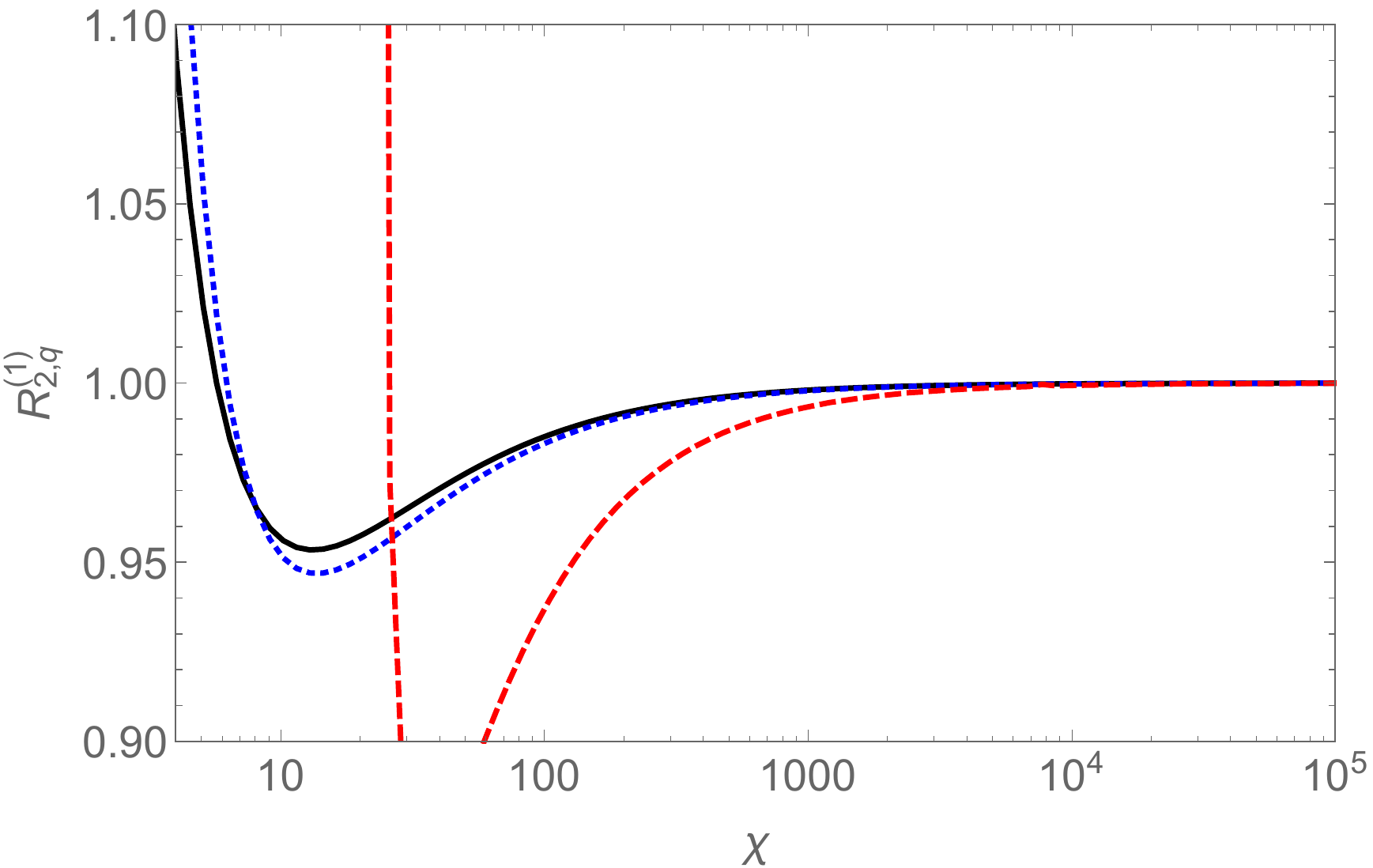}
\includegraphics[width=0.49 \linewidth]{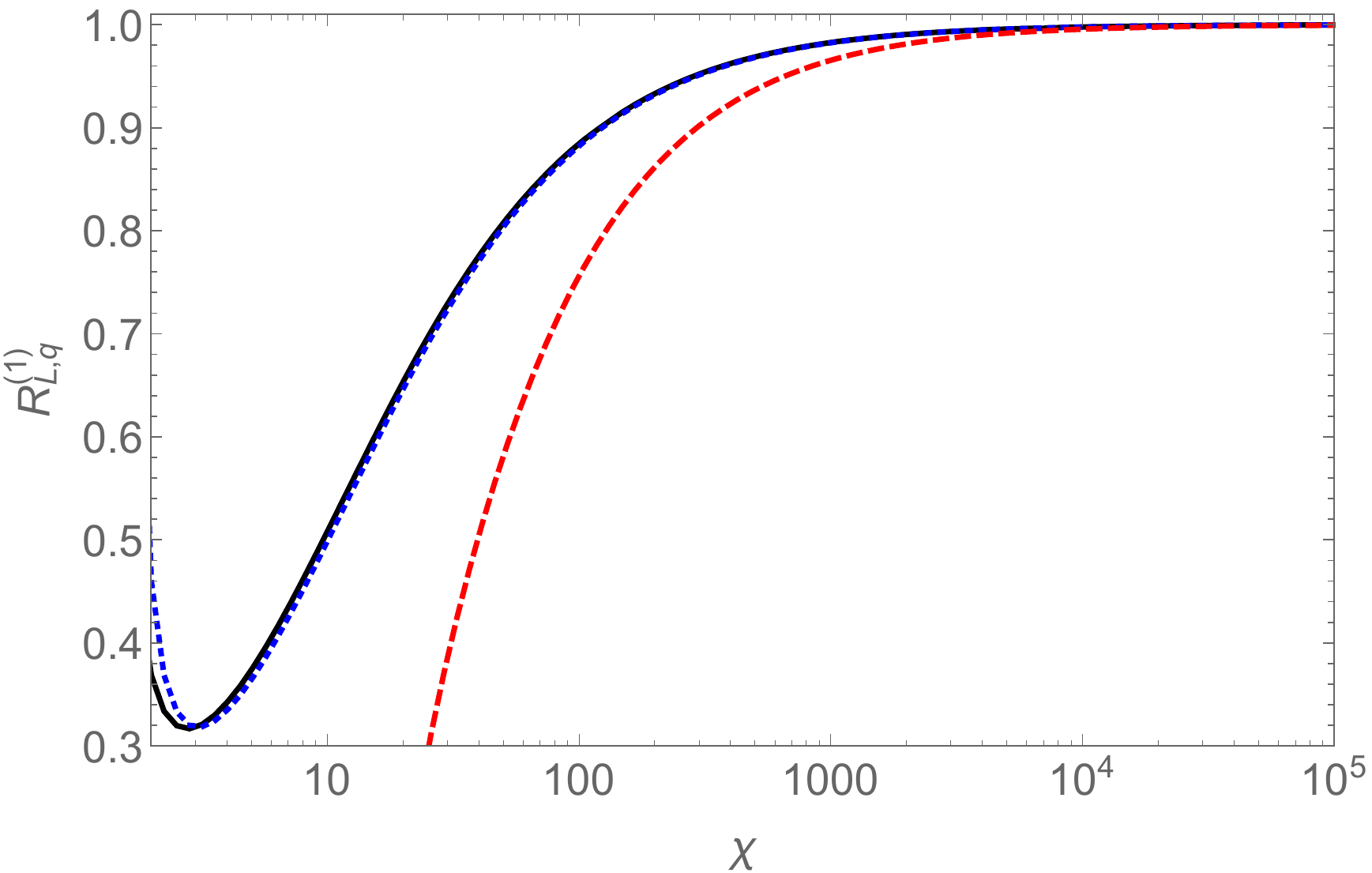}
\caption{\sf \small The ratios $R_{2,q}^{(1)}$ (left) and $R_{L,q}^{(1)}$ (right), Eq.~(\ref{eq:RAT1}),  as a function 
of $\chi=Q^2/m^2$. Solid line: $z=10^{-4}$; dotted line: $z=10^{-2}$; dashed line: $z=1/2$.}
\label{fig:RAT1}
\end{figure}

\begin{figure}[h!]
\includegraphics[width=0.49 \linewidth]{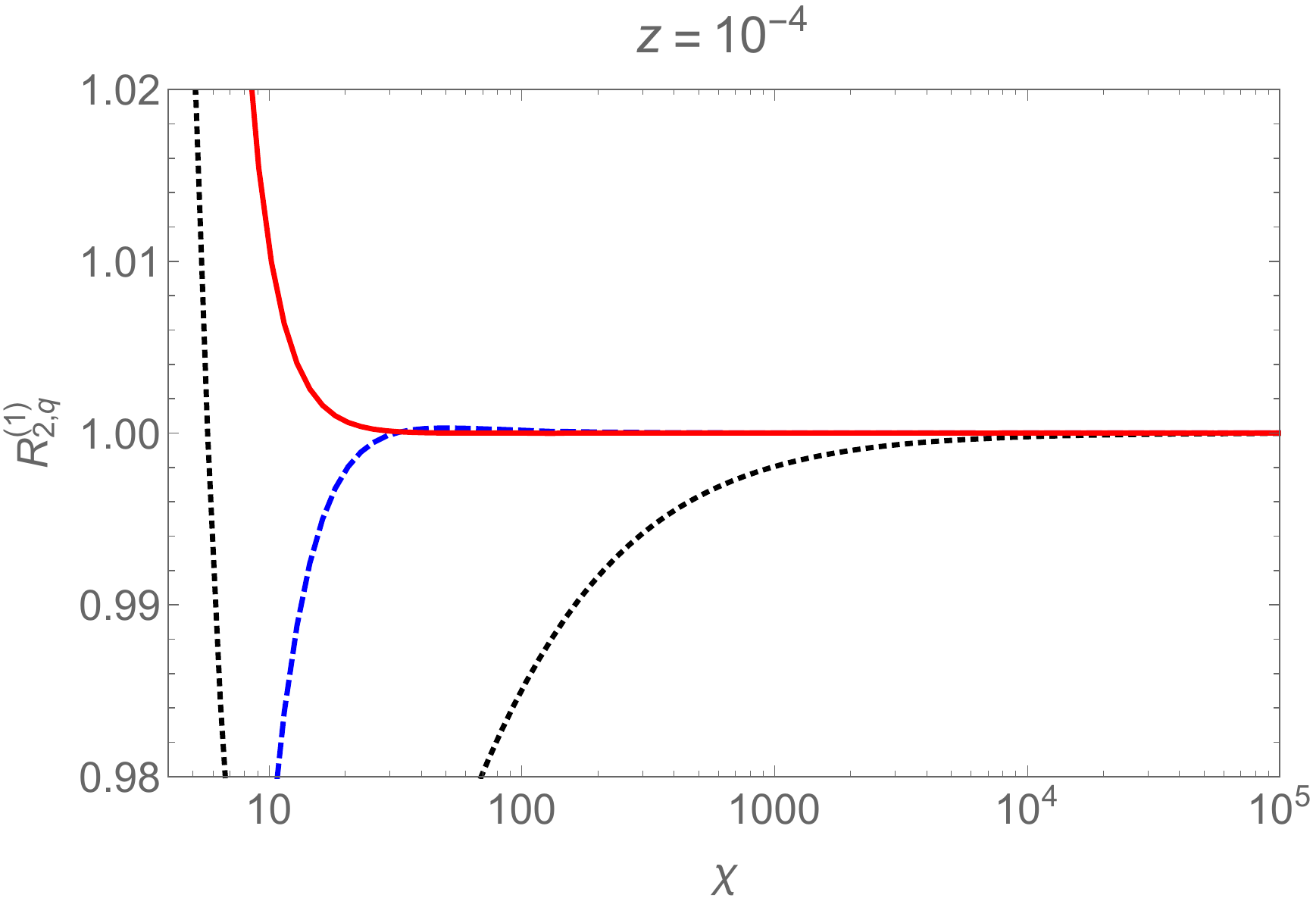}
\includegraphics[width=0.49 \linewidth]{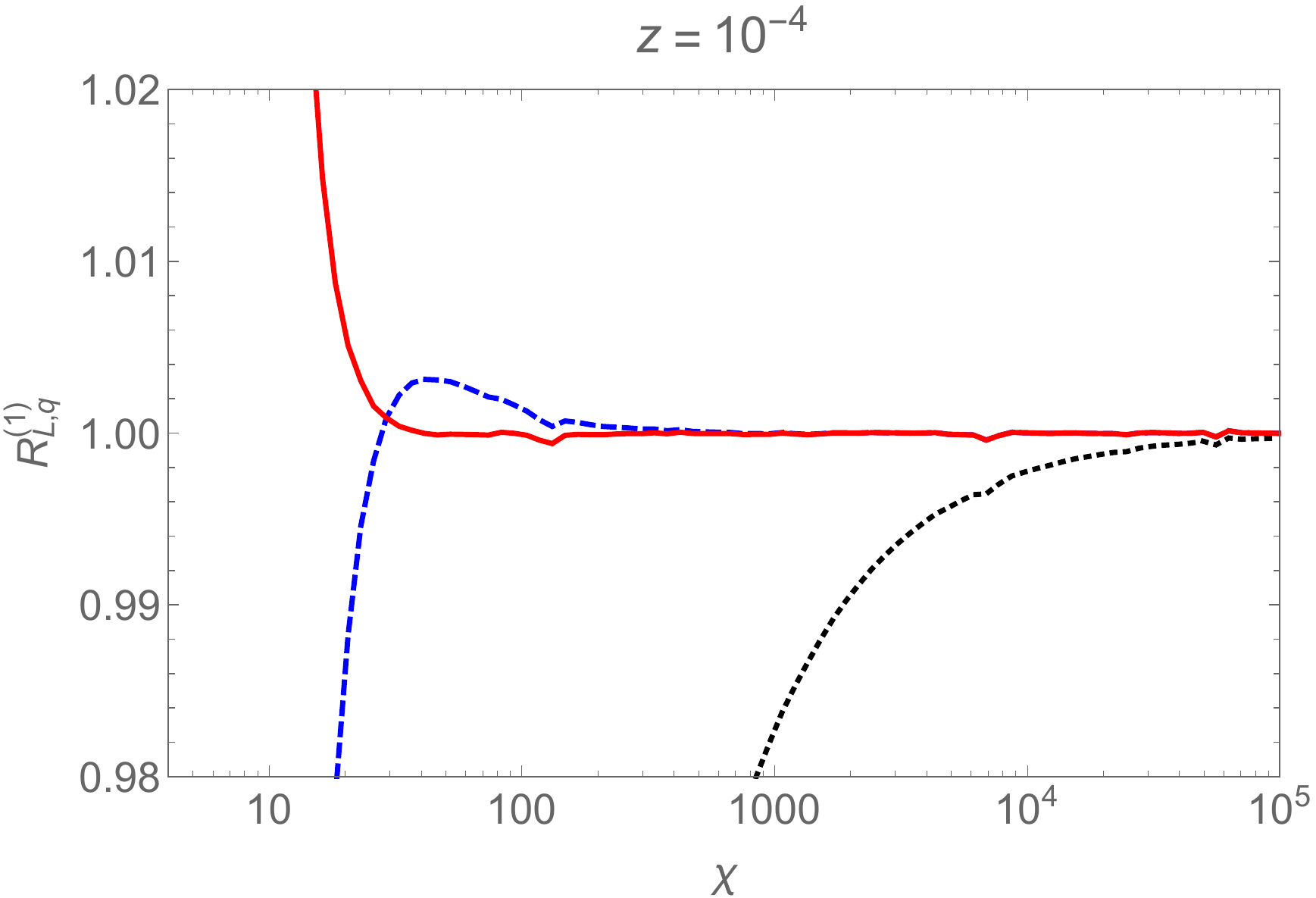}
\includegraphics[width=0.49 \linewidth]{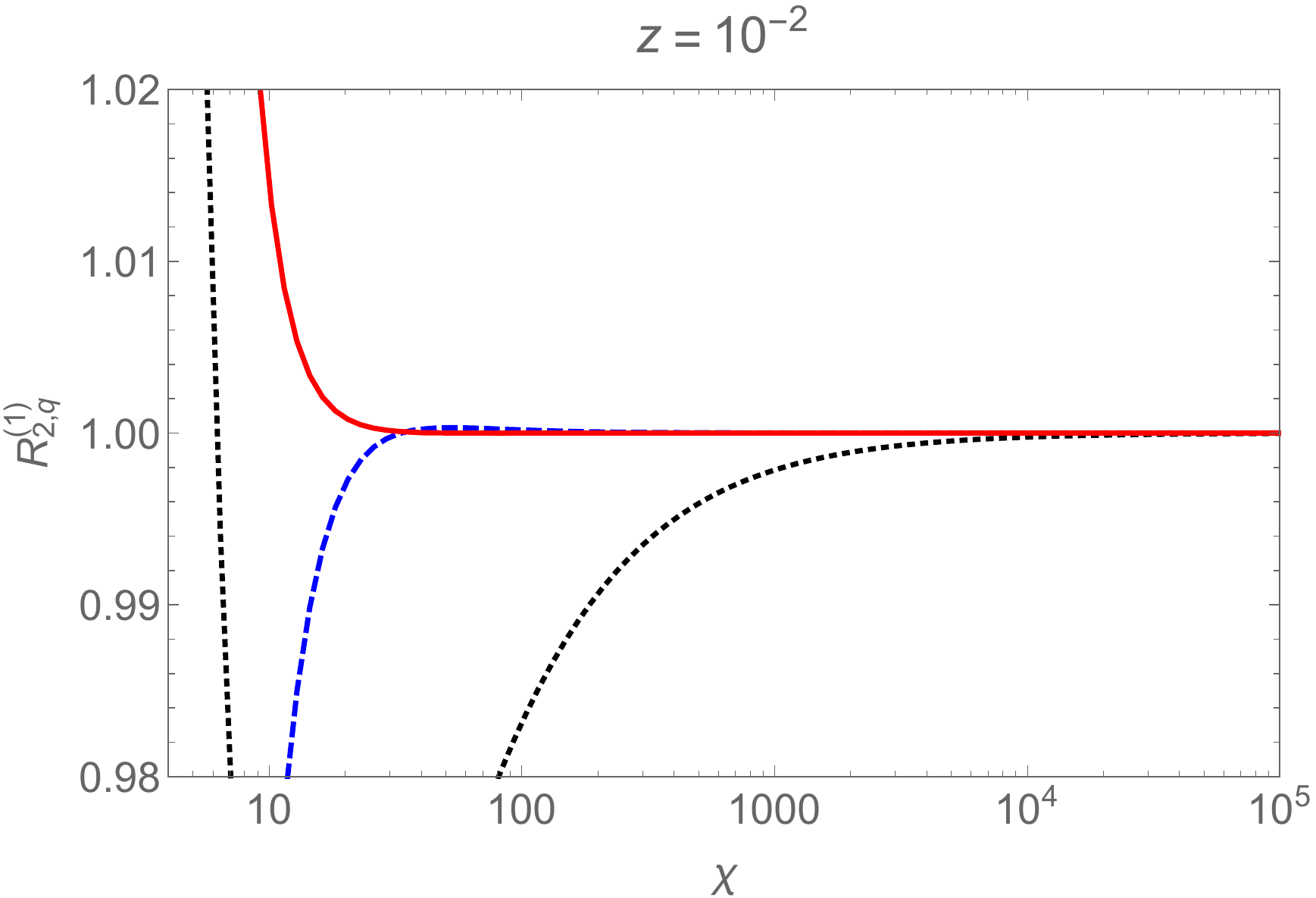}
\includegraphics[width=0.49 \linewidth]{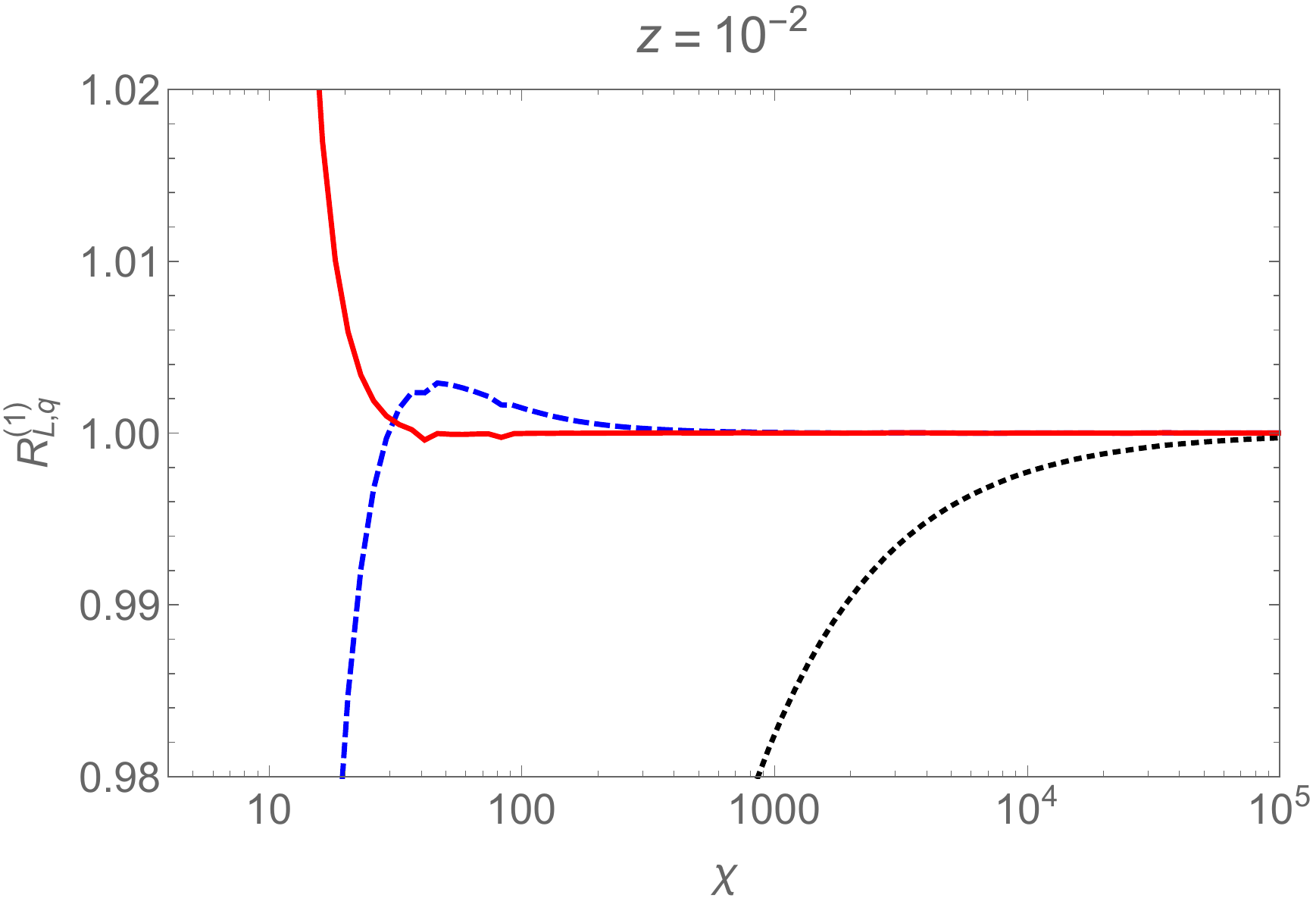}
\includegraphics[width=0.49 \linewidth]{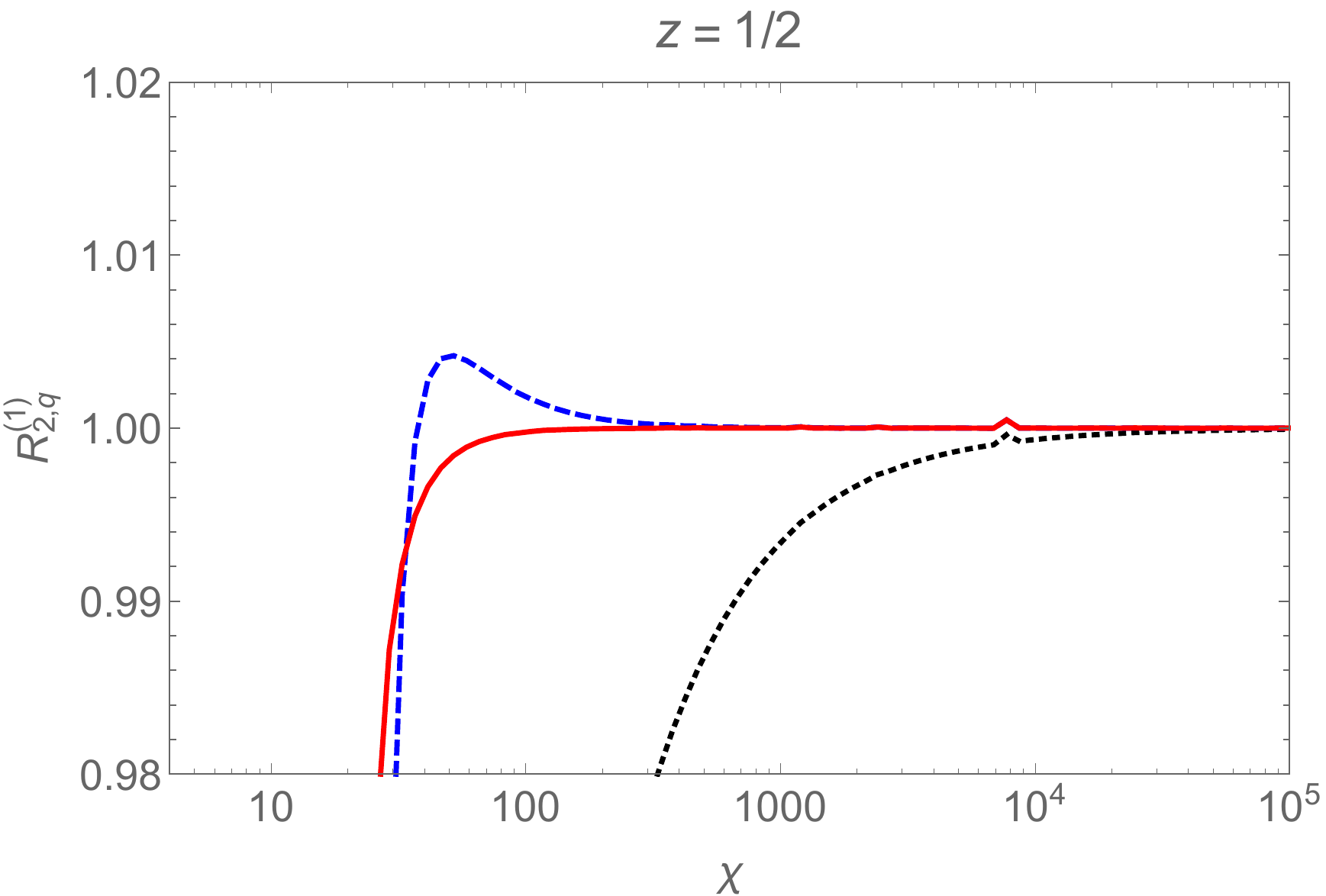}
\includegraphics[width=0.49 \linewidth]{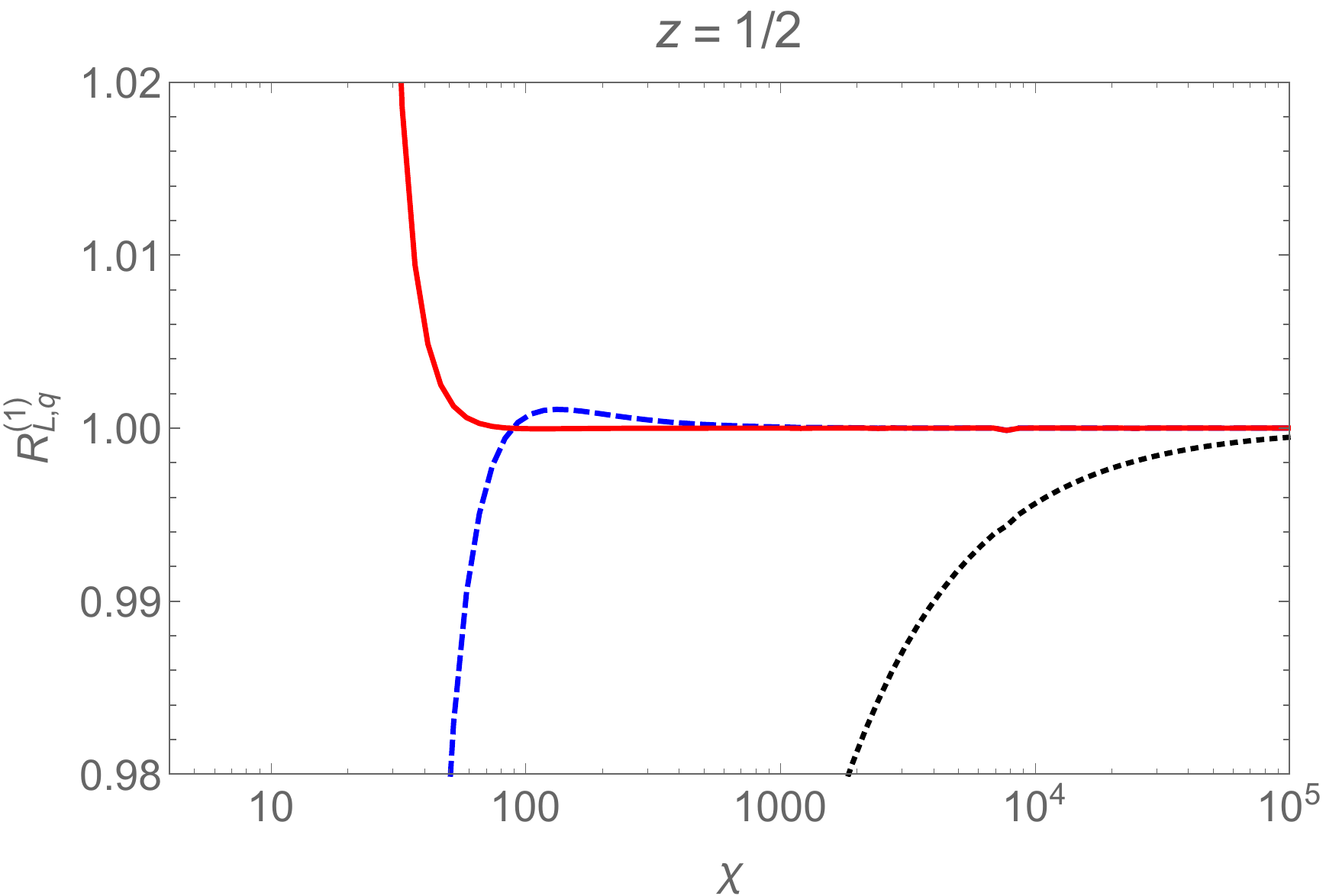}
\caption{\small \sf The ratios $R_{2,q}^{(1)}$ (left) and $R_{L,q}^{(1)}$ (right), Eq.~(\ref{eq:RAT1}),                    
as a function of $\chi=Q^2/m^2$ for different values of $z$ gradually improved with $\kappa$ suppressed terms.
Dotted lines: asymptotic result; dashed lines: $O(m^2/Q^2)$ improved; solid lines : $O((m^2/Q^2)^2)$ improved.}
\label{fig:RLandR2b}
\end{figure}

Next we study the ratios
\begin{eqnarray}
\label{eq:RAT1}
        R_{i,q}^{(1)} &=& \frac{H_{i,q}^{2,\rm PS}}{\tilde{H}_{i,q}^{2,\rm PS}} (\mu =\mu_F =m)~,
\end{eqnarray}
cf. also~\cite{Buza:1995ie}, comparing the full (\ref{eq:HFL}, \ref{eq:HF2}) and the asymptotic results, $\tilde{H}$,
(\ref{eq:AL}, 
\ref{eq:A2}) in Figure~\ref{fig:RAT1}.
For $H_{2,q}^{2,\rm PS}$ the asymptotic expansion agrees with the full calculation up to $Q^2/m^2 \equiv \chi = 100$ to about 
$2\%$ for the 
small values of $z=10^{-4},10^{-2}$. Extending the asymptotic representation down to $\chi=10$ does not introduce an 
error
larger than $5\%$ in this region. At larger $z$ (here $z=1/2$) the asymptotic representation begins to deviate significantly  
from the full calculation beginning at $\chi \sim 1000$. However, the Wilson coefficients are very small in this region.
As it was already noted earlier \cite{Buza:1995ie} the asymptotic representation for $H_{L,q}^{2,\rm PS}$ is only valid for 
much higher values of $\chi$. Demanding an agreement of $\leq 2\%$ requires $\chi > 900$ for the small values of $z$ and even 
higher values for larger $z$.
Similar to the ratio of the full and asymptotic Wilson coefficient we define the ratio 
\begin{eqnarray}
\label{eq:RAT2}
        R_{F_{i}} &=& \frac{ F_{i,q}^{(2), \rm PS} }{ \tilde{F}_{i,q}^{(2), \rm PS} },
\end{eqnarray}
where $\tilde{F}_{i,q}^{(2), \rm PS}$ is the structure function obtained by using the expansion of the
respective Wilson coefficient up the desired level. The corresponding results are depicted in Figure~\ref{fig:RLandR2b}.
We use the parameterization of the parton distribution \cite{Alekhin:2017kpj} at NNLO to better compare 
previous numerical results \cite{Ablinger:2014nga}. We used the {\tt LHAPDF} interface \cite{Buckley:2014ana}.
Demanding an agreement within $\pm 2\%$ for $F_2$ in the range $z \in [10^{-4}, 10^{-2}, 1/2]$ leads to values
$Q_0^2/m^2 \in [8, 9, 15]$ of the $O((m^2/Q^2)^2)$ improved result, $Q_0^2/m^2 \in [10, 12, 30]$ of the $O(m^2/Q^2)$ improved 
result, and $Q_0^2/m^2 \in [70, 80, 300]$ for the asymptotic result. For $F_L$ the corresponding values are
$Q_0^2/m^2 \in [15, 15, 30]$ of the $O((m^2/Q^2)^2)$ improved result, $Q_0^2/m^2 \in [15, 18, 40]$ of the $O((m^2/Q^2)$ 
improved   
result, and $Q_0^2/m^2 \in [200, 200, 700]$ for the asymptotic result. The values of $Q_0^2$ 
for $F_L$ are thus larger than those for $F_2$.

\begin{figure}[h!]
\centering
\includegraphics[width=0.49 \linewidth]{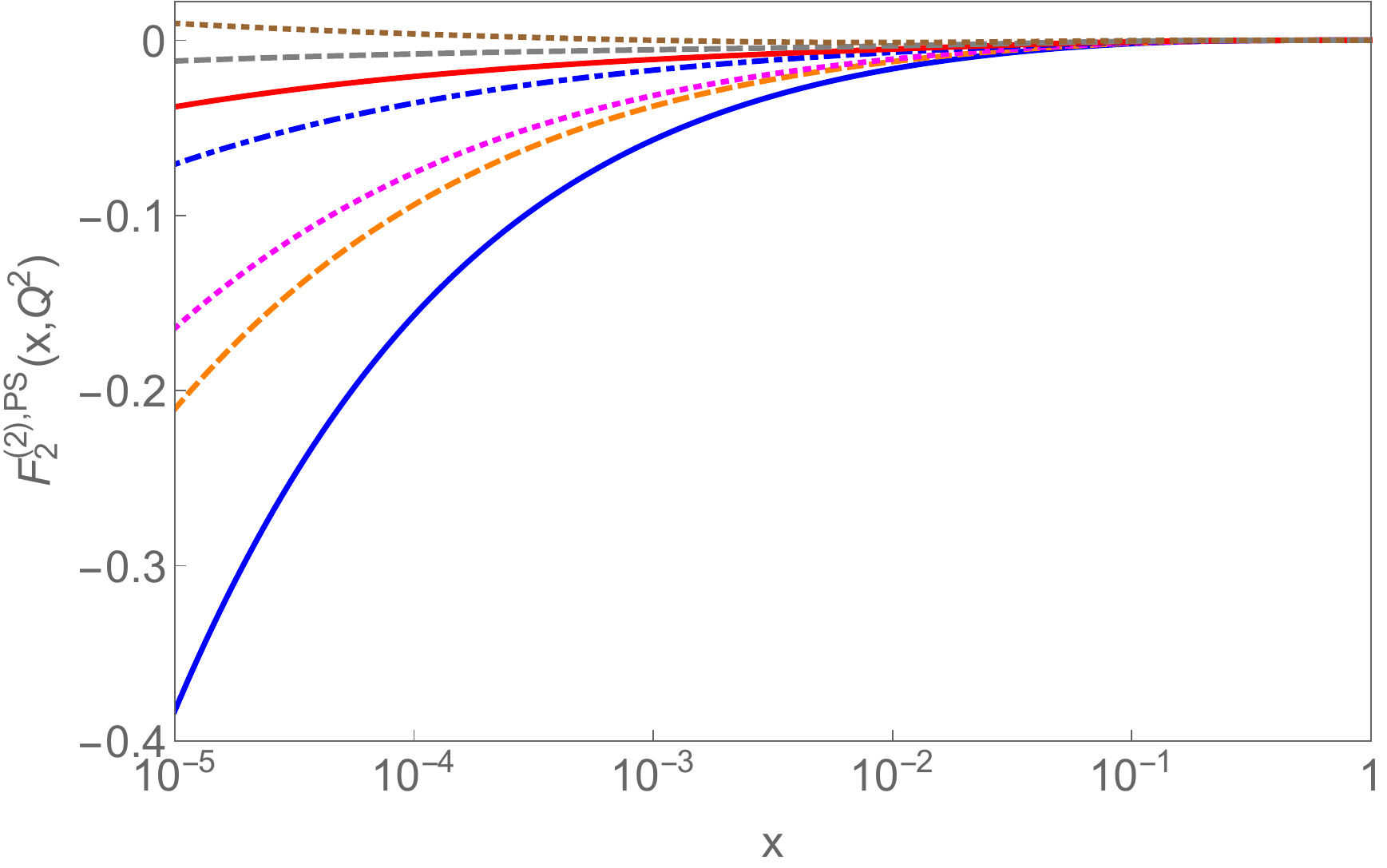}
\includegraphics[width=0.49 \linewidth]{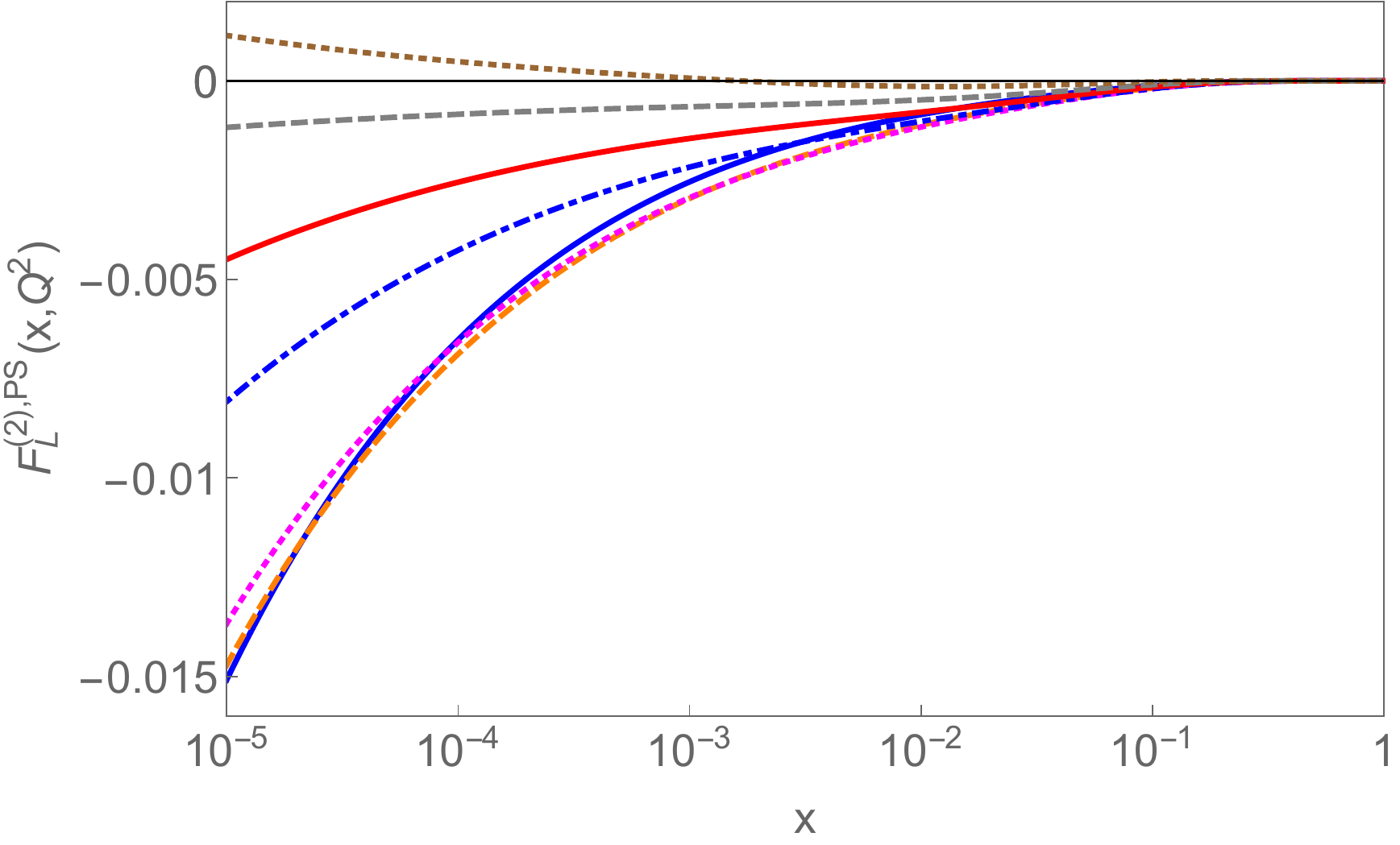}
\caption{\small \sf The pure singlet contributions $F_{2,q}^{2,\rm PS}$ (upper panel) and $F_{L,q}^{2,\rm PS}$ (lower panel) 
for different values of $Q^2$ and the scale choice $\mu^2 = \mu_F^2 = Q^2$. 
Full line (Blue):        $Q^2=10^4~\GeV^2$; 
dashed line (Orange):    $Q^2=10^3~\GeV^2$; 
dotted line (Magenta):   $Q^2=500~\GeV^2$; 
dash-dotted line (Blue): $Q^2=100~\GeV^2$; 
full line (Red):         $Q^2=50~\GeV^2$; 
dashed line (Gray):      $Q^2=25~\GeV^2$; 
dotted line (Brown):     $Q^2=10~\GeV^2$, 
using the parameterization of the parton distribution \cite{Alekhin:2017kpj}.}
\label{fig:FIG4}
\end{figure}

In Figures~\ref{fig:FIG4} we show the complete results for the two--loop pure singlet contributions to $F_2$ and $F_L$ 
as a 
function of $x$ for a series of $Q^2$-values. At large values of $Q^2$ the corrections are negative and turn to positive values 
around $Q^2 \sim 10~\GeV^2$. In the small $x$ region the corrections are large and grow with $Q^2$. The absolute corrections to 
$F_L$ are smaller in size than those to $F_2$.

\begin{figure}[h!]
\centering
\includegraphics[width=0.49 \linewidth]{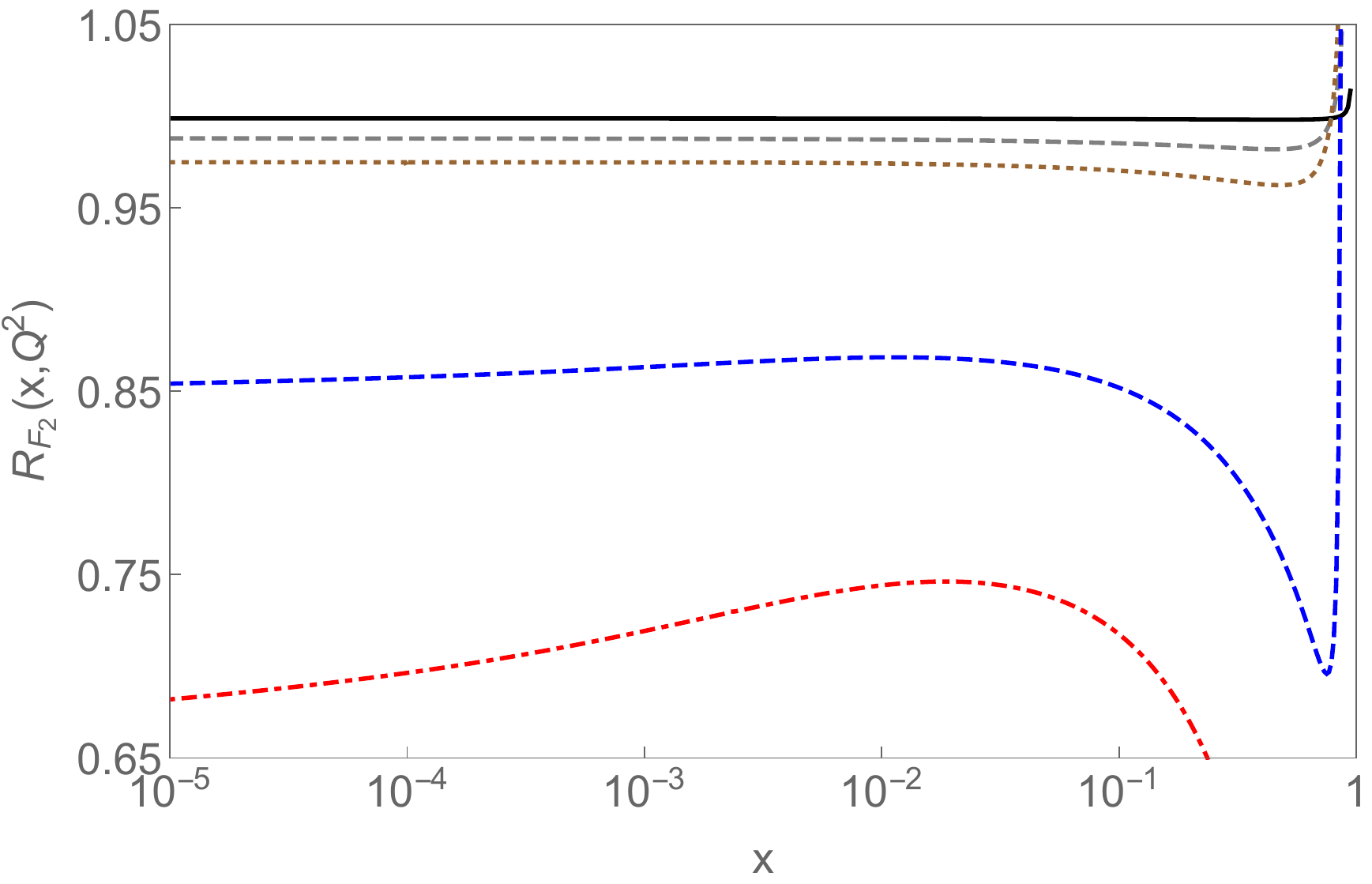}
\includegraphics[width=0.49 \linewidth]{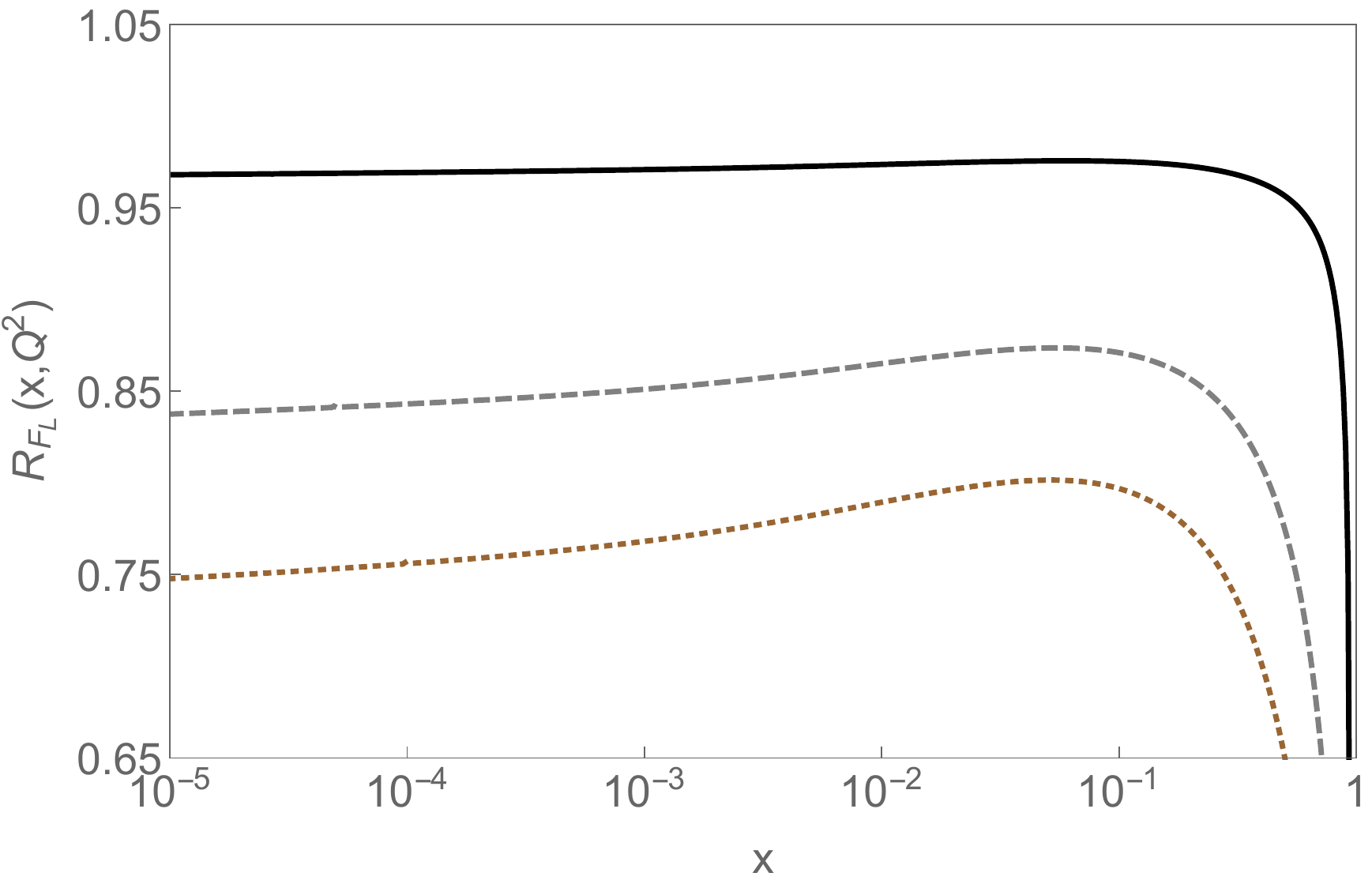}
\caption{\small \sf The ratios of the structure functions $F_{2,q}^{2,\rm PS}$ (left) and $F_{L,q}^{2,\rm PS}$ (right) 
in the full calculation over the asymptotic approximation for different values of $Q^2$ and the scale choice $\mu^2 = \mu_F^2 = 
Q^2$. 
Full line (Black):        $Q^2=10^4~\GeV^2$; 
dashed line (Gray):       $Q^2=10^3~\GeV^2$; 
dotted line (Brown):      $Q^2=500~\GeV^2$;
lower dashed line (Blue): $Q^2=100~\GeV^2$; 
dahs-dotted line (Red):   $Q^2=50~\GeV^2$, 
using the parameterization of the parton distribution \cite{Alekhin:2017kpj}}.
\label{fig:FIG5}
\end{figure}

In Figure~\ref{fig:FIG5} we illustrate the ratios Eq.~(\ref{eq:RAT2}) as a function of $x$ for different values of $Q^2$ for 
$F_2$ 
and $F_L$ comparing the asymptotic result to the full result. The corrections behave widely flat in $x$, turning to lower 
values in the large $x$ region. For $F_2$ the ratios are larger than $0.96$ for $Q^2 \geq 500~\GeV^2$. At $Q^2 = 100~\GeV^2$,
values of $\sim 0.85$ are obtained. For lower values of $Q^2$ the ratio is even smaller. 

For $F_L$ the corrections are generally larger. At $Q^2 = 10^4~\GeV^2$ one obtains a ratio of $0.96$, for $Q^2 = 10^3~\GeV^2$
~$0.85$, and for $Q^2 = 500~\GeV^2$ $\sim 0.75$, with even larger deviations from one for lower values of $Q^2$.

\begin{figure}[h!]
\centering
\includegraphics[width=0.49 \linewidth]{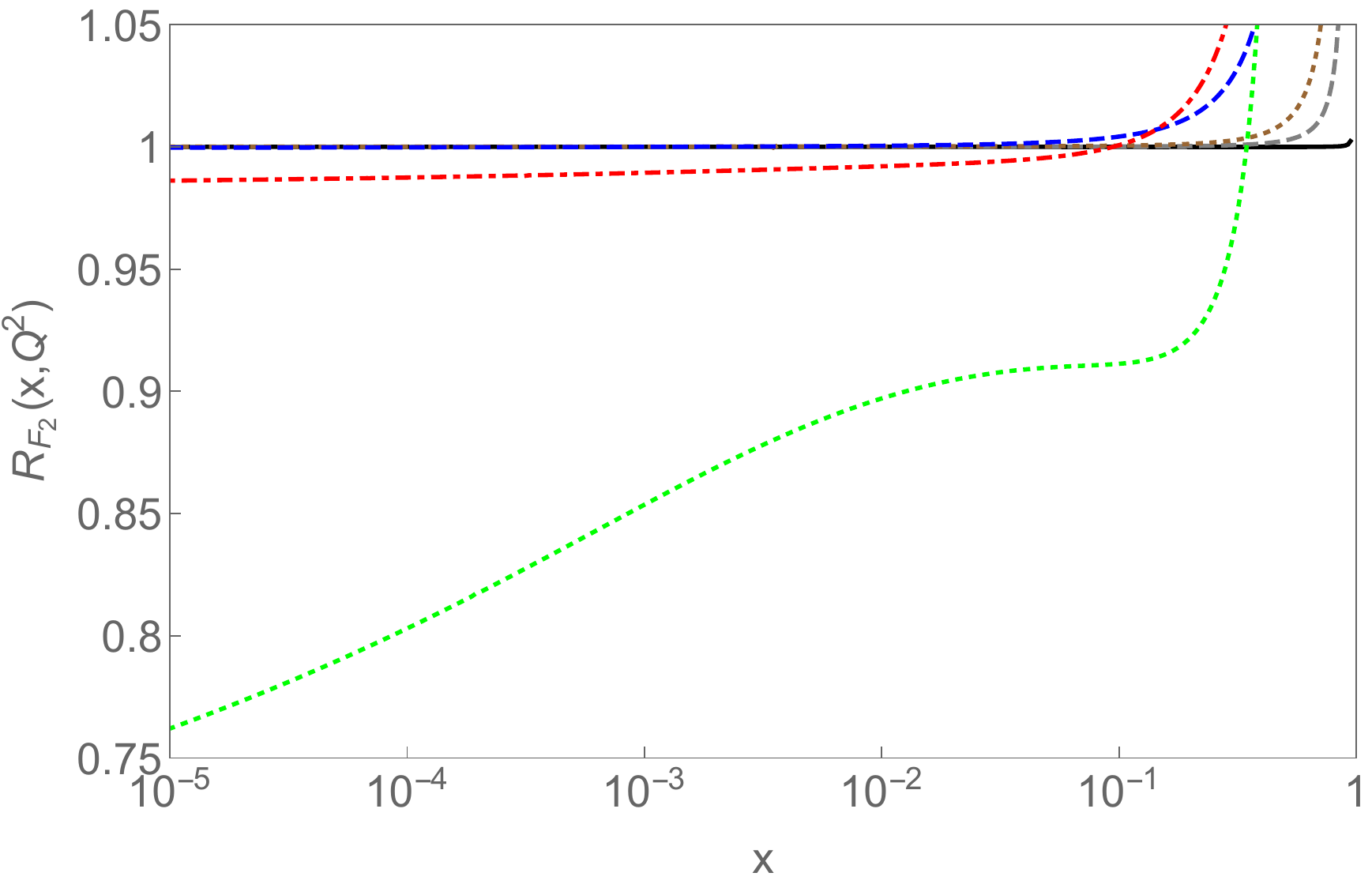}
\includegraphics[width=0.49 \linewidth]{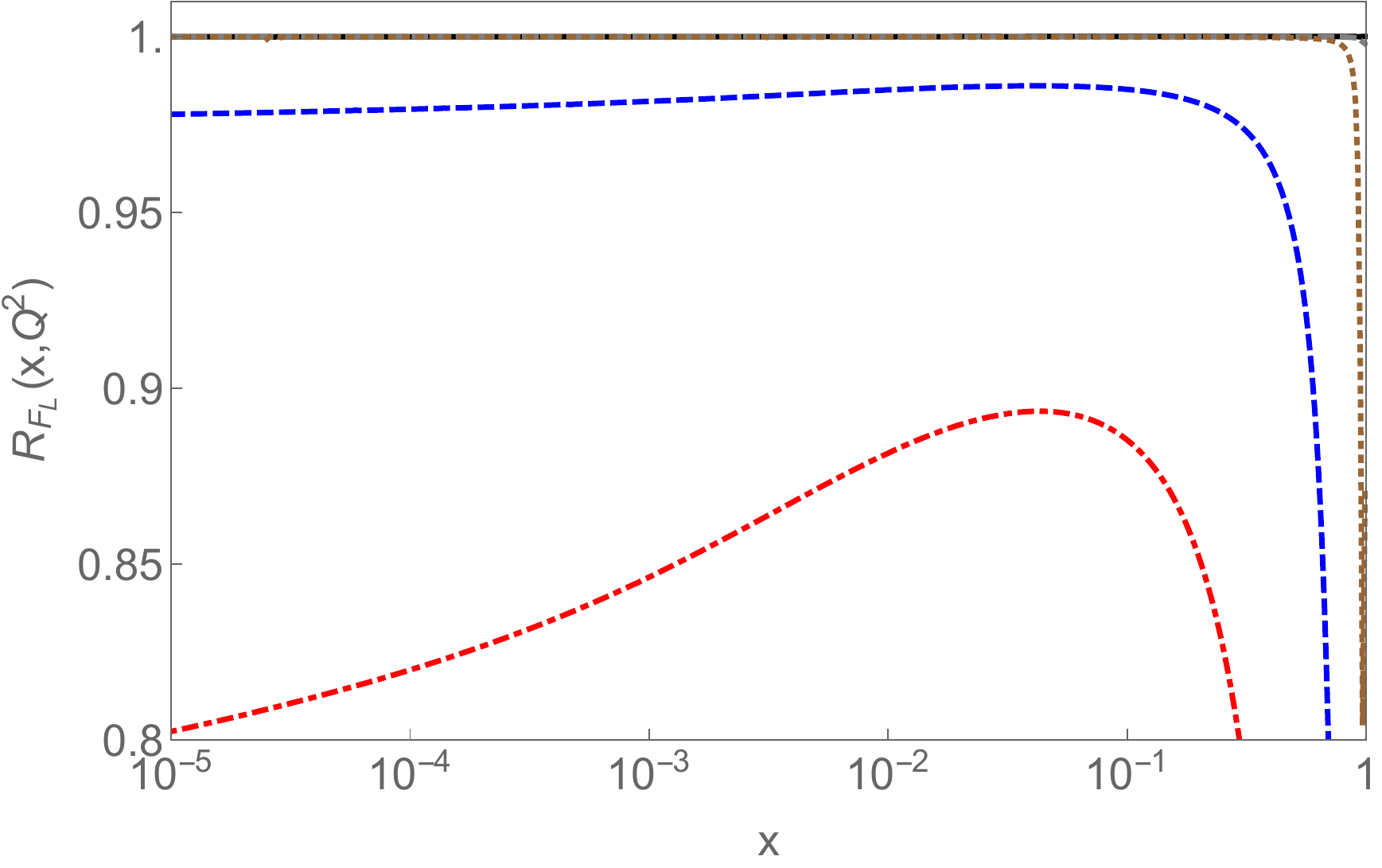}
\caption{\small \sf The ratios of the structure functions $F_{2,q}^{2,\rm PS}$ (left) and $F_{L,q}^{2,\rm PS}$ (right) 
in the full calculation over the $O((m^2/Q^2)^2)$ improved approximation for different values of $Q^2$ and the 
scale choice $\mu^2 = \mu_F^2 = Q^2$. 
Full lines (Black):         $Q^2=10^4~\GeV^2$; 
dashed lines (Gray):        $Q^2=10^3~\GeV^2$; 
dotted lines (Brown):       $Q^2=500~\GeV^2$; 
lower dashed lines (Blue):  $Q^2=100~\GeV^2$;
dash-dotted lines (Red):    $Q^2=50~\GeV^2$; 
lower dotted lines (Green): $Q^2=25~\GeV^2$, 
using the parameterization of the parton distribution \cite{Alekhin:2017kpj}.}
\label{fig:FIG6}
\end{figure}

In Figure~\ref{fig:FIG6} we depict the ratio of the full result over the $O((m^2/Q^2)^2)$ improved asymptotic
results for $F_2$ and $F_L$ as a function of $x$ for a series of $Q^2$-values. In the region $x < 0.1$ the ratios
for $F_2$ are larger than $0.98$ for $Q^2 > 50~\GeV^2$ and grow for larger values of $x$. Stronger deviations 
are observed for lower $Q^2$ values. For $F_L$ the corrections are larger. In the region $x < 0.3$ and $Q^2 > 
100~\GeV^2$ the ratio is larger than $0.97$, while for lower scales $Q^2$ the deviations are larger.
We limited the expansion to terms of $\sim O((m^2/Q^2)^2)$ in the present paper, but higher 
order terms can
be given straigtforwardly. The expanded expressions do also allow direct Mellin transforms and provide 
a suitable analytic basis for Mellin-space programmes.\footnote{In \cite{Alekhin:2003ev} precise numerical $N$-space 
implementations were given.}

\section{Conclusions}
\label{sec:8}

\vspace*{1mm}
\noindent
We have calculated the massless and massive two--loop unpolarized pure singlet Wilson coefficients 
of deep-inelastic scattering for the structure functions $F_2$ and $F_L$. In the massless case, we confirmed earlier 
analytic results in the literature, which can be expressed by harmonic polylogarithms. In the massive case,
the Wilson coefficients are calculated analytically for the first time. They are also given in terms of
iterative integrals, including now, however,  Kummer-elliptic integrals. The corresponding alphabets contain also elliptic 
letters. All integrals can be represented by classical (poly)logarithms with involed arguments with partly one
more (elliptic) letter iterated upon. This representation is very well suited to obtain numerical results.

We have studied systematic expansions in the ratio $m^2/Q^2$ in the asymptotic region and the velocity parameter 
$\beta$ in the threshold region. In the former case the leading asymptotic result has been recovered, known form
calculations based on massive OMEs and massless Wilson coefficients, proving asymptotic factorization in the present 
case. We have obtained a series of power corrections. Here the expansion coefficients are also spanned  
by harmonic polylogarithms. Retaining these terms extends the validity of the cross sections to lower scales of 
$Q^2$, which is relevant for experimental analyses. In particular, the predictions for the structure function
$F_L(x,Q^2)$ are significantly improved. In general, the Kummer-elliptic integarals, also obeying shuffling relations,
span a wide class of iterative integrals which play a role as well in other multi-scale calculations.

\appendix
\section{Details of the calculation}
\label{sec:A2}
Our calculation closely follows classical calculations in the literature, cf. e.g. 
\cite{Beenakker:1988bq,Matsuura:1988sm,Hamberg:1990np,Zijlstra:1992qd}.
Although these calculations are typically well documented, we encountered subtleties at several points
of our calculation.
Therefore, we provide a more detailed discussion of our calculation in the massless and massive case in
this Appendix.
First we will give the parametrization of the phase space we used in the massless and massive case, then
we will proceed by explaining the angular integration and give explicit results for the angular integrals
in $d$ dimensions.
In the end, we will comment on our resolution of the poles in $\ep$ and subtleties encountered in the 
massless case.
\subsection{Phase Space Parametrization}
\label{sec:A21}
\paragraph{The \boldmath{$2 \to 2$} Process} \ \\
In the $2 \to 2$ case in Figure~\ref{DIA1} we refer to the invariants
\begin{eqnarray}
s &=& (q+p)^2,~~t=(q-k_1)^2,~~u=(q-k_2)^2
\end{eqnarray}
with
\begin{eqnarray}
s + t + u = - Q^2 + 2m^2~~~\text{and}~~~Q^2 = -q^2.
\end{eqnarray}
We will also use the notation $\beta = \sqrt{1-4m^2/s}$.
In the cms of the outgoing particles, $\vec{k}_1 + \vec{k}_2 = 0,$ the scattering angle 
$\theta$ is defined by
\begin{eqnarray}
t = - Q^2 + m^2 - 2 q^0 k_{1}^0 + |\vec{k}_{1}||\vec{q}|  \cos(\theta) 
  = m^2 - \frac{Q^2}{2x}(1 - \beta \cos(\theta)),
\end{eqnarray}
with
\begin{eqnarray}
q^0 &=& \frac{s-Q^2}{2 \sqrt{s}}, ~~|\vec{q}| = \frac{s-Q^2}{2 \sqrt{s}}, \\
k_{1}^0 &=& \frac{\sqrt{s}}{2}, ~~~~~~ |\vec{k}_{1}| = \frac{\sqrt{s}}{2} \beta
\end{eqnarray}
and 
\begin{eqnarray}
\lambda(a,b,c) = (a-b-c)^2 - 4 bc.
\end{eqnarray}
\begin{figure}[H]
\centering
\includegraphics[width=0.6\textwidth]{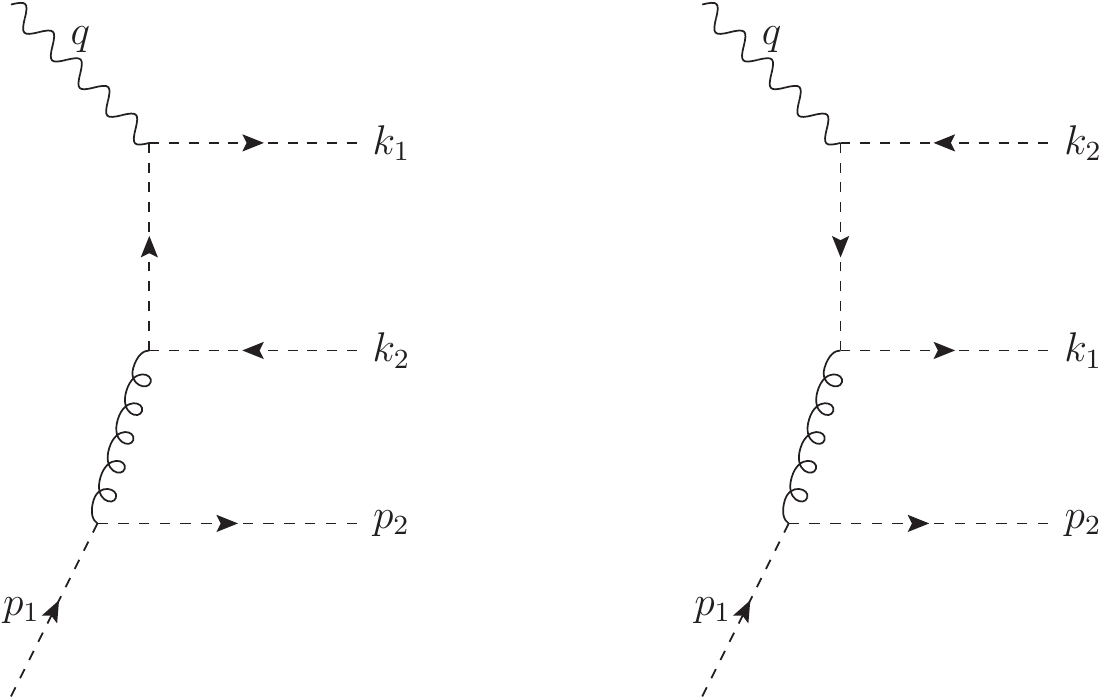}
\caption{\small \sf Diagrams of the $O(a_s^2)$ contributions to the pure singlet scattering cross section 
$\gamma^* + q \to Q + \overline{Q} + q$.}
\label{DIA2}
\end{figure}
The phase space integral is given by
\begin{eqnarray}
\int d \, \text{PS}_2 &=& 2^{4-2d} 
\frac{\pi^{1-d/2}}{\Gamma\left(\frac{d}{2}-1\right)} 
s^{d/2-2} \beta^{d-3} 
\int_0^\pi d\theta \sin^{d-3}(\theta).
\end{eqnarray}
The limit $m \to 0$ is easily obtained by setting $m=0$ and $\beta =1$.
\paragraph{The \boldmath{$2 \to 3$} Process} \ \\
The $2 \to 3$ process is slightly more involved. The contributing Feynman diagrams are shown in Figure~\ref{DIA2}.
We use
\begin{eqnarray}
\int d \, \text{PS}_3 &=& \int\frac{d^d p_2}{(2\pi)^{d-1}} \int\frac{d^d k_1}{(2\pi)^{d-1}} 
\int\frac{d^d k_2}{(2\pi)^{d-1}} 
\delta^{+}\left(p_2^2\right) \delta^{+}\left(k_1^2 - m^2\right) \delta^{+}\left(k_2^2 - m^2\right) \nonumber \\
&\times& (2\pi)^d \delta^{(d)}\left( p_1 + q - p_2 - k_1 - k_2 \right) \nonumber \\
&=& \frac{1}{(2\pi)^{2d-3}} \int d s_{12} \biggl\{ \int d^d p_2 \int d^d K \delta^{+}\left(p_2^2\right)
\delta^{+}\left(K^2 - 
s_{12} \right) \delta^{(d)}\left(p_1+q-p_2-K\right) \biggr\} \nonumber \\
&\times& \biggl\{ \int d^d k_1 \int d^d k_2 \delta^{+}\left(k_1^2 - m^2\right) \delta^{+}\left(k_2^2 - m^2\right) 
\delta^{(d)}\left(k_1+k_2-K\right) \biggr\}.
\end{eqnarray} 
Here 
\begin{align}
1 &= \int d s_{12} \int d^d K \delta^{+}\left(K^2-s_{12}\right) \delta^{(d)}\left(k_1+k_2-K\right)
\end{align}
was introduced to factorize the $2\to 3$ phase space into a $( 2 \to 2 ) \times ( 1 \to 2)$ phase space.
Both can now be calculated in the most appropriate system independent from each other.
Integrating the first factor in the cms system of the process and the second in the cms of the two heavy 
quarks one obtains
\begin{eqnarray}
\int d \, \text{PS}_3 &=& \frac{1}{(4\pi)^d} \frac{ (s-q^2)^{3-d} }{ \Gamma(d-3) } 
\int\limits_{s_{12}^-}^{s_{12}^+} d s_{12} 
\int\limits_{t^-}^{t^+} d t \int\limits_0^\pi d\theta \int\limits_0^\pi d\phi \left[ \sin(\theta) \right]^{d-3} 
\left[ \sin(\phi) \right]^{d-4} \nonumber \\
&\times& s_{12}^{d/2-2} \left[ 1 - \frac{4 m^2}{s_{12}} \right]^{d/2-3/2} \left[ (s-q^2)u 
- q^2 t \right]^{d/2-2} t^{d/2-2}, 
\end{eqnarray}
where we have chosen the kinematic invariants
\begin{align}
      t &= 2 p_1.p_2, & u &= 2 p_2.q, & s &= (p_1 + q)^2, & s_{12} &= s-t-u .
\end{align}
The phase space boundary is given by
\begin{eqnarray}
      s_{12}^{-} &=& 4 m^2 , ~~ s_{12}^{+} = s , \\
      t^- &=& 0 , ~~~~~~ t^+ = \frac{1}{s}(s-q^2)(s-s_{12}) .
\end{eqnarray}
We can use the following explicit parameterization of the vectors
\begin{eqnarray}
      k_1 &=& \left(k^0,0,\dots,| \vec{k} | \sin(\phi) \sin(\theta), | \vec{k} | \cos(\phi)\sin(\theta), | \vec{k} | \cos(\theta) \right) ,\\
      k_2 &=& \left(k^0,0,\dots,- | \vec{k} | \sin(\phi) \sin(\theta), - | \vec{k} | \cos(\phi)\sin(\theta),- | \vec{k} | \cos(\theta) \right),\\
      p_1 &=& \frac{s-t-q^2}{2\sqrt{s_{12}}} \left(1,\dots,0,0,1\right),\\
      p_2 &=& \frac{s-s_{12}}{2\sqrt{s_{12}}} \left(1,0,\dots,\sin(\chi),\cos(\chi)\right),\\
      q   &=& \frac{1}{2 \sqrt{s_{12}}} \left( q^2 + s_{12} + t,\dots, 0,0,(s-s_{12}) \sin(\chi) , 
      q^2+t-s+(s-s_{12})\cos(\chi) \right) , \nonumber\\ \\
      \cos(\chi) &=& 1-\frac{2s_{12}t}{(s-t-q^2)(s-s_{12})}, \\
      k^0 &=& \frac{\sqrt{s_{12}}}{2}, \\
      | \vec{k} | &=& \frac{\sqrt{s_{12}}}{2} \sqrt{ 1 - \frac{4 m^2}{s_{12}} }.
\end{eqnarray}
In the limit $m \to 0$, we recover the parameterization given in \cite{Zijlstra:1992qd}.

In a next step we want to introduce dimensionless variables with support over the unit cube.
Here it is advantageous to distinguish between the massless and the massive case.
In the massless case, we follow \cite{Zijlstra:1992qd} and introduce the new variables
\begin{eqnarray}
      x &=& - \frac{q^2}{s-q^2}, \nonumber\\
      u &=& [1-x-y-(1-x)(1-y)z](s-q^2), \nonumber\\
      t &=& y(s-q^2).
\end{eqnarray}
The massless three-particle phase space then reads
\begin{eqnarray}
      \int d \text{PS}_3 (m=0) &=& \frac{1}{(4\pi)^d} \frac{(s-q^2)^{3-d}}{\Gamma(d-3)} \int\limits_0^\pi 
d\theta \int\limits_0^\pi d\phi \left( \sin(\theta) \right)^{d-3} \left( \sin(\phi) \right)^{d-4} 
      \nonumber \\
      &\times& \int\limits_0^{s-q^2} dt \int\limits_{tq^2/(s-q^2)}^{s-t} du \ s_{12}^{d/2-2} t^{d/2-2} \left[ 
(s-q^2)u-q^2 t \right]^{d/2-2} 
\nonumber \\
      &=& \frac{1}{(4\pi)^d} \frac{(s-q^2)^{3-d}}{\Gamma(d-3)} (1-x)^{d-3} \int\limits_0^\pi d\theta 
\int\limits_0^\pi d\phi \left( \sin(\theta) \right)^{d-3} \left( \sin(\phi) \right)^{d-4} 
 \nonumber\\
      &\times& \int\limits_0^1 dy \int\limits_0^1 dz \, y^{d/2-2} (1-y)^{d-3} \left[ z(1-z) \right]^{d/2-2}. 
\end{eqnarray}
In the massive case the change to the following variables is useful
\begin{align}
      z  &= \frac{1}{\beta^2} \left( 1 - \frac{4m^2}{s_{12}} \right), & s_{12} &= \frac{4m^2}{1-\beta^2 z}, \\
      y &= \frac{ s t }{(s-q^2)(s-s_{12})} , & t &= (s-q^2)\beta^2 y \frac{1 - z}{1-\beta^2 z} .
\end{align}
The new parameterization then reads
\begin{align}
\int d \text{PS}_3 =& \frac{1}{(4\pi)^d} \frac{s^{d-3}}{\Gamma(3-d)} \beta^{3d-7} (1-\beta^2)^{d/2-1} 
\int\limits_0^1 dz 
\int\limits_0^1 dy \int\limits_0^\pi d\theta \int\limits_0^\pi d\phi \left[ \sin(\theta) \right]^{d-3} 
\left[ \sin(\phi) \right]^{d-4} \nonumber \\
&\times y^{d/2-2} (1-y)^{d/2-2} z^{d/2-3/2} (1-z)^{d-3} (1-\beta^2 z)^{3-3d/2} .
\end{align}
The limit $m \to 0$ is not easily recovered, because of the mass dependent transformation.

\vspace*{1mm}
\noindent

\subsection{Angular Integrals}
\label{sec:A22}
\paragraph{The massless case} \ \\
There are four angle dependent denominator structures appearing for the pure singlet process:
\begin{align}
      N_1 &= ( p_1 - k_1 )^2 = - 2 p_1.k_1 = a \left( 1 - \cos(\theta) \right) , \nonumber\\
      N_2 &= ( p_1 - k_2 )^2 = - 2 p_1.k_2 = a \left( 1 + \cos(\theta) \right) , \nonumber\\
      N_3 &= ( q - k_1 )^2 = q^2 - 2 q.k_1 = A + B \cos(\theta) + C \cos(\phi) \sin(\theta), \nonumber\\
      N_4 &= ( q - k_2 )^2 = q^2 + 2 q.k_1 = A - B \cos(\theta) - C \cos(\phi) \sin(\theta),
\end{align}
with 
\begin{align}
      a &= - \frac{s-t-q^2}{2} , \nonumber\\
      A &= \frac{1}{2} \left(q^2 - s_{12} - t\right), \nonumber\\
      B &= \frac{1}{2} \left[ q^2 -s + t + (s-s_{12})\cos(\chi) \right], \nonumber\\
      C &= \frac{s-s_{12}}{2} \sin(\chi).
      \label{eq:masslessCoeffs}
\end{align} 
Using partial fractioning we can express all angular integrals via
\begin{align}
I_{l,k} &= \int\limits_{0}^\pi d\theta \int\limits_{0}^{\pi} d\phi \frac{ \sin^{d-3}(\theta) }
{ a^l \left[ 1 - \cos(\theta) 
\right]^l } \frac{ \sin^{d-4}(\phi)}{ \left[A + B \cos(\theta) + C \sin(\theta) \cos(\phi) \right]^k }  .
\end{align}
We only encounter integrals with $k \leq 0$, however, it is possible to find closed form solutions for 
$k \leq 0$ and $l \leq 0$ in the massless case.
In the following we will list the result for these angular integrals in $d$-dimensions.
\ \\
\underline{$l$ negative:}\\
\begin{align*}
      I_{l,k} &= \sum\limits_{m=0}^k \sum\limits_{n=0}^{-l-m} \binom{-l}{m} \binom{-k-m}{n} 2^{2d-7} 
a^{-l} ( B^2 + C^2 )^{l/2} \left( B + \sqrt{B^2 + C^2} \right)^{-l-m-n} \\
&\times (-2B)^n \left( A - \sqrt{B^2+C^2} \right)^{-k} (2C)^m \frac{\Gamma^2(d/2-3/2)}{\Gamma(d-3)} {}_2 F_1 \left[ 
\begin{matrix} -m , d/2 - 3/2 \\ d-3 \end{matrix}, 2 \right] \\
&\times \frac{\Gamma(d/2-1+n+m/2)\Gamma(d/2-1+m/2)}{\Gamma(d-2+m+n)} {}_2 F_1 \left[ \begin{matrix} k , 
d/2 - 1 + n + m/2 \\ 
d-2+m+n \end{matrix}, - \frac{2 \sqrt{B^2 + C^2}}{A-\sqrt{B^2 + C^2}} \right].
\end{align*}  
For $l=0$ this reduces to
\begin{align}
      I_{0,k} 
      &= 2^{2d-7} \left[ A - \sqrt{B^2 + C^2} \right]^{-k} \frac{ \Gamma^2(d/2-3/2) }{ \Gamma(d-3) } 
\frac{ \Gamma^2(d/2-1) }{ 
\Gamma(d-2) } \ \nonumber \\
& \times {}_2 F_1 \left[ \begin{matrix} k , d/2 - 1 \\ d-2 \end{matrix}, 
- \frac{2 \sqrt{B^2 + C^2}}{A-\sqrt{B^2 + C^2}} 
\right].
\end{align}
\underline{$k$ negative:}\\
\begin{align*}
      I_{l,k} &= \sum\limits_{m=0}^{-k} \binom{-k}{m} \frac{2^{2d-7-l}}{a^l} ( A-B )^{-k-m} (-2z)^{m} 
\frac{\Gamma^2(d/2-3/2)}{\Gamma(d-3)} {}_2 F_1 \left[ \begin{matrix} -m , d/2 - 3/2 \\ d-3 \end{matrix}, 2 \right] 
\\
&\times \frac{\Gamma(d/2-1+m/2)\Gamma(d/2-1+m/2-l)}{\Gamma(d-2+m-l)}{}_2 F_1 \left[ \begin{matrix} m+k , 
d/2 - 1 + m/2 \\ 
d-2+m-l \end{matrix}, - \frac{2 B}{A-B} \right].
\end{align*}
For $k=0$ this reduces to 
\begin{align}
I_{l,0} 
      &= \frac{2^{2d-7-l}}{a^l} \frac{ \Gamma(d/2-1-l) \Gamma(d/2-1) }{ \Gamma(d-2-l) } \frac{ \Gamma^2(d/2-3/2) }
{ \Gamma( d-3 ) } .
\end{align}
Expanding these results around $\ep=d-4$ dimensions we recover the integrals given in \cite{Beenakker:1988bq}.

\paragraph{The massive case} \ \\
In the massive case the four denominator structures read
\begin{align}
      N_1 &= (p_1 - k_1)^2 = - 2 p_1.k_1 =  a + b \cos(\theta)  , \nonumber \\
      N_2 &= (p_1 - k_2)^2 = - 2 p_1.k_2 =  a - b \cos(\theta)  , \nonumber \\
      N_3 &= (q - k_1)^2 = q^2 - 2 q.k_1 = A + B \cos(\theta) + C \cos(\phi) \sin(\theta) \nonumber \\
      N_4 &= (q - k_2)^2 = q^2 - 2 q.k_2 = A - B \cos(\theta) - C \cos(\phi) \sin(\theta) ,
\end{align}
with 
\begin{align}
      a &= - \frac{s-t-q^2}{2} , \\
      b &= -\frac{1}{2} \sqrt{1 - \frac{4m^2}{s_{12}} } (q_2 - s -t ),\\ 
      A &= \frac{q^2 - s_{12} - t}{2}, \\
      B &= \frac{1}{2} \sqrt{1 - \frac{4m^2}{s_{12}} } \left( q^2 - s + t + (s-s_{12})\cos(\chi) \right), \\
      C &= \frac{1}{2} \sqrt{1 - \frac{4m^2}{s_{12}} } (s-s_{12}) \sin(\chi).
\end{align}
Therefore, we have to consider the more general angular integral
\begin{align}
I_{l,k} &= \int\limits_{0}^\pi d\theta \int\limits_{0}^{\pi} d\phi \frac{ \sin^{d-3}(\theta) }
{\left[ a + b \cos(\theta) 
\right]^l } \frac{ \sin^{d-4}(\phi)}{ \left[A + B \cos(\theta) + C \sin(\theta) \cos(\phi) \right]^k }  
\end{align}
in the following.
For $l \geq 0$ and arbitrary $k$ (the only case we encounter), we find:
\begin{align}
I_{l,k} &= \sum\limits_{n=0}^{-l} \sum\limits_{m=0}^n \sum\limits_{i=0}^{m} \binom{-l}{n}
\binom{n}{m} \binom{m}{i} \left( \frac{b C}{\sqrt{B^2+C^2}} \right)^{-l-n} a^{n-m}
\left( \frac{b B}{\sqrt{B^2+C^2}} \right)^{m} \left( A - \sqrt{B^2+C^2}\right)^{-k}
\nonumber \\
&\times 2^{2d -7-n-l+i} (-1)^{-n-l+m-i} \frac{\Gamma^2(d/2-3/2)}{\Gamma(d-3)}
\nonumber\\ &
\times
\frac{\Gamma(d/2-1-n/2-l/2+i)\Gamma(d/2-1-n/2-l/2)}{\Gamma(d-2-n+l+i)}
\nonumber \\
&\times 
{}_2 F_1 \left[ \begin{matrix} n+l , d/2 - 3/2 \\ d-3 \end{matrix}, 2 \right]
{}_2 F_1 \left[ \begin{matrix} k , d/2-1-n/2-l/2+i \\ d-2-n-l+i \end{matrix}, 
-\frac{2\sqrt{B^2+C^2}}{A-\sqrt{B^2+C^2}} \right].
\end{align}
\subsection{Regularization}
\label{sec:A23}
\vspace*{1mm}
\noindent
In order to perform the $\ep$-expansion of the functions we use a simple subtraction term for $y=0$.
However, there is a subtlety hiding in this limit.
The hypergeometric functions of interest are all of the argument
\begin{align}
      X &= - \frac{2 \sqrt{B^2+C^2}}{A-\sqrt{B^2+C^2}} .
\end{align}
Inserting the coefficients from Eqs.~\eqref{eq:masslessCoeffs}, we see that
\begin{align}
      X &= 1 + \mathcal{O}(y),
\end{align}
which means that there is a potential logarithmic singularity for $y \to 0$ in the massless case.
This divergence can be made explicit  by transforming the $_2F_1$'s from argument $x$ to $(1-x)$ 
\cite{HYPERGEOM} 
\begin{eqnarray}
      {}_2 F_1 \left[ \begin{matrix} a , b \\ c \end{matrix}, z \right] &=& \Gamma \left[ \begin{matrix} c,c-a-b \\ c-a,c-b \end{matrix} \right] {}_2 F_1 \left[ \begin{matrix} a , b \\ a+b-c+1 \end{matrix}, 1-z \right] \nonumber \\
&+& (1-z)^{c-a-b} \Gamma \left[ \begin{matrix} c,a+b-c \\ a,b \end{matrix} \right] {}_2 F_1 \left[ \begin{matrix} c-a , c - b 
\\ c-a-b+1 \end{matrix}, 1-z \right].
\end{eqnarray}
The new hypergeometric functions have Taylor expansions around $y=0$.
The only singular behavior can now occur for $y \to 0$.
This means that we can resolve the divergences via
\begin{align}
      F(x) &= \int\limits_0^1 dz \int\limits_0^1 dy y^{-2+\ep/2} f(x,y,z) \\
      &= \int\limits_0^1 dz \int\limits_0^1 dy y^{-2+\ep/2} \left[ f(x,y,z) - f^{(0)}(x,0,z) - y f^{(1)}(x,0,z) \right]
      \nonumber \\
      &- \int\limits_0^1 dz \int \limits_0^1 dy y^{-2+\ep/2} \left[ f^{(0)}(x,0,z) + y f^{(1)}(x,0,z) \right]
      \nonumber \\
      &\equiv (A) - (B) ,
\end{align}
where we used the notation
\begin{align}
      f(x,y,z) &= \sum\limits_{i=0}^{\infty} y^i f^{(i)}(x,0,z) .
\end{align}
In the massive case we have
\begin{align}
      X &= - \frac{\sqrt{B^2+C^2}}{A-\sqrt{B^2+C^2}} = \frac{2 \beta \sqrt{z}}{1+\beta \sqrt{z}} + \mathcal{O}(y),
\end{align}
which means that this divergence is regulated by the quark mass.
The subtraction term $(B)$ can be trivially integrated over $y$, which will lead to poles in $\ep$.
In the massless case the expansion in $\ep$ can be performed afterwards and the last integration over $z$ can be carried out.
In the massive case there can be additional singularities hiding in the $z \to 1$ limit.
Therefore, term $(B)$ has to be regularized accordingly.
Term $(A)$ is not singular in the limit $y \to 0$ and can be expanded in $\ep$ and then
integrated over $y$ and $z$.

\section{Contributing expressions due to renormalization}
\label{sec:A3}
\vspace*{1mm}
\noindent
In the following we list some Mellin-convolutions, which occurred in Eqs.~(\ref{eq:HFL}, \ref{eq:HF1}). These are
convolutions with leading order splitting functions, using the parameter $\kappa = m^2/Q^2$.
\begin{eqnarray}
        P_{gq}^{(0)} \otimes h_{L,g}^{(1)} &=& 
        C_F T_F\Biggl\{64 \beta  (1-z) \frac{1 + 6 \kappa - (8 \kappa +2) z - (8 \kappa +2) z^2 }{3 z (1 + 4 \kappa )}
        - \frac{64}{3} z ( 3 + 4 \kappa  z ) \ln \left( \frac{1-\beta }{1+\beta} \right) 
\nonumber \\ &&
        + \frac{64}{3} \frac{ 4 \kappa  (1 + 3 \kappa)
        - 6 \kappa ( 1 + 4 \kappa ) z
        + 3 ( 1 + 4 \kappa )^2 z^2}{z (1 + 4 \kappa )^{3/2}} \ln \left(\frac{\sqrt{1 + 4 \kappa} 
- \beta }{\sqrt{1 + 4 \kappa} + \beta }\right)\Biggr\}
        ~,
\\
        P_{gq}^{(0)} \otimes \bar{b}_{L,g}^{(1)} &=&
C_F T_F \Biggl\{-\frac{32 (1-z) \big( 3 - 4 z - 6 z^2 \big) \beta }{3 z}
+\frac{8}{3} z (3 + 4 z \kappa ) \ln^2 \left( \frac{1-\beta }{1+\beta } \right)
\nonumber \\ &&
- \frac{64}{3} z (3+4 z \kappa ) \bigl[ {\Li}_2\big(\frac{1-\beta }{2}\big) - {\Li}_2(1-\beta ) - {\Li}_2(-\beta ) \bigr]
\nonumber \\ &&
- \frac{8}{3 z (1+4 \kappa )^{5/2} } 
        \biggl[
        2 \kappa ^2 (1+\kappa )
        -3 z \kappa ^2 (1+4 \kappa )
        +3 z^2 (1 + 4 \kappa )^2 
        \big(
                \kappa 
                +\sqrt{1+4 \kappa }
        \big)
\nonumber \\ &&
        +4 z^3 \kappa  (1+4 \kappa )^{5/2}
\biggr] \ln^2(1-z)
-\frac{8 \kappa R_3 }{3 z (1+4 \kappa )^{5/2}} 
\biggl[
\ln^2 \left( \frac{\sqrt{1+4 \kappa }-1}{\sqrt{1+4 \kappa }+1} \right)
\nonumber \\ &&
+ \ln^2 \left( \frac{ \sqrt{1+4 \kappa } - \beta }{ \sqrt{1+4 \kappa } + \beta } \right)
- 4 \ln \big( \kappa \big) \ln \left( \frac{ \sqrt{1+4 \kappa } - 1 }{ \sqrt{1+4 \kappa } + 1} \right)
- 8 {\Li}_2 \left( \frac{1}{1-\sqrt{1+4 \kappa }} \right)
\nonumber \\ &&
+ 8 {\Li}_2 \left( \frac{1}{1+\sqrt{1+4 \kappa }} \right)
+ 8 {\Li}_2 \left( \frac{ \sqrt{1+4 \kappa } - 1 }{ \sqrt{1+4 \kappa } + 1 } \right)
- 8 \ln(2) \ln \left( \frac{ \sqrt{1+4 \kappa } -1 }{ \sqrt{1+4 \kappa } + 1} \right)
\nonumber \\ &&
+ 8 {\Li}_2 \left( \frac{\beta - \sqrt{1+4 \kappa }}{\beta +\sqrt{1+4 \kappa }} \right)
- 8 {\Li}_2 \left( \frac{ \big( \sqrt{1+4 \kappa } - 1 \big) \big( \sqrt{1+4 \kappa } - \beta \big)}{\big(1+\sqrt{1+4 \kappa }\big)\big(\beta +\sqrt{1+4 \kappa } \big)} \right)
\nonumber \\ && 
- 2 \ln(1-z)  \ln \left( \frac{\sqrt{1+4 \kappa } - 1}{\sqrt{1+4 \kappa } + 1} \right)
\biggr]
+ \frac{64}{3} z (3+4 z \kappa ) \ln (\beta ) \ln (2)
\nonumber \\ &&
+\frac{16 R_7}{3 z (1+4 \kappa )^{5/2} } 
\ln \left(\frac{\sqrt{1+4 \kappa }-1}{\sqrt{1+4 \kappa }+1}\right) 
\ln \left(\frac{\sqrt{1+4 \kappa }-\beta}{\sqrt{1+4 \kappa }+\beta}\right) 
\nonumber \\ &&
+\frac{32 R_5}{3 z (1+4 \kappa )^{5/2}} \biggl[
  {\Li}_2 \left( \frac{ \sqrt{1+4 \kappa } - \beta }{ \sqrt{1+4 \kappa } + 1} \right)
+ {\Li}_2 \left( \frac{ \sqrt{1+4 \kappa } - 1 }{ \sqrt{1+4 \kappa } + \beta } \right)
\biggr]
\nonumber \\ &&
-\frac{32 R_4}{3 z (1+4 \kappa )^{3/2} } 
\ln \left(\frac{ \sqrt{1+4 \kappa } - \beta }{ \sqrt{1+4 \kappa } + \beta }\right)
- \frac{32 R_2}{3 z (1+4 \kappa )^{3/2} } \biggl[
  2 {\Li}_2 \left (- \frac{\beta }{\sqrt{1+4 \kappa }} \right)
\nonumber \\ &&
- 2 {\Li}_2 \left( \frac{\beta }{\sqrt{1+4 \kappa }} \right)
+   {\Li}_2 \left( \frac{ \sqrt{1+4 \kappa } - 1}{ \sqrt{1+4 \kappa } - \beta } \right)
+   {\Li}_2 \left( \frac{ \sqrt{1+4 \kappa } + \beta }{ \sqrt{1+4 \kappa} + 1} \right)
\nonumber \\ &&
- 2 \ln(\beta) \ln \left(\frac{\sqrt{1+4 \kappa } -\beta }{\sqrt{1+4 \kappa } + \beta} \right)
\biggr]
+\frac{32}{3 z (1+4 \kappa )^{5/2} } \biggl[
         6 \kappa ^2 (1+\kappa )
        -9 z \kappa ^2 (1+4 \kappa )
\nonumber \\ &&
        +3 z^2 (1 + 4 \kappa )^2 \big(3 \kappa -\sqrt{1+4 \kappa }\big)
        -4 z^3 \kappa  (1+4 \kappa )^{5/2}
\biggr] \zeta_2
+ \frac{32 \beta  R_1}{3 z (1+4 \kappa )} \ln (1-z)
\nonumber \\ &&
+\frac{16 R_6}{3 z (1+4 \kappa )^{5/2} } 
\ln (1-z) \ln \left( \frac{ \sqrt{1+4 \kappa } - \beta }{ \sqrt{1+4 \kappa } + \beta } \right)               
- \frac{16}{3} z (3+4 z \kappa ) \biggl[        
        \ln \left(\frac{1-\beta }{1+\beta }\right)
\nonumber \\ &&
        - \ln (z)
        + 2 \ln (\beta )
        - \ln (\kappa )
\biggr] \ln (1-z)
-\frac{32 \beta  R_1}{3 z (1+4 \kappa )} \ln (z)
\nonumber \\ &&
+ \frac{16}{3} z (3+4 z \kappa) \biggl[
        \ln \left(\frac{1-\beta }{1+\beta }\right)
        + 2 \ln (\beta )
        - \ln (\kappa )
\biggr] \ln (z)
-\frac{8}{3} z (3+4 z \kappa) \ln^2(z)
\nonumber \\ &&
+ \frac{64 \beta  R_1}{3 z (1 + 4 \kappa)} \ln (\beta )
- \frac{32}{3} z (3+4 z \kappa) \biggl[
        \ln \left(\frac{1-\beta }{1+\beta }\right)        
        - \ln (\kappa )
\biggr] \ln (\beta )
\nonumber \\ &&
-\biggl[
        \frac{32}{3} \left( 3 - 6 z - 4 z^2 \kappa - \frac{1 + 6 \kappa }{z( 1 + 4 \kappa )} \right)
        +\frac{16}{3} z (3+4 z \kappa) \ln (\kappa )
\biggr] \ln \left( \frac{1-\beta }{1+\beta } \right)
\nonumber \\ &&
-\frac{8}{3} z (3+4 z \kappa) \ln^2(\kappa )\Biggr\}
        ~,
\end{eqnarray}
where we introduced the polynomials 
\begin{eqnarray}
        R_1 &=& 6 \kappa +(8 \kappa +2) z^3-(14 \kappa +3) z+1 ~,
\\ 
        R_2 &=& 4 \kappa  (1 + 3 \kappa )+3 ( 1 + 4 \kappa )^2 z ^2-6 \kappa  (1 + 4 \kappa ) z~,
\\ 
        R_3 &=& 2 \kappa  (1 + \kappa )+3 (1 + 4 \kappa )^2 z ^2 -3 \kappa  (1 + 4 \kappa ) z~,
\\ 
        R_4 &=& 24 \kappa ^2+12 \kappa -3 (1 + 4 \kappa )^2 z+6 (1 + 4 \kappa )^2 z ^2 +1~,
\\ 
        R_5 &=& 4 \kappa  \left(11 \kappa ^2+6 \kappa +1\right)-6 \kappa  \left(12 \kappa ^2+7 \kappa +1\right) z
+3 (1 + 2 \kappa ) (1 + 4 \kappa )^2 z^2 ~,
\\ 
        R_6 &=& 2 \kappa  \left(23 \kappa ^2+13 \kappa +2\right)-3 \kappa  \left(28 \kappa ^2+15 \kappa +2\right) z+3 (1 + 3 
\kappa) (1 + 4 \kappa )^2 z^2 ~,
\\ 
        R_7 &=& 2 \kappa  \left(25 \kappa ^2+15 \kappa +2\right)-3 \kappa  \left(36 \kappa ^2+17 \kappa +2\right)z +3 (1 + 5 
\kappa) (1 + 4 \kappa )^2 z^2 ~.
\end{eqnarray}
For $F_1$ the corresponding quantities read
\begin{eqnarray}
        P_{gq}^{(0)} \otimes h_{1,g}^{(1)} &=&
C_F T_F \Biggl\{(1+z-2 z \kappa ) \biggl[
        - 32 \ln^2\left(\frac{1-\beta }{1+\beta }\right)
        - 64 {\Li}_2\left(\frac{1-\beta }{2}\right)
        + 64 {\Li}_2\left(\frac{1+\beta }{2}\right)
\nonumber \\ &&
        - 64 {\Li}_2\left(\frac{\beta +1}{1-\sqrt{1+4 \kappa }}\right)
        + 64 {\Li}_2\left(\frac{\beta -1}{\sqrt{1+4 \kappa }-1}\right)
        + 64 {\Li}_2\left(\frac{1-\beta }{1+\sqrt{1+4 \kappa }}\right)
\nonumber \\ &&
        - 64 {\Li}_2\left(\frac{1+\beta }{1+\sqrt{1+4 \kappa }}\right)
+\biggl(
        - 64  \ln \left(1+\beta \right)
        -128  \ln \left(1+\sqrt{1+4 \kappa }\right)
\nonumber \\ &&
        +128  \ln \left(\beta +\sqrt{1+4 \kappa }\right)
        - 64  \ln \left(\frac{\sqrt{1+4 \kappa }-1}{\sqrt{1+4 \kappa }+1}\right)
        + 64  \ln \left(\frac{\sqrt{1+4 \kappa }-\beta}{\sqrt{1+4 \kappa }+ \beta}\right)
\nonumber \\ &&       
        + 64 \ln (2)
\biggr) \ln \left(\frac{1-\beta }{1+\beta }\right)
\biggr]
-\frac{64 (1-z) \beta }{3 z (1 + 4 \kappa )} \big(
         3 z (1+4 \kappa )
        +2 z^2 (1-2 \kappa ) (1+4 \kappa )
\nonumber \\ &&       
        +2 (1+7 \kappa )
\big)
-\frac{32}{3} \big(3-3 z-4 z^2 (1-2 \kappa ) (1+2 \kappa )\big) \ln \left(\frac{1-\beta }{1+\beta }\right)
\nonumber \\ &&       
-\frac{128}{3 z (1+4 \kappa )^{3/2} } 
\big(
                1+9 (1-z) \kappa +2 (7-18 z) \kappa ^2
\big) \ln \left(\frac{\sqrt{1+4 \kappa } -\beta}{\sqrt{1+4 \kappa }+\beta}\right)\Biggr\}
        ~,
\end{eqnarray}
\begin{eqnarray}
        P_{gq}^{(0)} \otimes \bar{b}_{1,g}^{(1)} &=&
C_F T_F \Biggl\{\frac{2 (1+k)^3 R_8}{3 k^4 z}
\biggl[
         k \HA_{w_{1}}
        -k \HA_{w_{2}}
        + \ln (1-k^2)
        - \ln (1-z)
\biggr] \HA_0
+\frac{32 R_9}{3 z} \bigl( \HA_{w_{1}} 
\nonumber \\ &&
+ \HA_{w_{2}} \bigr)
-\frac{R_{10}}{6 k^2 z} \ln (1-k) \HA_{w_{2}}
+\frac{R_{11}}{6 k^2 z} \bigl[ \ln (1-k) \HA_{w_{1}} + \ln (1+k) \HA_{w_{2}} \bigr]
+\frac{8  R_{12}}{3 z}  
\nonumber \\ &&
\times        \biggl[\HA_{w_{1},-1} 
      - \HA_{w_{2},1} 
      + \HA_{w_{2},-1} 
      + 2 \ln(k) \bigl( 
              \HA_{w_{1}} 
            + \HA_{w_{2}} 
      \bigr) 
\biggr]
+\frac{96 k z (1+z)}{3 z} \bigl( \HA_{w_{1},-1} - \HA_{w_{2},1} 
\nonumber \\ &&
- \HA_{w_{2},-1} \bigr)
+\frac{-16 R_{12}}{3 z} \biggl( \HA_{w_{1},0} + \HA_{w_{2},0} + \frac{1}{2} \HA_{w_{1},1} \biggr)
+\frac{96 k z (1+z)}{3 z} \HA_{w_{1},1}
\nonumber \\ &&
-\frac{\big(1-3 k^2\big) R_{13}}{6 k^3 z} \biggl[
        \ln ^2(1-k) 
      - \ln ^2(1+k)
      - \ln (1-z) \bigl\{ \ln (1-k) - \ln (1+k) \bigr\}
\biggr]
\nonumber \\ &&
+\frac{R_{14}}{6 k^2 z} \ln (1+k) \HA_{w_{1}}
+\frac{16 R_{15}}{3 k^4} \HA_1 \HA_{-1}
+ \frac{16 R_{16}}{3 k^4} \biggl[
        2 \HA_{0,1}
      - 2 \HA_{-1,0}
      - 2 \HA_1 \HA_0
\nonumber \\ &&
      - \bigl[ \ln (1-k^2) - 2 \ln (k) \bigr] \bigl( \HA_1 + \HA_{-1} \bigr)
\biggr]
+\frac{16 (1-z) \beta  R_{17}}{3 k^2 z}
-\frac{8 R_{18}}{3 k^4 z} \ln (2) \biggl[
        \ln (1-z)
\nonumber \\ &&
      - \ln (1-k^2)
      - k \bigl( \HA_{w_{1}} - \HA_{w_{2}} \bigr)
\biggr]
+\frac{16 R_{19}}{3 k^4 z (1-\beta )} \big(z-k^2 (1-(1-z) \beta)\big) \biggl[ \ln (1-k^2) 
\nonumber \\ &&
 - 2 \ln (k) \biggr]
-\frac{32 (1-z) \beta  R_{20}}{3 k^2 z} \HA_0
-\frac{8 R_{21}}{3 k^4 z} \HA_1
+\frac{8 R_{22}}{3 k^4 z} \HA_{-1}
-\frac{8}{3} \biggl[3+9 z
\nonumber \\ &&
-\frac{\big(1+k^2\big)\big(1-3 k^2\big) z^2}{k^4}\biggr] \bigl( \HA_1^2 - \HA_{-1}^2 \bigr)
+\frac{32}{3} \biggl[9+3 z+\frac{\big(1+k^2\big)\big(1-3 k^2\big) z^2}{k^4}\biggr] \HA_{-1,1}
\nonumber \\ &&
+\big(\frac{16 z}{k}-16 k (2+3 z)\big) \biggl[
      - 2 \HA_{w_{1},1,0}
      -   \HA_{w_{1},1,1}
      +   \HA_{w_{1},1,-1}
      - 2 \HA_{w_{1},-1,0}
      -   \HA_{w_{1},-1,1}
\nonumber \\ &&
      +   \HA_{w_{1},-1,-1}
      + 2 \HA_{w_{2},1,0}
      +   \HA_{w_{2},1,1}
      -   \HA_{w_{2},1,-1}
      + 2 \HA_{w_{2},-1,0}
      +   \HA_{w_{2},-1,1}
\nonumber \\ &&
      -   \HA_{w_{2},-1,-1}
      - \bigl( \zeta_2 - \ln ^2(2) \bigr) \bigl( \HA_{w_{1}} - \HA_{w_{2}} \bigr)
      - \bigl( \HA_{w_{1},1} 
            + \HA_{w_{1},-1} 
            - \HA_{w_{2},1} 
            - \HA_{w_{2},-1} 
      \bigr) 
\nonumber \\ && \times
\bigl\{ \ln (1-k^2) - 2 \ln(k) \bigr\}
\biggr]
+ \big(2+\big(3-\frac{1}{k^2}\big) z\big)
\biggl[ 
      -\frac{8}{3} \bigl( \HA_{-1}^3 + \HA_1^3 \bigr)
      - 32 \HA_{-1,1} \HA_{-1}
\nonumber \\ &&
      + 32 \HA_{-1,0,1}
      + 64 \HA_{-1,1,0}
      + 64 \HA_{-1,1,1}
      + 32 \HA_{-1,-1,0}
      + 64 \HA_{-1,-1,1}
      - 32 \HA_{0,1,1}
\nonumber \\ &&
      + 16 \bigl[ \ln (1-z) - \ln( 1 - k^2 ) \bigr]  \bigl( \ln ^2(2) - \zeta_2 \bigr)
      + 8 \bigl( \HA_{-1} - 2 \HA_0 \bigr) \HA_1^2
      + 8 \bigl( \HA_{-1}^2 
\nonumber \\ &&
+ 4 \HA_{0,1} - 4 \HA_{-1,0} - 4 \HA_{-1,1} \bigr)  \HA_1
      - 8 \bigl[ \ln(1-k^2) - 2 \ln(k) \bigr] 
      \bigl\{
                    2 \HA_{-1}\HA_1
                  - 4 \HA_{-1,1}
\nonumber \\ &&
                  + \HA_1^2
                  - \HA_{-1}^2
      \bigr\}            
\biggr]\Biggr\}
        ~,
\end{eqnarray}
with the polynomials
\begin{eqnarray}
R_8&=&99 k^6-297 k^5+270 k^4-18 k^3-77 k^2+39 k-8,
\\
R_9&=&k^4+k^2 (3 z+2)+6 z-3,
\\
R_{10}&=&9 k^8+48 k^6 (3 z-2)+k^4 (214-552 z)+48 k^2 (9 z-5)-24 z+17,
\\
R_{11}&=&9 k^8+48 k^6 (3 z-4)+6 k^4 (4 z+57)-16 k^2 (9 z+1)-24 z+17,
\\
R_{12}&=&3 k^4-2 k^2 (9 z+2)+18 z-7,
\\
R_{13}&=&3 k^6+k^4 (48 z-47)+k^2 (77-72 z)+24 z-17,
\\
R_{14}&=&-9 k^8-48 k^6 (3 z-2)+k^4 (552 z-214)-48 k^2 (9 z-5)+24 z-17,
\\
R_{15}&=&3 k^4 \bigl(z^2-z-3\bigr)+2 k^2 z^2-z^2,
\\
R_{16}&=&3 k^4 \bigl(z^2+z-1\bigr)+2 k^2 z^2-z^2,
\\
R_{17}&=&2 k^4+k^2 \bigl(2 z^2+9 z+12\bigr)-2 z^2,
\\
R_{18}&=&9 k^4 z (z+3)+2 k^2 \bigl(3 z^2-9 z+5\bigr)-3 z^2+3 z-2,
\\
R_{19}&=&3 k^4-k^2 \bigl(6 z^2+6 z+7\bigr)+2 z^2,
\\
R_{20}&=&-3 k^4+k^2 \bigl(6 z^2+6 z+7\bigr)-2 z^2,
\\
R_{21}&=&6 k^6 (\beta  (z-1)+1)+k^4 \bigl(14 (\beta -1)-2 (6 \beta -5) z^3+3 z^2-2 (\beta -15) z\bigr)
\nonumber \\ &&
+k^2 z^2 (-4 \beta +4 (\beta -1) z+3)+2 z^3,
\\
R_{22}&=&6 k^6 (\beta  (z-1)-1)-k^4 \bigl(-14 (\beta +1)+2 (6 \beta +5) z^3+3 z^2+2 (\beta +15) z\bigr)
\nonumber \\ &&
+k^2 z^2 (-4 \beta +4 (\beta +1) z-3)-2 z^3~.
\end{eqnarray}

\section{Remarks on the encountered iterated integrals}
\label{sec:A4}
\vspace*{1mm}
\noindent
In this calculation a large number of generalized iterated integrals appear.
If no elliptic letter is present, it is possible to represent them using harmonic polylogarithms when the letters 
do not involve kinematic variables or polylogarithms at involved arguments. The expressions become large already 
in simple situations. In total about 1050 logarithms, di- and trilogarithms contribute. In a series of cases a further
elliptic letter is integrated over these structures. 

A few examples are given in the following. Let us refer to the letters $f_{w_9}$ and $f_{w_6}$.
The corresponding iterated integral reads
\begin{eqnarray}
H_{w_9,w_6}(\beta) &=&
\frac{1-\beta ^2 (1-z)}{2 k (1-z)^2 z (z+1)}
   \Biggl\{
-\text{Li}_2\Biggl[\frac{\sqrt{z+1} (k+z)}
   {z\sqrt{z+1}+k\left((1-z)\sqrt{z \beta ^2+1}+\sqrt{z+1}\right)}
\Biggr]
\nonumber\\ &&
+\text{Li}_2
   \Biggl[\frac{\sqrt{z+1} ((z-1)
   \beta  k+k+z)}
{z\sqrt{z+1}+k\left((1-z)\sqrt{z \beta ^2+1}+\sqrt{z+1}\right)}
\Biggr]
\nonumber\\ &&
-\text{Li}_2
   \Biggl[\frac{\sqrt{z+1}
   (k+z)}
{z\sqrt{z+1}-k\left((1-z)\sqrt{z \beta ^2+1}-\sqrt{z+1}\right)}
\Biggr]
\nonumber\\ &&
+\text{Li}_2
   \Biggl[\frac{\sqrt{z+1} ((z-1)
   \beta  k+k+z)}
{z\sqrt{z+1}-k\left((1-z)\sqrt{z \beta ^2+1}-\sqrt{z+1}\right)}
\Biggr]
+\ln (k+z)
   \Biggl\{
-\ln \left(1-\beta ^2\right)
\nonumber\\ &&
-\ln \left(-\frac{k(z-1) \sqrt{\beta ^2 z+1}}{k\left(-z \sqrt{\beta ^2z+1}+\sqrt{\beta ^2
   z+1}+\sqrt{z+1}\right)+\sqrt{z+1
   } z}\right)
\nonumber\\ &&
-\ln \left(\frac{k
   (z-1) \sqrt{\beta ^2 z+1}}{k
   \left(-(1-z) \sqrt{\beta ^2
   z+1} + \sqrt{z+1}\right)+z\sqrt{z+1
   }}\right)
+\ln \left(\beta ^2
   z+1\right)\Biggr\}
\nonumber\\ &&
+\ln (\beta  k
   (z-1)+k+z) 
\nonumber\\ && \times
\Biggl\{\ln
   \left(-\frac{k (z-1)
   \left(\sqrt{\beta ^2 z+1}+\beta 
   \sqrt{z+1}\right)}{k \left((1-z)
   \sqrt{\beta ^2
   z+1}+\sqrt{z+1}\right)+z\sqrt{z+1
   }}\right)
\nonumber\\ &&
+\ln \left(\frac{k
   (z-1) \left(\sqrt{\beta ^2
   z+1}-\beta  \sqrt{z+1}\right)}{k
   \left(-(1-z) \sqrt{\beta ^2
   z+1} + \sqrt{z+1}\right)+z\sqrt{z+1
   }}\right)\Biggr\}
\Biggr\}.
\end{eqnarray}
Examples of the contributing functions are
\begin{eqnarray}
&&\Li_2\left(
\frac{\sqrt{1 + z} (k + z)}{
 z \sqrt{1+z} + 
  k \left(\sqrt{1 + z} - \sqrt{1 + z \beta^2} + z \sqrt{1 + z \beta^2}\right)}
\right),
\\
&&\Li_2\left(
\frac{ k \sqrt{1-z^2}(-z+k(1+(1-z)\beta)}
{-z k  \sqrt{1-z^2} +  k(k \sqrt{1-z^2} + \sqrt{k^2-z^2}(1-z))}
\right),
\\
&&\Li_3\left(
-\frac{2 (1-k) z \beta}
      {(1-\beta) (z-k (1+(1-z) \beta))}\right)
\end{eqnarray}
and logarithms of similar arguments.

Finally, we expand one of the iterated integrals, containing an elliptic letter, in the ratio $m^2/Q^2$. 
While the asymptotic expansion of the functions in Appendix~\ref{sec:A3} is straight forward after the
integration into polylogarithmic expressions, the asymptotic expansion of the Kummer-elliptic integrals
is more involved.
Here we rely heavily on the techniques developed in the context of Ref.~\cite{Blumlein:2019srk} for the
expansion of massive iterative integrals in the Drell--Yan process.
The main idea is to perform the first integration analytically and then regularize the
integrand in the limit $Q^2 \gg m^2$ before the expansion.
Since we aim for a deeper expansion in this paper, the term for the regularization turns out to be a
power series in $\kappa$.
For example, we find
\begin{eqnarray}
\HA_{w_{10},w_7}(\beta) &=& 
\frac{1}{1-z}
\biggl\{
 \frac{1}{4} \ln^2\left( \textcolor{blue}{\frac{m^2}{Q^2}} \right)
+\frac{1}{2} \bigl(
         \ln (1-z)
        -\ln(2)
        -2 \ln \big(1-\sqrt{z}\big)
\bigr) \ln \left( \textcolor{blue}{\frac{m^2}{Q^2}} \right)
\nonumber \\ &&
+\biggl(
        2 \ln (1-z)
        -\frac{5}{4} \ln(z)
\biggr) \ln \big(1-\sqrt{z}\big)
-\frac{3}{4} \ln^2\big(1-\sqrt{z}\big)
-\ln^2(1-z)
\nonumber \\ &&
+\frac{1}{2} \ln(1-z) \ln(z)
-\frac{1}{16} \ln^2(z)
-\text{Li}_2\big(1-\sqrt{z}\big)
-\text{Li}_2\big(\sqrt{z}\big)
-\frac{1}{2} \text{Li}_2\left(\frac{2 \sqrt{z}}{1+\sqrt{z}}\right)
\nonumber \\ &&
-\text{Li}_2\left(\frac{1}{2} \big(1-\sqrt{z}\big)\right)
-\frac{1}{2} \text{Li}_2\left(-\frac{1-\sqrt{z}}{2 \sqrt{z}}\right)
+\frac{11}{4} \zeta_2
+ \frac{1}{4} \biggl(
          6 \ln (1-z)
\nonumber \\ &&
        - 6 \ln \big(1-\sqrt{z}\big)
        - \ln (z)
\biggr) \ln(2)
-\frac{1}{4} \ln^2(2)
+ \textcolor{blue}{\frac{m^2}{Q^2}}  
\biggl[
         \frac{1}{2} \ln^2\left( \textcolor{blue}{\frac{m^2}{Q^2}} \right)
\nonumber \\ &&
        -\biggl(
                  \frac{5-10 \sqrt{z}-3 z}{4 \big(1-z \big)}
                +2 \ln \big(1-\sqrt{z}\big)
                - \ln (1-z)
                + \ln (2)
        \biggr) \ln \left( \textcolor{blue}{\frac{m^2}{Q^2}} \right)
\nonumber \\ &&
        +\frac{1-8 \sqrt{z}+z}{4 \big( 1 - z \big)}
        +\biggl(
                \frac{5+6 \sqrt{z}-3 z}{2 \big(1 - z \big)}
                +4 \ln (1-z)
                -\frac{5}{2} \ln(z)
        \biggr) \ln \big(1-\sqrt{z}\big)
\nonumber \\ &&
        -\frac{3}{2} \ln^2\big(1-\sqrt{z}\big)
        -\biggl(
                 \frac{5+22 \sqrt{z}-3 z}{4 \big( 1 - z \big)}
                -\ln (z)
        \biggr) \ln (1-z)
        -2 \ln^2(1-z)
\nonumber \\ &&
        -\frac{1}{8} \ln^2(z)
        -2 \text{Li}_2\big(1-\sqrt{z}\big)
        -2 \text{Li}_2\big(\sqrt{z}\big)
        -\text{Li}_2\left(\frac{2 \sqrt{z}}{1+\sqrt{z}}\right)
\nonumber \\ &&
        -2 \text{Li}_2\left(\frac{1}{2} \big(1-\sqrt{z}\big)\right)
        -\text{Li}_2\left(-\frac{1-\sqrt{z}}{2 \sqrt{z}}\right)
        +\frac{2}{\big( 1 - z \big)} \sqrt{z} \ln (z)
        +\frac{11}{2} \zeta_2
\nonumber \\ &&
        +\biggl(
                \frac{3+10 \sqrt{z}-z}{2 \big( 1 - z \big)}
                -3 \ln \big(1-\sqrt{z}\big)
                +3 \ln (1-z)
                -\frac{1}{2} \ln(z)
        \biggr) \ln (2)
        -\frac{1}{2} \ln ^2(2)
\biggr]
\nonumber \\ &&
+ \left( \textcolor{blue}{\frac{m^2}{Q^2}} \right)^2 
\biggl[
        -\frac{1}{2} \ln^2\left( \textcolor{blue}{\frac{m^2}{Q^2}} \right)
        +\biggl(
                -\frac{15(1+z^2)-6 z-100 \sqrt{z}(1+z)}{32 \big( 1 - z \big)^2} 
                +2 \ln \big(1-\sqrt{z}\big)
\nonumber \\ &&
                -\ln (1-z)
                +\ln (2)
        \biggr) \ln \left( \textcolor{blue}{\frac{m^2}{Q^2}} \right)
        +\biggl(
                \frac{15-6 z+15 z^2+28 \sqrt{z}+28 z^{3/2}}{16 \big( 1-z \big)^2}
                -4 \ln (1-z)
\nonumber \\ &&
                +\frac{5}{2} \ln(z)
        \biggr) \ln\big(1-\sqrt{z}\big)
        +\frac{3}{2} \ln^2\big(1-\sqrt{z}\big)
        +\biggl(
                -\frac{3 \big(5-2 z+5 z^2+52 \sqrt{z}+52 z^{3/2}\big) }{32 \big( 1 - z \big)^2} 
\nonumber \\ &&
                -\ln (z)
        \biggr) \ln (1-z)
        +2 \ln^2(1-z)
        +\frac{1}{8} \ln^2(z)
        +2 \text{Li}_2\big(1-\sqrt{z}\big)
        +\text{Li}_2\left(-\frac{1-\sqrt{z}}{2 \sqrt{z}}\right)
\nonumber \\ &&
        +2 \text{Li}_2\big(\sqrt{z}\big)
        +\text{Li}_2\left(\frac{2 \sqrt{z}}{1+\sqrt{z}}\right)
        +2 \text{Li}_2\left(\frac{1}{2} \big(1-\sqrt{z}\big)\right)
        +\frac{2 (1+z)}{\big(1 - z \big)^2} \sqrt{z} \ln(z)
\nonumber \\ &&
        +\frac{97-202 z+33 z^2-324 \sqrt{z}+316 z^{3/2}}{64 \big( 1 - z \big)^2} 
        +\biggl(
                \frac{7(1+z^2)+10 z+60 \sqrt{z}(1+z)}{16 \big( 1 - z \big)^2} 
\nonumber \\ &&
                +3 \ln \big(1-\sqrt{z}\big)
                -3 \ln (1-z)
                +\frac{1}{2} \ln(z)
        \biggr) \ln (2)
        -\frac{11}{2} \zeta_2
        +\frac{1}{2} \ln^2(2)
\biggr] 
\biggr\}
\nonumber\\ &&
+ O ( \kappa^3 \ln^2(\kappa) ),
\end{eqnarray}
and similar expressions for the other Kummer-elliptic integrals. When calculating the complete expansion
all dependence on $\sqrt{z}$ drops out of the Wilson coefficients. We did not exploit here the well-known
relations for the dilogarithm of different arguments \cite{LEWIN}.

\vspace*{3mm}
\noindent
{\bf Acknowledgment.}\\
We thank would like to thank J.~Ablinger, A.~Behring, and C.~Schneider for discussions. This work has been funded in part by 
EU TMR network SAGEX agreement No. 764850 (Marie Sk\l{}odowska-Curie),  COST action CA16201: Unraveling new physics 
at the LHC through the precision frontier and from the Austrian FWF grants P 27229 and P 31952 in part.
The Feynman diagrams have been drawn using {\tt Axodraw 2} \cite{Collins:2016aya}.



\begin{thebibliography}{100}
%
\bibitem{Laenen:1992zk}
  E.~Laenen, S.~Riemersma, J.~Smith and W.L.~van Neerven,
  Nucl.\ Phys.\ B {\bf 392} (1993) 162--228.
%
\bibitem{Laenen:1992xs}
  E.~Laenen, S.~Riemersma, J.~Smith and W.L.~van Neerven,
  Nucl.\ Phys.\ B {\bf 392} (1993) 229--250.
%
\bibitem{Riemersma:1994hv}
  S.~Riemersma, J.~Smith and W.L.~van Neerven,
  Phys.\ Lett.\ B {\bf 347} (1995) 143--151
  [hep-ph/9411431].
%
\bibitem{Hekhorn:2018ywm}
  F.~Hekhorn and M.~Stratmann,
  Phys.\ Rev.\ D {\bf 98} (2018) no.1,  014018
  [arXiv:1805.09026 [hep-ph]].
%
\bibitem{Buza:1995ie}
  M.~Buza, Y.~Matiounine, J.~Smith, R.~Migneron and W.~L.~van Neerven,
  Nucl.\ Phys.\ B {\bf 472} (1996) 611--658
  [hep-ph/9601302].
%
\bibitem{Blumlein:2016xcy}
  J.~Bl\"umlein, G.~Falcioni and A.~De Freitas,
  Nucl.\ Phys.\ B {\bf 910} (2016) 568--617
  [arXiv:1605.05541 [hep-ph]].
%
\bibitem{Gehrmann:2001pz}
T.~Gehrmann and E.~Remiddi, { Comput. Phys. Commun.} {\bf 141} (2001) 296--312,
[hep-ph/0107173].
%
\bibitem{Maitre:2005uu} 
D.~Maitre, { Comput. Phys. Commun.} {\bf 174} (2006) 222--240,
[hep-ph/0507152].
%
\bibitem{Ablinger:2018sat}
  J.~Ablinger, J.~Bl\"umlein, M.~Round and C.~Schneider,
  arXiv:1809.07084 [hep-ph], Comput. Phys. Commun. (2019) in print.
%
\bibitem{Vermaseren:1998uu}
  J.A.M.~Vermaseren,
  Int.\ J.\ Mod.\ Phys.\ A {\bf 14} (1999) 2037--2076
  [hep-ph/9806280].
%
\bibitem{Blumlein:1998if}
  J.~Bl\"umlein and S.~Kurth,
  Phys.\ Rev.\ D {\bf 60} (1999) 014018
  [hep-ph/9810241].
%
\bibitem{Blumlein:2006mh}
  J.~Bl\"umlein, A.~De Freitas, W.L.~van Neerven and S.~Klein,
  Nucl.\ Phys.\ B {\bf 755} (2006) 272--285
  [hep-ph/0608024].
%
\bibitem{Bierenbaum:2009mv}
  I.~Bierenbaum, J.~Bl\"umlein and S.~Klein,
  Nucl.\ Phys.\ B {\bf 820} (2009) 417--482
  [arXiv:0904.3563 [hep-ph]].
%
\bibitem{Ablinger:2010ty}
  J.~Ablinger, J.~Bl\"umlein, S.~Klein, C.~Schneider and F.~Wi\ss{}brock,
  Nucl.\ Phys.\ B {\bf 844} (2011) 26--54
  [arXiv:1008.3347 [hep-ph]].
%
\bibitem{Ablinger:2014vwa}
J.~Ablinger, A.~Behring, J.~Bl\"umlein, A.~De Freitas, A.~Hasselhuhn, A.~von Manteuffel,
M.~Round, C.~Schneider, and F.~Wi\ss{}brock,
  Nucl.\ Phys.\ B {\bf 886} (2014) 733--823
  [arXiv:1406.4654 [hep-ph]].
%
\bibitem{Ablinger:2014nga}
  J.~Ablinger, A.~Behring, J.~Bl\"umlein, A.~De Freitas, A.~von Manteuffel and C.~Schneider,
  Nucl.\ Phys.\ B {\bf 890} (2014) 48--151
  [arXiv:1409.1135 [hep-ph]].
%
\bibitem{Ablinger:2014lka}
  J.~Ablinger, J.~Bl\"umlein, A.~De Freitas, A.~Hasselhuhn, A.~von Manteuffel, M.~Round,
  C.~Schneider and F.~Wi\ss{}brock,
  Nucl.\ Phys.\ B {\bf 882} (2014) 263--288
  [arXiv:1402.0359 [hep-ph]].
%
\bibitem{AGG}
J.~Ablinger, A.~Behring, J.~Bl\"umlein, A.~De Freitas, A.~von Manteuffel, and
C.~Schneider, DESY 15--112.
%
\bibitem{Behring:2014eya}
  A.~Behring, I.~Bierenbaum, J.~Bl\"umlein, A.~De Freitas, S.~Klein and F.~Wi\ss{}brock,
  Eur.\ Phys.\ J.\ C {\bf 74} (2014) no.9,  3033
  [arXiv:1403.6356 [hep-ph]].
%
\bibitem{Ablinger:2017ptf}
  J.~Bl\"umlein, J.~Ablinger, A.~Behring, A.~De Freitas, A.~von Manteuffel, and C.~Schneider,
  PoS (QCDEV2017) 031
  [arXiv:1711.07957 [hep-ph]].
%
\bibitem{Ablinger:2017err}
  J.~Ablinger, J.~Bl\"umlein, A.~De Freitas, A.~Hasselhuhn, C.~Schneider and F.~Wi\ss{}brock,
  Nucl.\ Phys.\ B {\bf 921} (2017) 585--688
  [arXiv:1705.07030 [hep-ph]].
%
\bibitem{Ablinger:2017xml}
  J.~Ablinger, J.~Bl\"umlein, A.~De Freitas, C.~Schneider and K.~Sch\"onwald,
  Nucl.\ Phys.\ B {\bf 927} (2018) 339--367
  [arXiv:1711.06717 [hep-ph]].
%
\bibitem{Ablinger:2018brx}
  J.~Ablinger, J.~Bl\"umlein, A.~De Freitas, A.~Goedicke, C.~Schneider and K.~Sch\"onwald,
  Nucl.\ Phys.\ B {\bf 932} (2018) 129--240
  [arXiv:1804.02226 [hep-ph]].
%
\bibitem{Kazakov:1987jk}
  D.I.~Kazakov and A.V.~Kotikov,
  { Nucl.\ Phys.}\ B {\bf 307} (1988) 721--762
   [Erratum-ibid.\ B {\bf 345} (1990) 299].
%
\bibitem{Kazakov:1990fu}
  D.I.~Kazakov, A.V.~Kotikov, G.~Parente, O.A.~Sampayo and J.~Sanchez Guillen,
  { Phys.\ Rev.\ Lett.}\  {\bf 65} (1990) 1535--1538
   [Erratum-ibid.\  {\bf 65} (1990) 2921].
%
\bibitem{SanchezGuillen:1990iq}
  J.~Sanchez Guillen, J.~Miramontes, M.~Miramontes, G.~Parente and O.A.~Sampayo,
  { Nucl.\ Phys.}\ B {\bf 353} (1991) 337--345.
%
\bibitem{vanNeerven:1991nn}   
  W.L.~van Neerven and E.B.~Zijlstra,
  {Phys.\ Lett.}\ B {\bf 272} (1991) 127--133.
%
\bibitem{Zijlstra:1992kj}
  E.B.~Zijlstra and W.L.~van Neerven,
  {Phys.\ Lett.}\ B {\bf 297} (1992) 377--384.
%
\bibitem{Larin:1991fv} 
  S.A.~Larin and J.A.M.~Vermaseren,
  {Z.\ Phys.}\ C {\bf 57} (1993) 93--98.
%
\bibitem{Moch:1999eb}
  S.~Moch and J.A.M.~Vermaseren,
  {Nucl.\ Phys.}\ B {\bf 573} (2000) 853--907
  [hep-ph/9912355].
%
\bibitem{Vermaseren:2005qc}
  J.A.M.~Vermaseren, A.~Vogt and S.~Moch,
  Nucl.\ Phys.\ B {\bf 724} (2005) 3--182
  [hep-ph/0504242].
%
\bibitem{DUDE}
  A.~Devoto and D.W.~Duke,
  Riv.\ Nuovo Cim.\  {\bf 7N6} (1984) 1--39.
%
\bibitem{LEWIN}
L.~Lewin, {\sf Dilogarithms and associated functions}, (Macdonald, London, 1958);
\\
L.~Lewin, {\sf Polylogarithms and associated functions}, (North Holland, New York, 1981).
%
\bibitem{RaabRegensburger}
 C.G.~Raab and G.~Regensburger,
 \textit{The fundamental theorem of calculus in differential rings},
 in preparation.
%
\bibitem{Ablinger:2014bra}
  J.~Ablinger, J.~Bl\"umlein, C.~G.~Raab and C.~Schneider,
  J.\ Math.\ Phys.\  {\bf 55} (2014) 112301
  [arXiv:1407.1822 [hep-th]].
%
\bibitem{Bierenbaum:2007qe}
  I.~Bierenbaum, J.~Bl\"umlein and S.~Klein,
  Nucl.\ Phys.\ B {\bf 780} (2007) 40--75
  [hep-ph/0703285].
%
\bibitem{FORM}
  J.A.M.~Vermaseren,
  {\it New features of FORM},
  math-ph/0010025;\\
  M.~Tentyukov and J.A.M.~Vermaseren,
  Comput.\ Phys.\ Commun.\  {\bf 181} (2010) 1419--1427
  [hep-ph/0702279].
%
\bibitem{Schneider:2007a}
C.~Schneider, { S\'em.~Lothar. Combin.\/} {\bf 56} (2007) 1--36,
{article B56b}.
%
\bibitem{Schneider:2013a}
 C.~Schneider, in: {\sf Computer Algebra in Quantum Field Theory: Integration,
  Summation and Special Functions}, Texts and Monographs in Symbolic
  Computation eds. C.~Schneider and J.~Bl{\"u}mlein (Springer, Wien, 2013),
  325--360 [arXiv:1304.4134 [cs.SC]].
%
\bibitem{Ablinger:2010pb}
  J.~Ablinger, J.~Bl\"umlein, S.~Klein and C.~Schneider,
  Nucl.\ Phys.\ Proc.\ Suppl.\  {\bf 205-206} (2010) 110--115
  [arXiv:1006.4797 [math-ph]].
%
\bibitem{Schneider:2013zna}
  C.~Schneider,
  J.\ Phys.\ Conf.\ Ser.\  {\bf 523} (2014) 012037
  [arXiv:1310.0160 [cs.SC]].
%
\bibitem{Ablinger:2014rba}
  J.~Ablinger,
  PoS LL {\bf 2014} (2014) 019
  [arXiv:1407.6180 [cs.SC]].
%
\bibitem{Ablinger:2010kw}
  J.~Ablinger,
  {\sf A Computer Algebra Toolbox for Harmonic Sums Related to Particle Physics}, Master Thesis, JKU Linz,
  arXiv:1011.1176 [math-ph].
%
\bibitem{Ablinger:2013hcp}
  J.~Ablinger,
  {\sf Computer Algebra Algorithms for Special Functions in Particle Physics}, PhD Thesis, JKU Linz,
  arXiv:1305.0687 [math-ph].
%
\bibitem{Ablinger:2011te}
  J.~Ablinger, J.~Bl\"umlein and C.~Schneider,
  J.\ Math.\ Phys.\  {\bf 52} (2011) 102301
  [arXiv:1105.6063 [math-ph]].
%
\bibitem{Ablinger:2013cf}
  J.~Ablinger, J.~Bl\"umlein and C.~Schneider,
  J.\ Math.\ Phys.\  {\bf 54} (2013) 082301
  [arXiv:1302.0378 [math-ph]].
%
\bibitem{Ablinger:2017Mellin}
J.~Ablinger, PoS (RADCOR2017) 001 [arXiv:1801.01039 [cs.SC]].
%
\bibitem{Landshoff:1971xb}
  P.V.~Landshoff and J.C.~Polkinghorne,
  Phys.\ Rept.\  {\bf 5} (1972) 1--55.
%
\bibitem{Jackson:1989ph}
  J.D.~Jackson, G.G.~Ross and R.G.~Roberts,
  Phys.\ Lett.\ B {\bf 226} (1989) 159--166.
%
\bibitem{Roberts:1996ub}
  R.G.~Roberts and G.G.~Ross,
  Phys.\ Lett.\ B {\bf 373} (1996) 235--245
  [hep-ph/9601235].
%
\bibitem{Blumlein:1996tp}
  J.~Bl\"umlein and N.~Kochelev,
  Phys.\ Lett.\ B {\bf 381} (1996) 296--304
  [hep-ph/9603397].
%
\bibitem{Blumlein:2003wk}
  J.~Bl\"umlein, V.~Ravindran and W.L.~van Neerven,
  Phys.\ Rev.\ D {\bf 68} (2003) 114004
  [hep-ph/0304292].
%
\bibitem{Witten:1975bh}
  E.~Witten,
  Nucl.\ Phys.\ B {\bf 104} (1976) 445--476.
%
\bibitem{Babcock:1977fi}
  J.~Babcock, D.W.~Sivers and S.~Wolfram,
  Phys.\ Rev.\ D {\bf 18} (1978) 162--181.
%
\bibitem{Shifman:1977yb}
  M.A.~Shifman, A.I.~Vainshtein and V.I.~Zakharov,
  Nucl.\ Phys.\ B {\bf 136} (1978) 157--176
   [Yad.\ Fiz.\  {\bf 27} (1978) 455--469].
%
\bibitem{Leveille:1978px}
  J.~P.~Leveille and T.~J.~Weiler,
  Nucl.\ Phys.\ B {\bf 147} (1979) 147--173.
%
\bibitem{Gluck:1980cp}
  M.~Gl\"uck, E.~Hoffmann and E.~Reya,
  Z.\ Phys.\ C {\bf 13} (1982) 119--130.
%
\bibitem{Remiddi:1999ew}
E.~Remiddi and J.A.M.~Vermaseren,
{{ Int. J. Mod. Phys.}
  {\bfseries A15} (2000) 725--754},
{{[hep-ph/9905237]}}.
%
\bibitem{Zee:1974du}
  A.~Zee, F.~Wilczek and S.B.~Treiman,
  Phys.\ Rev.\ D {\bf 10} (1974) 2881--2891.
%
\bibitem{Furmanski:1981cw}
  W.~Furmanski and R.~Petronzio,
  Z.\ Phys.\ C {\bf 11} (1982) 293--314, and references given therein.
%
\bibitem{Zijlstra:1992qd}
  E.B.~Zijlstra and W.L.~van Neerven,
  {Nucl.\ Phys.}\ B {\bf 383} (1992) 525--574.
%
\bibitem{Gross:1974cs}
  D.J.~Gross and F.~Wilczek,
  Phys.\ Rev.\ D {\bf 9} (1974) 980--993.
%
\bibitem{Georgi:1951sr}
  H.~Georgi and H.D.~Politzer,
  Phys.\ Rev.\ D {\bf 9} (1974) 416--420.
%
\bibitem{Blumlein:2012bf}
  J.~Bl\"umlein,
  Prog.\ Part.\ Nucl.\ Phys.\  {\bf 69} (2013) 28--84
  [arXiv:1208.6087 [hep-ph]].
%
\bibitem{Blumlein:2018cms}
  J.~Bl\"umlein and C.~Schneider,
  Int.\ J.\ Mod.\ Phys.\ A {\bf 33} (2018) no.17,  1830015
  [arXiv:1809.02889 [hep-ph]].
%
\bibitem{RAAB1}
 C.G.~Raab, unpublished.
%
\bibitem{KOUTSCHAN}
 C.~Koutschan,
 \textit{HolonomicFunctions (User's Guide)}.
 Technical report no. 10-01 in RISC Report Series, University of Linz, Austria, Jan.~2010.\\
 \url{http://www.risc.uni-linz.ac.at/publications/download/risc_3934/hf.pdf} 
%
\bibitem{Raab}
 C.G.~Raab,
 \textit{On the arithmetic of d'Alembertian functions},
 in preparation.
%
\bibitem{GuoRegensburgerRosenkranz}
 Li Guo, G.~Regensburger, and M.~Rosenkranz,
 J. Pure and Applied Algebra {\bf 218} (2014) 456--473.
%
\bibitem{KUMPO}
E.E.~Kummer,
Journal f\"ur die reine und angewandte Mathematik (Crelle)
{\bf 21} (1840) 74--90;\\
H.~Poincar\'{e},
{Acta Math.}  {\bf 4} (1884)  201--312.
%
\bibitem{ELLI1}
F.G.~Tricomi, {\sf Elliptische Funktionen}, (Geest \& Portig, Leipzig, 1948); \"ubersetzt und
bearbeitet von M.~Krafft;\\
E.T.~Whittaker and G.N.~Watson, {\sf A Course of Modern Analysis}, (Cambridge University Press, Cambridge, 1996); 
reprint of the 4th edition (1927).
%
\bibitem{ELLI2}
A.~Sabry,
Nucl. Phys. {\bf 33} (1962) 401--430;\\
  D.J.~Broadhurst,
  Z.\ Phys.\ C {\bf 47} (1990) 115--124;\\
  S.~Laporta and E.~Remiddi,
  Nucl.\ Phys.\ B {\bf 704} (2005) 349--386
  [hep-ph/0406160];\\
S.~Bloch and P. Vanhove,
J. Number Theor. {\bf 148} (2015) 328--364
[arXiv:1309.5865 [hep-th]];\\
  L.~Adams, C.~Bogner and S.~Weinzierl,
  J.\ Math.\ Phys.\  {\bf 56} (2015) no.7,  072303
  [arXiv:1504.03255 [hep-ph]];\\
  J.~Ablinger, J.~Bl\"umlein, A.~De Freitas, M.~van Hoeij, E.~Imamoglu, C.~G.~Raab,
  C.-S.~Radu and C.~Schneider,
  J.\ Math.\ Phys.\  {\bf 59} (2018) no.6,  062305
  [arXiv:1706.01299 [hep-th]];
J.~Br\"odel, C.~Duhr, F.~Dulat, and L.~Tancredi,
\newblock JHEP {\bf 05} (2018) 093, [arXiv:1712.07089].
%
\bibitem{Alekhin:2017kpj}
  S.~Alekhin, J.~Bl\"umlein, S.~Moch and R.~Placakyte,
  Phys.\ Rev.\ D {\bf 96} (2017) no.1,  014011
  [arXiv:1701.05838 [hep-ph]].
%
\bibitem{Buckley:2014ana}
  A.~Buckley, J.~Ferrando, S.~Lloyd, K.~Nordstr\"om, B.~Page, M.~R\"ufenacht, M.~Sch\"onherr and G.~Watt,
  Eur.\ Phys.\ J.\ C {\bf 75} (2015) 132
  [arXiv:1412.7420 [hep-ph]].
%
\bibitem{Alekhin:2003ev}
  S.I.~Alekhin and J.~Bl\"umlein,
  Phys.\ Lett.\ B {\bf 594} (2004) 299--307
  [hep-ph/0404034].
%
\bibitem{Beenakker:1988bq}
  W.~Beenakker, H.~Kuijf, W.L.~van Neerven and J.~Smith,
  Phys.\ Rev.\ D {\bf 40} (1989) 54--82.
%
\bibitem{Matsuura:1988sm}
  T.~Matsuura, S.~C.~van der Marck and W.~L.~van Neerven,
  Nucl.\ Phys.\ B {\bf 319} (1989) 570-622.
%
\bibitem{Hamberg:1990np}
  R.~Hamberg, W.L.~van Neerven and T.~Matsuura,
  Nucl.\ Phys.\ B {\bf 359} (1991) 343--505
   Erratum: [Nucl.\ Phys.\ B {\bf 644} (2002) 403].
%
\bibitem{HYPERGEOM}
 F.~Klein, {\sf Vorlesungen \"uber die hypergeometrische Funktion},
  Wintersemester 1893/94, Die Grundlehren der Mathematischen Wissenschaften
  {\bf 39}, (Springer, Berlin, 1933);\\
 W.N. Bailey, {\sf Generalized Hypergeometric Series}, (Cambridge University
  Press, Cambridge, 1935);\\
 L.J.~Slater, {\sf Generalized hypergeometric functions}, (Cambridge University
 Press, Cambridge, 1966).
%
\bibitem{Blumlein:2019srk}
  J.~Bl\"umlein, A.~De Freitas, C.G.~Raab and K.~Sch\"onwald,
  Phys.\ Lett.\ B {\bf 791} (2019) 206--209
  [arXiv:1901.08018 [hep-ph]].
%
\bibitem{Collins:2016aya}
  J.C.~Collins and J.A.M.~Vermaseren,
  {\it Axodraw Version 2},
  arXiv:1606.01177 [cs.OH].
\end{thebibliography}
\end{document}